\input amstex
\magnification1200
\loadbold
\input arrow.tex
\documentstyle{amsppt}
\topmatter
\title
Lectures on Topological Aspects of Theoretical Physics
\endtitle
\author
Leonid D. Lantsman
\endauthor
\affil
Wissenschaftliche Gesellschaft bei \linebreak
Judische Gemeinde zu Rostock \linebreak
Wilhelm-K\" ulz Platz, 6.\linebreak
18055 Rostock.\linebreak
e-mail: llantsman\@freenet.de
\endaffil
\abstract
This series of lectures is planned as a generalization of author's large
(more than fifteen years) experience of work in the theoretical physics.
The successful work of physicist-theorist is unthinkable without of good
understanding of topological aspects of physical theory.
The modern theoretical physics is based on the group-theoretical approach
which generates the formalism of the principal fibre bundles and
the instanton approach. The latter is based on the Pontrjagin's degree
of map theorem and this theorem is the original ``bridge'' between
homology and cohomology theories. This is the rough sketch of connections
of modern physics with modern topological theories. Indeed, these
connections are very complicated and very interesting and are the
object of these lectures. The author plans to devote his two first
lectures to fibre bundle theory: this is the foundation on which the
modern physics rests -- the theory of gauge groups and the Yang-Mills
fields. The idea of connection and curvature (first of all for the
principal fibre bundles) will be given also. The lectures are devoted
to the Pontrjagin's degree of map theorem, to the theories of monopoles
and instantons, to the theory of the topological index of the elliptical
operator. The information accumulated to this moment allows us to apply
these theories to some questions of conformal anomalies and to the
topological aspects of QCD. More comprehensive questions of modern
topology (for example the algebraic (co)homology theories, the theory
of the spectral sequences) will be expounded in the further lectures.
On the author's opinion, these lectures will be useful for the
physicists-theorists of all directions of modern physics.
\endabstract
\endtopmatter
\pagebreak
\document
\vsize8.7truein
\hsize6.5truein
\head {\bf Introduction}
\endhead
The author stated already his purposes in the abstract. He wants to add
here that these lections are the extract of the beautiful
monographs (on the author's  opinion these books are the best
in the sphere of  the modern mathematics in the XX
century) of the authors which names are the actually legendary names
in the modern mathematics: M.M. Postnikov, A.N. Maltsev, A.S. Schwartz,
S. Kobayashi, K. Nomizu, R.M. Switzer, F. Hirzebruch, R.S. Palais,
R. Solovay, M.F. Atiyah, R.T. Seely and many others.
\head {\bf 1. Fibre Bundles}
\endhead
The main principle of gauge field theories  is  the invariancy of the
action under the gauge (local) transformations.
These transformations form as a rule the Lee group with the algebra
Lee.Therefore
we should  begin our story from the general definition \it of the
 Lee group and the Lee algebra\rm .
\definition {\bf Definition 1.1}[1] The Lee group this is the group
which is  in the same time is the smooth manifold such that the group
operation
$ (a,b)\to ab^{-1} \in G$ ; $ a,b \in G$ is the smooth map from
$G\times G$ in G.
\enddefinition
Let us denote as $ L_a$ ($ R_a$ correspondingly) the left (the right
correspondingly) displacements on G realised with the element
$ a \in G$ :$ L_a x = ax$ ($ R_a x = xa$ correspondingly) for every
$x \in G $. Then we can  define the inner authomorphism ad a for
$ a \in G$ as a $ (ad a)x =axa^{-1}$ and  for every $x \in G $.
This authomorphism is called the \it adjoint authomorphism \rm .
\definition{\bf Definition1.2} The tangential vector field X on G
is called the left-invariant (the right-invariant correspondingly) if
it is invariant with respect  to all the left displacements $ L_a$
(the all right displacements $ R_a$ correspondingly).
\enddefinition
We always consider the above vector field X as a differentiable field.
\definition{\bf Definition 1.3} We define the Lee algebra g  of the
group Lee G as a set of the all left-invariant vector fields
on G with the usual adding, with the multiplication on the scalar and
with the Lee bracket. As a vector space g  is isomorphic to the
tangent space $ T_e (G)$ in  u .
 This isomorphism is given with the map which associates the vector
 $ X_e$, the value X in
e to the field $X \in  g$.
\enddefinition
 One can consider also [2,p17] the notion of the {\it topologic group}.
 It is sufficiently then to consider the maps in the definition 1.1
 as a
continuous maps only. The category of the topologic groups is the more
weak category then the Lee groups.\par
 We define in  conclusion of our "Lee group" theme what it is the
 \it free (\it effective\rm ) action of the
 Lee group on the manifold M.\par
We say that G acts free(correspondingly effective) if $ R_a x=x$
for all $x\in M$ (for the some $x\in M$ correspondingly)
involves a=e.
After our definitions of the Lee groups and the Lee algebras we can
devote ourselves to the study of the fibre bundle theory.
\definition{\bf Definition 1.4} [2,p.13] The  triad
$$
\zeta = (E,\pi,B) \tag 1.1
$$
where E and B are the topologic spaces and $\pi :E\to B$ is the
continuous map (the map "on") is called the fibre bundle over B.
The space
B is called the base of the  fibre bundle $\zeta$, the  space E
is called the total space of the
fibre bundle $\zeta$ ,and the map $\pi $ is called the (canonical)
projection of the  fibre bundle $\zeta$.
\enddefinition
\definition {\bf Definition 1.5} The counterimage of the arbitrary
point $ b \in B :{\Cal F} = \pi^{-1} (b)$ is called the fibre of
the fibre
bundle $\zeta $ over the point b .
\enddefinition
\definition{\bf Definition 1.6} The fibre bundle  $\zeta = (E,\pi,B)$
is called the fibre bundle with the typical fibre if the
fibres
${\Cal F}_1,{\Cal F}_2 $
where $b_1,b_2 \in B$ are homeomorphic.
\enddefinition
Now we give the classification of the fibre bundles which are very
important for the physicist- theorist.\par
\it The trivial fibre bundle\rm .
\definition{\bf Definition 1.7}The fibre bundle
$\zeta = (E,B,{\Cal F}, \pi,)$ is called the  trivial fibre bundle
if it is
equivalent to the some direct product,
i.e. it exists the  topologic map $\lambda : E \to Bx
{\Cal F}$ where ${\Cal F}$  is the topologic  space; and the  fibre
over the point $b\in B$ in the fibre bundle
$(E,B,{\Cal F},\pi $ turns into the  fibre  over the same point in the
fibre bundle
$(Bx {\Cal F},B,  {\Cal F},\pi_1)$ .
\enddefinition
\it The principal fibre bundle\rm[1] .\par
Let M  be the manifold and let G be the  Lee group. \it The principal
fibre bundle over \rm M  \it with the structural group \rm G
consists of the
 manifold P and the action of the group  G on P
 which  satisfies to the following conditions :
\roster
\item G acts \it  free \rm on P to the right (the free action of
the group G it is the such action of the group G  when $ R_a x = x $
 for the some
 $x \in M$  involves a=e ). We can write down this action
as a
$$
(u,a) \in P\times G \to ua = R_a \in P
\tag 2.1
$$
\item M is the factor-space for P \it by the equivalence relation \rm
which is induced with the group G.Thus M is divided onto the
 equivalence classes with respect to  the  group G. These equivalence
  classes
are called \it the orbits \rm of the  group G. This notion has the
great  importance in theoretical physics, since it  is translated on
 the language
of  the Hamiltonian formalism [3] as a \it  classes
of the equivalent trajectories \rm which  set  forms \it the physical
 sector of the theory\rm.One can write down  M= P/G. We consider usually
the case of the smooth manifold M, hence the
 canonical projection $ \pi : P\to M$ is smooth ;
\item P is \it local  trivial \rm ,i.e. the every point $ x \in M$ has
 the neighbourhood U such that $\Psi=\pi ^{-1}(U)\to U\times G $ is the
diffeomorphism,which is called
\it the trivialisation\rm. If we
define the map
$$
\phi (u): =\pi ^{-1}(u)\to G
\tag 3.1
$$
which satisfies the  conditions
$$
\phi  (ua) = \phi  (u) a
\tag 4.1
$$
for every $ u \in =\pi ^{-1}( U) $ and $ a \in G $,
that $ \Psi (U) = ( \pi (U), \phi  (U) ) $. \par One denotes the
principal fibre bundle usually as $ P(M,G, \pi ) $ or P(M,G) or P.
\endroster
The   fibre of the principal fibre bundle  is the orbit of the group G
 over the fix point $a \in M $. \par
Basing on this definition of the principal fibre bundle we can[1] now
give the definition of the \it trivial principal fibre bundle
\rm .
\par Let again the  manifold M and the group G  be given,
and G acts again free  on $ P = M \times G $ to the  right.\par The map
$ (x,a) \in M \times G \to (x,ab) \in  M\times G $ exists for every
$b \in G$  with $ R_b$. The such obtained principal fibre bundle  is
called the
\it principal  trivial fibre bundle \rm .\par
The definition of the principal trivial fibre bundle  allow us to
discuss briefly (the complete discussion will follow after the
introduction of the
 connections on the principal trivial  fibre bundle
 ). So,we have according to the definition of the principal trivial
 fibre bundle the direct product of the (smooth) manifold M and the
 Lee group G
acting on M as a (local) gauge transformations
which  would  leave the action of the theory \it invariant \rm.
Let the manifold M be  for simplicity \it  the Minkowsky space\rm.
This is the
 situation of the all gauge  theories which not
include the gravitation. The structure of the direct product induces
on the total space of the principal trivial fibre bundle the some topology
 different from the topology  of the flat
 Minkowsky space. One can construct always the some metrical space
 isomorphic to this direct  product. The latter has its local metrics
 in the
 every point which is determines in fact
 with the  structure of the Lee group G .\par
The very beautiful historical example of the such approach to the
principal trivial fibre bundle is the  Kalutza-Klein theory which
was the one
from the attempts to unite the general
relativity with the electromagnetism into  the one theory.
The electromagnetic field was considered in this  theory as a
fifth supplementary
co-ordinate such that \it the circle of the radius
of  Planck length
intersects the  Minkowsky space\rm .We have in this case in the some
neighbourhood of the fixed point of the Minkowsky space
the some direct product of this neighbourhood (which is  isomorphic
to the open ball with the centre in above point) on the sphere $ S^1$
 and this  direct product  generates the following metrics :\par
With account of the   fifth supplementary co-ordinate  we use the
indexes $\hat { \mu} :\hat { \mu} =0,1,2,3,5$ for these co-ordinates.
 Here as usual $ \mu =0,1,2,3 $ the Minkowsky indexes and
$ \mu =5 $ is the index correspond to the fifth co-ordinate.\par
Then according to the Oscar Klein's supposition (1926 y.) we can
represent the  metrics of this \it five-dimensional world \rm as a
$$
\hat {g}_{\hat \mu \hat \nu}=\left ( \matrix g_ {\mu \nu}-
 \phi A_{\mu} A_{\nu} & -\phi A_{\mu} \\ -\sigma  & A_{\nu} -
\phi
\endmatrix
\right )
\tag 5.1
$$
where $A_{\mu}$ are the vector electromagnetic potential and $ \phi $
is the scalar potential. Thus we can see as the structure
 $M \times S^1$ generates in this model the non-trivial metrics
of  the four-dimensional world. We ought to pay  here the especial
attention on the character of the components $ \hat {g}_{\mu,5}$
and $ \hat {g}_{5,\mu }$ which are the projections
of the fifth co-ordinate on  the four-dimensional world.\par
This theory is somewhat naive, it has a lot of the shortcomings,
but this conception of  the direct product ( and the sphere $ S^1$
is isomorphic to the U(1) \it group of the two-dimensional
rotations \rm) is
applied to the principal trivial fibre bundle as for the   U(1)
gauge theory,- for the electromagnetism ,as for the nonabelian
Yang-Mills theories.
The idea of the sphere $ S^1$
of the Planck radius was the first  example  in the history of
modern physics of the application of  the compact extra dimensions to
the description of the gauge field theory.This
idea of the \it compactification \rm found its application later on:
with the development of the strengs and p-branes theories.\par
 The author is familiar with the people which apparently
were the pioneers in the investigation of the \it spontaneous
compactification \rm of the extra dimensions in the striengs and
p-branes theories.
This is the scientific leader of my postgraduate studentship V.I. Tkach.
 The above investigations are the fruit of the joint work of  D.V.Volkov
which to his
death in 1996 y. was at the head of the Kharkov school of the
theoretical physics , V.I. Tkach  and D.P.Sorokin. It will useful to
recommend  our reader the works [4],[5] of these authors or
the very interesting work[6] of I.P.Volobuev,Ju.A.Kubyshin ,
J.M.Mourao
and G.Rudolph. I want to note in conclusion of this theme that
the understanding of the  four-dimensional structure of the total
space E of the principal fibre bundle is the main part of such theories.
\par
The author want to dwell also on the global symmetries.
There are the such transformations which group parameters $\epsilon$ are
\it  not depend on the space co-ordinates\rm.\ It is
means, on the language of the principal trivial fibre bundle, that
we (for example  for the Minkowsky space) deals with the direct
product of the base space M and \it the some constant
\rm . If we fix the some point $ p\in  M$ and consider the open
ball-shaped  neighbourhood $U_r$ of this point then the typical
fibre over p is
 the constant and we have the topology of the \it
general cylinder \rm over  $U_r$. This is the geometrical sense  of
the global symmetries. \par
If the principal fibre bundle P(M,G) is given that the action of the
group G on P generates the homomorphism between the Lee algebra of the
 group G  and the Lee algebra of the vectors
fields ${\Cal X }(p)$. \par We hope that our reader is acquainted
with the features of the vectors. Then we can devote ourselves to the
following consideration.\par
Let $\phi$ be the some gauge
transformation on M. On the other hand  exists always the \it integral
 curve \rm $ x(t)\in M$ to which the vector X is touched
 in the point $ x_0\ x(t)$ where t is the parameter of this curve.
 The
fixing of the point  $ x_0 $  is equivalent to the choice of the
integration constant, i.e \it to the fixing of the  Cauchy
condition \rm . We shall denote this  vector X as a $ X{_x(t_0)} $.\par
Let us introduce now the \it one-parameter group $\phi _t$\it  of the
\rm (\it smooth \rm )  \it transformations \rm in M as a such
map
 {\bf R} $\times M \to M : (t,p )\in $ {\bf R} $\times x M \to
 \phi _t (p) \in M $
which satisfied  the t one-parameter group $\phi _t$\it  of the
 \rm (\it smooth \rm )  \it transformations \rm in M as a such
map
 {\bf R} $\times M \to M : (t,p )\in $ {\bf R} $\times x M \to
  \phi _t (p) \in M $
which satisfied  the following conditions:
\roster
\item The map $ \phi_t : p\to \phi_t (p)$ is the transformation
 in M  for every $ t\in ${\bf R} ;
\item $\phi_{t+s}(p) =\phi_t(\phi_s (p) )$  for every $ t,s  \in $
 {\bf R}  and $p \in M$.
\endroster
The every one-parameter group $\phi _t$ generates the vector field
X on the following way. $X_ p$ is the vector tangential to the curve
 $x(t)=\phi _t (p)$ for the every point $p \in M$. This
curve as we already this know is \ it the orbit \rm of the
 point p in $p=\phi _0 (p)$. The  orbit $\phi _t(p)$ is the \it integral
curve\rm of the field X issued from  p.\par
The \it local one-parameter group of the  local transformations \rm
 can be defined analogous with the additional condition that $\phi _t(p)$
is defined for the t in the neighborhood of 0 and p
belongs to the \it open set \rm in M. More precisely: let $I_\epsilon$ be
the open interval $(- \epsilon,\epsilon )$ and U is the
open set in M.
\definition{\bf Definition 1.8} The local one-parameter group of the
local transformations defined on $I_\epsilon \times U$ is the
map $ I_\epsilon \times U \to M$ which satisfied  the  following
conditions:\par
1a. $ \phi _t :p \to \phi _t(p)$ is the diffeomorphism U onto the
open set $\phi _t (U)$ in M for every $ t\in  I_\epsilon $ \par
2a. if $t,s,t+s \in I_\epsilon $ and $ p,\phi _s \in U $ then
$$
\phi _{t+s}(p) = \phi_t(\phi_s (p) )
\tag 6.1
$$
\enddefinition
The  local one-parameter group of the  local transformations as in
the case of the  one-parameter group of the  transformations
induces the vector field X defined on U.\par
It is very interesting to prove the \it contrary statement\rm.
\proclaim{\bf Theorem1.1} Let X be the vector field on the manifold M.
There exist the neighbourhood U, the positive number
$\epsilon$ and the local one-parameter group of the  local
transformations $\phi_t :U \to M, t\in I_\epsilon$  for every point
$p_0 \in M$ which generate this X.
\endproclaim
\remark {\bf Remark}
We shall say that X generates  the local one-parameter group of the
local transformations $\phi_t$ in the neighbourhood of
the point $ p_0$. If the (global)one-parameter group of the  local
transformations  generated  X  exists on M  that we say
that X is the \it complete field \rm .
If $ \phi_t (p)$ is defined on $I_\epsilon \times M$ for the some
$\epsilon$ that X is complete.
\endremark
\demo{Proof} Let $u^1,...u^n $ be  the local co-ordinate system in
the neighbourhood W of the point
$ p_0$ such that $ u^1(p_0) =...=  u^n(p_0) =0$. Let
$ X=\sum  \xi ^i  (u^1,...u^n)(\partial/ \partial  u^i) $be  in W.
Let us consider the following system of the usual differential equations:
$$
d f^i/dt = \xi ^i (f^1(t),...f^n(t)),\qquad i=1,...,n
\tag 7.1
$$
with the unknown functions $f^1(t),...f^n(t)$. According to the
basic theorem for the systems of the usual differential equations
it exists the
only set of the functions $f^1(t,u),...f^n(t,u)$ defined for
$u=(u^1,...u^n) $ with $\vert u^j \vert <\delta_1$ and for
$\vert t\vert< \epsilon_1 $
which forms the solution of the differential equation for the every
fixed u and satisfies the initial conditions:
$$
f^i(0;u)= u^i
\tag 8.1
$$
Let us put  $\phi_t(u) = (f^1(t,u),...,f^n(t,u))$ for
$\vert t\vert< \epsilon_1 $ and $ u \in U_1=(u;\vert u^i \vert
<\delta_1)$.
 If $\vert t\vert ,\vert s \vert$ and $\vert t+s \vert $ all less
than $ \epsilon_1 $ and both u and $\phi_ s(u)$ are in U that the
functions
 $ g^i(t)=f^i(t+s;u)$ as it is easy to see are the solutions of the
 differential equation with the initial data $g^i(0)=f^i(s;u)$.
 Because of the
 unique solution we have $ g^i(t)=f^i(t;\phi_s(u))$.
 So we proved that $\phi_t (\phi_s(u)) =\phi_{t+s}(u))$.
 Since $\phi_0$ is the identical transformation in $U_1$ then there
 exist  $\delta >0$
and $\epsilon >0$ that $ fi_t(u) \subset U_1$ for
$U=u:\vert u^i \vert <\delta)$ and $\vert t  \vert < \epsilon$.
Hence $ \phi_{-t}( \phi_t (u))=\phi_t( \phi_{-t}(u)= \phi_0(u)=u$.
Thus $\phi_t$ is the diffeomorohism on U for  $\vert t  \vert <\epsilon$
 and
therefore $\phi_t$ is the local one-parameter group of the  local
transformations defined on $I_ \epsilon\times U$ according to the
definition 1.8. It is evident from the construction of
$fi_t$ that  $fi_t$  generates the vector field  X in U. \qed
\enddemo
The every physicist-theorist know the very important example of
the such local one-parameter group of the unitary local
transformations U(1)  which we seen already when we
considered the Kalutza-Klein theory  . This is the group \it  of
the rotations in the flat \rm (x,y)
or on the complex flat $C^1$. This group is isomorphic  [7] to the
circle $S^1$ which is characterised with the angle
$ \phi ,0\leq \phi_\leq 2\pi$
  and the unitary matrix $ e^{i\phi } $ corresponds
to this angle. Namely this matrix is the function $\phi_t$ of the
theory considered by us above.\par
We shall now adduce (without of proof) the following result which
defines ,in fact, \it the Lee derivative \rm for the  local
one-parameter group of the local transformations
(we hope by this  that our reader know the basic features of the
differential forms;else he can study our further course where
these features
will stated).
\proclaim {\bf Theorem 1.2} Let X and Y be the vector fields on M.
If  X induces the local one-parameter group of the local
transformations $  \phi _t$ then
$$
[X , Y]= \lim _{t\to 0} \frac {1}{t} [Y- (\phi _t)_* Y]
\tag 9.1
$$
where $\phi*$ is the authomorphism  of the algebra ${\Cal D}(M)$ of
the differential forms on M\par
More precisely,
$$
[X , Y]_p = \lim_{t\to 0} \frac {1}{t} [Y_p - (( (\phi _t)_* Y)_p ],
\qquad  p\in M
\tag 10.1
$$
We take this limit  with respect to the natural topology of the
tangential vector space $T_p(M) $.
\endproclaim
Let us now define \it the tensor fields \rm on M.\par
Let $T_x =T_x (M)$ be the tangential  space to the manifold M in
the point x and \bf T\rm (x) is the standard  tensor algebra over
$T_x $ : $ T_x =\sum $\bf T \rm  $^r_s (x)$ where
 \bf T \rm $ ^r_s (x)$  is the tensor space of the   (r,s)
 type over T(x).
\definition{\bf Definition 1.9} The tensor field of the (r,s)
type on the subset $ N\subset M$ is the juxtaposition of the tensor
$K_x \in $\bf T \rm $ ^r_s (x)$ to the every point $x\in M$.
We take $X_i=  \partial /\partial x_i, i=1,...,n$ as a basis for
the every tangential  space $ T_x ,x\in U$ in the co-ordinate
neighbourhood U  with the local co-ordinate system $x^1,...x^n$ ;
 we
introduce also the \it dual \rm basis $\omega^i =d x^i :i=1,...,n$
 as a dual  basis in $ T^*_x$. (so we  introduce first \it
the differential form \rm upon which we shall much dwell in the future).
\enddefinition
The tensor field K of the  (r,s)  defined on U then expressed as
$$
K_x = \sum K^{i_1 ...i_r}_ {j_1...j_s} X_{i_1}\bigotimes
 ...\bigotimes X{i_r} \bigotimes \omega^ {i_1}\bigotimes ...\bigotimes
\omega^{j_s}
\tag 11.1
$$
where $ K^{i_1 ...i_r}_ {j_1...j_s} $ are the functions on U which
are called \it the components \rm for K with respect  to the local
co-ordinate
system $x^1,...x^n$. We say that
K is the  field of the $C^k$ class if  all its components are  the
functions of the $C^k$ class;of course it is necessary to check that
this notion is
 not depends on the local
co-ordinate system . It is easy to do.\par
Let us transform the above basis $X_i$ and its dual basis $\omega^i $
as
$$
X_i = \sum _j A^j_i\bar {X}_ j
\tag 12.1
$$
(this transformation generates the corresponding  transformation of
 the dual basis). Then the components of the
$K^{i_1 ...i_r}_ {j_1...j_s}$ transform as
$$
\bar K^{i_1 ...i_r}_ {j_1...j_s} = \sum A^{i_1}_ {j_1}...A^{i_r}_{j_s}
B ^{m_1} _{j_1}... B^{m_s} _{j_s}
K^{k_1...k_r}_{m_1... m_s}
\tag 13.1
$$
by this transformation. If we substitute now the matrix $ (A^i_j)$
 from (12.1) onto the \it Jacobian matrix \rm of the two  local
 co-ordinate
 systems then we shall prove this. We shall understand
the tensor field of the  $C^\infty$  class as a tensor field by the
further consideration. The  tensor of the  r-type is called
\it the contravariant tensor \rm and the  tensor of the  s-type is
called \it the covariant tensor \rm .Any vector is the tensor of the
r=1 or s=1 type.\par
We now introduce  the \it Lee  derivative \it in the terms of the
tensor field \rm K. Let X be the vector field on M and $\phi_t$ is
the local
 one-parameter group of the local transformations induced
X (we suppose that our reader knows the features of the tensor algebra).
\definition{\bf Definition1.10} Let us suppose for simplicity that
$\phi_t$ is the global one-parameter group of the  transformations
on M. Then $\phi_t$ is the authomorphism of the tensor algebra
${\Cal T}(M)$ for every t. Let us set
$$
(L_X K)_x = \lim_{t\to 0} \frac {1}{t} [K_x -({\tilde \phi}K)_x]
\tag 14.1
$$
The map $L_X $ of  ${\Cal T}(M)$ on itself  which moves K in $ L_X K$
is called \it the Lee differentiation
with respect to  \rm X.
\enddefinition
Let us  prove the following features of the Lee derivative.
\proclaim {\bf Theorem 1.3} The  Lee differentiation  $L_X $ with respect
to the vector field  X satisfied the following conditions:
\roster
\item $L_X $ is the differentiation  for ${\Cal T}(M)$, i.e. it is
linear and satisfied the equality
$$
L_X (K \bigotimes K')=  (L_X K )\bigotimes  K' +K \bigotimes ( L_X K')
\tag 15.1
$$
for all $ K,K'  \in {\Cal T}(M)$ ;
$ L_X $ preserves  the type of the tensor ;
\item $ L_X $  commutes with the every contraction of the tensor
field;
\item $  L_X f =Xf$ for  every function f ;
\item  $ L_X Y =[X,Y] $   for  every vector  field Y.
\endroster
\endproclaim
\demo{\it Proof} It is evident that  $ L_X $ is linear.
Let $\phi_t $ be the local one-parameter group of the
local  transformations generated with the  field X. Then
$$ L_X (K \bigotimes K')= \lim_{t\to 0} \frac {1}{t} [K\bigotimes
K' -{\tilde \phi_t}(K\bigotimes K')] $$
$$ = \lim _{t\to 0} \frac {1}{t} [K \bigotimes K' - (
{\tilde \phi} _tK)\bigotimes  {\tilde \phi_t } K') ]$$
$$ = \lim _{t\to 0} \frac {1}{t} [K \bigotimes K' - ( {\tilde \phi}_tK)
\bigotimes  K')]$$
$$ + \lim _{t\to 0} \frac {1}{t} [( {\tilde \phi} _t K')-
({\tilde \phi} _t K)
 \bigotimes ( {\tilde \phi} _t K')] $$
$$ = ( \lim _{t\to 0 }\frac {1}{t} [K-({\tilde \phi} _t K) ] )
\bigotimes  K' $$
$$ + \lim_{t\to 0}  ({\tilde \phi} _t K)\bigotimes  \frac {1}{t}
[K'- ({\tilde \phi} _t K) ] ) $$
$$ = (L_X K)\bigotimes K' + K\bigotimes (L_X K') $$
Since $ {\tilde \phi} _t $ preserves the type of the tensor and
commutes with the every contraction  then $L_X $ has the same
features. If  f is the function on M then
$$
(L_X f)(x) = \lim _{t\to 0} \frac {1}{t} [f(x)-f(\phi^{-1} _t(x))]
= - \lim _{t\to 0} \frac {1}{t} [f(\phi^{-1} _t x)-f(x)]
\tag 16.1
$$
If we note that $\phi^{-1}_t = \phi_{-t}$ is the local one-parameter
group of the local  transformations generated with the
field  X  that we obtain $L_X f = -(-X) f =X f$. And the last point
of the theorem is the reformulation of the  theorem 1.2.
\qed
\enddemo
We shall prove now the some more theorem which will very useful for
the  Lee groups theory.
\proclaim{\bf Theorem 1.4} Let $\phi$ be the transformation on M. If
the vector field X induces the local one-parameter group of the local
transformations $\phi_t$ then the vector field $\phi* X$ where$\phi*$
is the authomorphism  of the algebra ${\Cal D}(M)$ of the differential
forms on M, induces $ \phi\circ \phi_t \circ \phi^{-1}$.
\endproclaim
\demo{\it Proof}It is evident that $ \phi\circ \phi_t \circ \phi^{-1}$
is the local one-parameter group of the local transformations. This
should
 show that it generates the vector field $\phi* X$ . Let us consider
 the arbitrary point p in M and $q=\phi^{-1}(p)$. Since $\phi_t$
 induces X
 the vector $X_q \in T_q(M) $   touches  with the curve
 $x(t)=\phi_t (q)$ in q=x(0).Hence the vector
$$
(\phi* X)_p = \phi* (X_q) \in T_ p(M)
\tag 17.1
$$
touches with the curve $y(t)=\phi \circ \phi_t (q)=\phi \circ
\phi_t \circ \phi^{-1}(p) $\qed
\enddemo
\proclaim{\bf Corollary 1.5} The vector field X is invariant with
respect to the action of $ \phi$,i.e. \linebreak $\phi* X=X$ when
 $ \phi$ is commutes with $\phi_t $ only .
\endproclaim
Let us return again to the Lee groups and the  Lee algebras.\par
The every $A\in g $ generates the (global)one-parameter group of
the transformations in G. Really if $\phi_t$ is the
local one-parameter group of the local  transformations generated
with A (as this was explained in the theorem 1.1)and $\phi_t e$,
where e is the unit element of the Lee group G is defined for
$\vert t \vert <\epsilon $ then one can define
$\phi_t a$ at $\vert t \vert <\epsilon $ for every $a\in G$ as
$L_a(\phi_t e)$ (this definition is correct because of the
corollary 1.5 and since $A \in  g $ is the left-invariant vector.
Since $\phi_t a$ is defined for $\vert t \vert <\epsilon $ and for
every $a \in G$ then $\phi_t a$ is defined at  $\vert t \vert \infty$
for every $a \in G$ .Let us set $a_t=\phi_t e$. Then
$a_{t+s} =a_t  a_s $ for every $t,s \in $ {\bf R}
according to the definition of the one-parameter group of the local
transformations.We shall call $a_t$ \it the one-parameter
subgroup in\rm G \it generated with the element \rm A. The other
characteristic of $a_t$ is the fact that it is the unique curve in G
such that its tangential vector $\dot a_t$ in $a_t$ is equal to
$L_{a_t} A_e$ and $ a_0 =e$.In other words this is \it the unique
solution of the differential equation \rm $a{-1}_t \dot a_t =A_e$
with the initial condition $a_0 =e$. Let us denote $a_1=\phi_1 e$
as  $\exp A$.Then $\exp tA =a_t$ for all t. The map $A \to\exp A$
from g  into G is called \it the exponential map \rm. The U(1)
group is the one example of the such construction.\par
The every authomorphism $\phi$ of the Lee group G induces the
authomorphism $\phi*$ of the Lee algebra g. Really if  $A\in g$  then
  $ \phi* A$ is again the left-invariant vector field and
  $\phi*[A,B] = [\phi* A,\phi* B] $ for $A,b \in {\Cal g}$.
  In particular the above map
 $ ad a:x\to axa{-1} $ generates the authomorphism in ${\Cal g}$
  for every $a\in G$. We denote this authomorphism also as ad a.
The representation $a \to ad a$, $a\in G$ is called \it the adjoint
representation \rm of the  Lee group G in ${\Cal g}$.
We have $(ad a) A=(R_{a^{-1}})_* A$ for every $a\in G$ and $A\in{\Cal g}$,
since $ axa^{-1}=L_a R_{a^{-1}} x$ and A is left-invariant.
Let $A,B \in {\Cal g}$ and $\phi _t$ is the one-parameter group of the
transformations in G induced with A. Let us set
$a_t=\exp tA= \phi _t (e)$
 as above.Then $\phi _t(x)=x a_t$ for $x\in G$.
According to  theorem 1.2
$$
[B,A] =\lim _{t\to 0} \frac {1}{t}[(\phi _{t*} B-B]
= \lim _{t\to 0} \frac {1}{t}[ad (a^{-1}_t )B -B]
\tag 18.1
$$
Let now the Lee group G acts from the right on the manifold M,
then we introduce the field A* which is such connected with the field
$ A\in g $. The action of the  local one-parameter subgroup of the
local  transformations $a_t=\exp tA $ in the general Lee group G
induces as this was explained in the theorem 1.1 the vector field A* on M.
\proclaim{\bf Theorem 1.6}  The Lee group G acts now from the right on the
 manifold M. The map $\sigma : A\in g \to A* \in {\Cal X}(M) $
(where ${\Cal X}(M) $  is the standard vector algebra on M) is the
homomorphism of the Lee algebras.If G acts effective on M that $\sigma$
is the monomorphism $ g\to {\Cal X}(M)$. If G acts free on M that
$\sigma(A)$ is nowhere equal to zero on M for every non-zero
$A\in  g $ .
\endproclaim
\demo{\it Proof} Let us note firstly that we can define $\sigma$ as
following.Let $\sigma_x$ be the map $a\in G \to xa \in M$ for every
$ x \in M $.Then $(\sigma_x)_* A_e=(\sigma A)_x$ .Hence $\sigma$ is
the linear map from  g in ${\Cal X}(M)$.  That we should  show
that $\sigma$ commutes with the Lee bracket.  Let us suppose that
$ A,B \in  g $ and $ A* =\sigma A ; B*=\sigma B ; a_t=\exp tA $.
According to  theorem 1.2. we have
$$
[A*,B*]=\lim _{t\to 0 }\frac {1}{t}[B*- R_{a_t} B*]
\tag 19.1
$$
Since $R_{a_t }\circ \sigma _{xa_t^{-1}} (c)$ for $c\in G$,
we obtain denoting the differential with the same letter:
$$
(R _{a_t} B*)_x =R _{a_t} \circ \sigma _ {xa^{-1}_t} B_e =
\sigma _ x (ad (a^{-1}_t) B_e)
\tag 20.1
$$
whence
$$
\lbrack A*,B*\rbrack=\lim _{t\to 0} \frac {1}{t}[\sigma _ x
B_e -\sigma _ x (ad (a^{-1}_t) B_e)]$$
$$=\sigma _ x(\lim _{t\to 0} \frac {1}{t}[B_e -ad (a^{-1}_t) B_e])$$
$$ =\sigma _ x([A,B]_e) = (\sigma[A,B])_x $$
because of the formula for [A,B] in the terms of the ad G.
Thus we proved that $ \sigma $ is the homomorphism from the Lee algebra
g into  the Lee algebra ${\Cal X}(M)$. Let us suppose that $\sigma A=0$
everywhere on M. This means that the one-parameter group of the
transformations $R _{a_t}$ is trivial,i.e.$R _{a_t}$ is the identical
transformation on M for every t. If G is effective on M then
$a_t =e$ for every t,hence A=0 and $\sigma$ is indeed the monomorphism.
In conclusuon  we should to prove the last point
of the theorem. Let us suppose that $\sigma A$ is equal to zero
in some point $x\in M$. Then $R _{a_t}$ leaves x immovable for
every t. If G acts free on M then $a_t =e$ for every t,hence A=0.
\qed
\enddemo
Thus we described briefly the principal fibre bundles and
ascertained the connection between the Lee algebras g and ${\Cal X}(M)$.
This will need us  by study of connections at this fibre bundle.\par
The following type of the fibre bundles which we shall study is the
\it vector fibre bundle \rm. We cite here the monograph [2] of
M.M.Postnikov.\par
Let K be the field of the \par
a.  real numbers {\bf R}; \par
b. or the complex numbers {\bf C};\par
c. or the quaternions H ;\par
d. or the octaves O .\par
(These fields are the object of the most interest for the
physicist-theorist). \par
\definition {\bf Definition 1.11} The triad
$$
\zeta = (E,\pi,B)
\tag  21.1
$$
which consists of the topologic spaces E,B and of the continuous map
$$
\pi : E\to B
\tag 22.1
$$
is called the vector fibre bundle over the field K if:\par
a. the set
$$
{\Cal F}=\pi^{-1}(b)
\tag 23.1
$$
i.e.the fibre over the arbitrary point $b\in B$ is the linear vector
space over K;\par
b. (the condition of the local triviality).Always exists the open
covering  {\bf U } of the space B and the such homeomorphism
$\phi_\alpha :U_\alpha \times ${\bf R}$^n \to {\Cal E}_{U_\alpha} $
where ${\Cal E}_{U_\alpha} =\pi ^{-1} U_\alpha $ that the diagram
$$
\sarrowlength=.42 \harrowlength
\commdiag{ U_\alpha \times${\bf R}$^n & \mapright^{\fam6 \phi_\alpha }
        & {\Cal E}_{U_\alpha}\cr
& \arrow (1,-1)\lft{} \quad \arrow (-1,-1)\rt{} \cr
& U_ \alpha  \cr
}
\tag 24.1
$$
is commutative for every point $(b,x)\in U\alpha \times ${\bf R}$^n $
and $\phi_\alpha (b,x) \in {\Cal F} _b $
(lied in the fibre over the point b).;\par
c. the map
$\phi_ {a,b}:${\bf R}$^n \to {\Cal F} _b $
which is defined with the formula
$\phi_ {a,b}(${\bf x}$)= \phi_ \alpha (b,${\bf x}),\linebreak{\bf x}
$\in ${\bf R}$^n $
is the isomorphism of the linear spaces.
\enddefinition
\remark{\bf Remark}
The left arrow in this commutative diagram is the natural projection
$(b,x)\to b $ of the direct product
$ U_ \alpha \times ${\bf R}$^n $ on the first factor of this direct
product and the
right arrow(the map $\pi_\alpha $) is the restriction of the projection
$ \pi  $ on the fibre ${\Cal F}_b $.
\endremark
We have the evident parallel between the principal and the vector
fibre bundles.The above interpretation of the direct product
$U_ \alpha \times ${\bf R}$^n $ (where we now take the open ball
${\Cal B}_r$ of the radius r (in the usual metric space!) as a such
neighbourhood $ U_ \alpha $ ) as a cylinder constructed in this
neighbourhood is acceptable in the both cases.
\definition{\bf Definition 1.12} The dimension n of the vector
fibre bundle $\zeta $ is called the rank of this bundle. We shall denote
 it as
dim  $\zeta $ or  {dim} $_K\zeta $.
\enddefinition
\definition{\bf Definition1.13} The above homeomorphism $\phi_ \alpha$
is called the trivialisation  of the vector fibre bundle
$\zeta $ over the open set $U_ \alpha $. The latter is called the
neighbourhood of the trivialisation. The pair
$(U_ \alpha,\phi_ \alpha $) is also called  the trivialisation in
the some literature. The
covering which consists of the neighbourhoods of the trivialisation
is called the covering of the trivialisation. The family
$\lbrace U_ \alpha,\phi_ \alpha \rbrace ; \alpha \in I$ where I is
the family of the indexes is called  the atlas of the
trivialisation.
\enddefinition
\definition{\bf Definition 1.14} The map
$$
S:B\to E
\tag 25.1
$$
which satisfies the relation
$$
\define \Id {\operatorname{Id}}
\pi \circ s =\Id
\tag 25.1a
$$
where Id is the identical map ,is called the section of the fibre
bundle $\zeta $ .
\enddefinition
This definition is universal for all kinds of the fibre bundles.\par
It is evident that the map $S:B\to E$ is the section of the vector
fibre bundle $\zeta $ if and only if
$ s(b) \in {\Cal F}_b$ for the arbitrary point $b\in B$, i.e. if we
choose the vector s(b) in every fibre ${\Cal F}_b$.
This is the reason to call the sections of the vector fibre bundle
$\zeta $ \it the \rm $\zeta $-\it vector fields\rm on B.\par
The section of the vector fibre bundle $\zeta $  has the following
properties:
a. the formulas
$$
(S_1+s_2)(b) =S_1(b)+s_2(b)
\tag 26.1
$$
$$
(\lambda s)(b) =\lambda s (b)
\tag 26.1a
$$
define the sections $s_1+s_2$ and $ \lambda s $ of the vector fibre
bundle $\zeta $ correspondingly for the arbitrary point $b\in B$
and $\lambda \in K$. Therefore the set $\Gamma \zeta $
of all sections of the vector fibre bundle $\zeta $ is\it  the
linear space over the  field \rm K (all this is evident from the
definition 1.11 of the vector fibre bundle);
b.the formula
$$
(fs)(b)= f(b)s(b)
\tag 27.1
$$
defines the section $fs \in \Gamma \zeta $ for the arbitrary
continuous function f on B, and the \it lineal \rm
 $\Gamma
\zeta $ is \it the module  over \rm B \it and over the algebra
{\bf F}$_K B$ \it  of all continuous functions with their values
 on
\rm K.
\remark{\bf Remark} We want to remind our reader the standard
features of the module  over the field K (loo ,for example the
monograph [8]
 of A.I.Maltsev).\par
The module over the some ring K is the abelian group A for which
one introduces the multiplication on the elements of the ring K has
the following properties
( for example this is the left module with respect to the ring K)
$$
\lambda (a+b) =\lambda a +\lambda b
\tag 28.1a
$$
$$
(\lambda +\mu )a =\lambda a +\mu a
\tag 28.1b
$$
$$
(\lambda\mu )a =\lambda (\mu )a
\tag 28.1c
$$
for $\lambda,\mu \in K $ and $ a,b \in A $.\par
The vectors is one of examples of the modulees .\par
\endremark
The module is called \it unitary \rm if $1.a =a, a\in A $.
\definition{\bf Definition 1.15} The triad $(E_u,\pi_u, B)$ where
$E_u =\pi^{-1}_u ;\pi_u =\pi \vert _u $ is evidently also the vector
 fibre bundle for every subspace $U \subset B$. This triad is called
 the part of the vector fibre bundle $\zeta $ over U and is
denoted as $\zeta \vert _u $.
If  U is the neighbourhood of the trivialisation then  every
trivialisation $\phi :U\times K^n \to {\Cal E}_U $ defines the sections
$s_1,...s_n$ in $\Gamma (\zeta \vert _u ) $ which operate according to
the formula
$$
s_i (b) =\phi (b,e_i),\qquad  i=1,...,n
\tag 29.1
$$
where $e_1,...,e_n$ is the standard basis of the space $K^n$.
Since the vectors $s_i (b)$ form the basis of the linear space
${\Cal F}_b$  the every section $s:U \to {\Cal E}_U$ sets the
functions $s^1,...s^n$ on U satisfied the formula
$$
s (b) = \sum_{i=1}^n s^i(b) s_i
\tag 30.1
$$
\enddefinition
The now stated theory together with the definition 1.11 of the
vector fibre bundle allow the immediate interpretation  in the vector
analysis and in the theory of the differential operators.\par
For example if we interpret now  the functions $s_i (b)$ as a
$\partial / \partial x_i$ then we can interpret s(b) as a
\it linear combination of the partial derivatives\rm.
This interpretation  will stand us in good stead by study of the
topological index theory.\par
We shall give now the two very important  examples of the vector
fibre bundles.
\example{\bf Example 1}\it The trivial vector fibre bundle\rm.\par
The triad $ (B\times V,\pi , B)$ for the
arbitrary topologic space B and for the arbitrary linear n-dimensional
space V over the field K where $ \pi :B\times V \to B$ is the
 projection of this direct product of the first factor is the vector
 fibre bundle from the definition 1.11.
The covering of the trivialisation {\bf U} consists now of the U = B
and the trivialisation $\phi :B\times K^n \to B\times V $
is defined with the choice  of the standard basis $e_1,...,e_n$in V
and is given with the formula
$\phi (b,$ {\bf x})=(b,$\alpha^{-1} (${\bf x}))
for $b \in B$ and {\bf x}$\in ${\bf R}$^n $ where $\alpha :V\to K^n$
is the co-ordinate isomorphism corresponded to the basis
$e_1,...,e_n$.
\endexample
\example{\bf Example 2}\it The tangential vector fibre bundle\rm .\par
Let X is the smooth n-dimensional manifold.{\bf T} X is the manifold
of the tangential vectors on X and $\pi:$ {\bf T} $X \to X$ is the
natural projection which compares the point $p\in X$ to  every vector
$A\in $ {\bf T} X. In definition the fibre $\pi^{-1} (p)$, $p\in X$ of
the projection $\pi$  is the tangential space {\bf T}$_p X$ and the
every \it chart \rm (U,h) of the manifold X where U belongs to the
 covering of X   defines the chart ({\bf T}U,{\bf T}h  of the
 manifold {\bf T}X  for which
{\bf T}U $=\bigsqcup _{p\in U}${\bf T}$_p X=\pi^{-1} U$
and the map {\bf T}h :{\bf T}U $\to$ {\bf R}$^{2n}$ is given with
the formula
{\bf T}h (A)=$(x^1,...,x^n,a^1,...,a^n)$
for $ A\in ${\bf T}U  where $x^1,...,x^n$ are the co-ordinates of
the point $p=\pi (A)$ in the
chart (U,h) and $a^1,...,a^n$ are the co-ordinates of the vector A
in the basis
$$
(\frac {\partial}{\partial x^1})_p,...,(\frac {\partial}{\partial x^n})_p
\tag 31.1
$$
of the space {\bf T}$_p X$ It is convenient to replace the map {\bf T}h
onto the map \par
$(h^{-1}\times id)\circ ${\bf T}h :{\bf T}U $\to U\times
${\bf R}$^n $ \par acting according to the formula
$$
A \to (p,\bar a) ;\quad p =\pi (A) ; \quad {\bar a}= ( a^1,...,a^n)
\tag 32.1
$$
Let
$\phi_h :U\times ${\bf R}$^n \to ${\bf T}U
be the opposite map:\par
$\phi_h (p,\bar a) = a^i \frac {\partial}{\partial x^i}_p ;\quad p\in U
 ;\quad {\bar a} \in ${\bf R}$^n$ \par
 Note that the formula (31.1)coincides in fact with the formula (30.1)
 for the linear combination of the partial derivatives.\par
The map $\phi_h$ is the homeomorphism which closes the commutative
diagram
$$
\sarrowlength=.42 \harrowlength
\commdiag{
 U\times ${\bf R}$^n & \mapright^{\fam6\phi_h}           &
       $   {\bf T}$U    \cr
& \arrow (1,-1)\lft{} \quad \arrow (-1,-1)\rt{} \cr
              &        U    &}
\tag 33.1
$$
Thus $\phi_h$ is the \it trivialisation \rm of the fibre bundle
({\bf T}X,$\pi ,X)$ over  the neighbourhood U.\par
So the triad $\tau_X = $({\bf T}X,$\pi ,X)$ is the vector fibre
bundle of the rank n ,and  it is denoted as $\tau X$ or $\tau (X)$.\par
 We  also grounded  in  this example  the correctness of the formula
 for the partial derivative.
\endexample
In definition  exists always the open covering {\bf U } of the base
space B  in the some vector fibre bundle
$\zeta =(E, \pi ,B)$ consisted of
the neighbourhoods of  the  trivialisation. Let now $U_\alpha$ and
$U_\beta$ be the two crossed elements of  this  open covering. Then
the map
$$
\phi_{ \beta \alpha}= \phi^{-1}_{\beta,b}\circ \phi_{\alpha,b} :
K^n \to K^n
\tag 34.1
$$
where $ \phi_{\alpha,b} $ and $ \phi_{\beta ,b} $  are the maps
{\bf  R}$^n \to ${\bf  F}$_b$ induced with the trivialisations
$\phi_\alpha :U_\alpha
\times  K^n \to  ${\bf  E}$_{U_\alpha }$ and  $\phi_\beta:U_\beta
\times  K^n \to $ {\bf  E}$_{U_\beta }$ from the definition 1.11 is
defined. This
map is \it  linear and has always the opposite map\rm . These matrices
(and all the matrixes in general!) forms the linear group GL(n;K) over
 the field K. Therefore the formula
$$
\phi_{ \beta \alpha}: b\to \phi_{ \beta \alpha}(b)
\tag 35.1
$$
sets the some map
$$
\phi_{ \beta \alpha}: U_\alpha \bigcap U_\beta \to GL(n;K)
\tag 35.1a
$$
which is called \it  the map \rm (\it  or the function \rm ) \it
of the transition \rm from $\phi_\alpha$ to $\phi_\beta$.
\proclaim{\bf Lemma 1.7} The map
$$
\phi : U \to GL(n;K)
\tag 36.1
$$
of the topologic space U into the group GL(n;K) is continuous if
and only if  the map
$$
\hat \phi :U\times K^n \to K^n
\tag 36.1a
$$
given with the formula
$\hat \phi (b,${\bf x})$ = \phi  (b)${\bf x} ;$ b \in U;
\quad ${\bf x} $\in K^n $
is continuous.
\endproclaim
\demo{\it Proof}It is evident that if the map $\phi $ is the map
$\hat \phi $ is  then the map $\hat \phi $ is also continuous.
On the conrary
let the map $\hat \phi $ is continuous. Then  all the maps
$U \to K^n $ of the form
$$
\phi _i :b \to  \phi (b) {\Cal E}_i ,\qquad i= 1,...,n
\tag 37.1
$$
where as usually ${\Cal E}_1,..,{\Cal E}_n$ is the standard basis
of the space $K^n $.Hence all the maps $U \to K$ of the form
$$
\phi ^j_i :b \to  \phi ^j_i (b) ,\qquad i,j= 1,...,n
\tag 38.1
$$
where $\phi ^j_i (b) $ are the components  of the vector $\phi
 (b){\Cal E}_i$ are also continuous. The remark that the numbers
 $\phi ^j_i (b) $
form  the matrix $\phi  (b) \in GL(n;K) $ completes this lemma.\qed
\enddemo
The map $\hat \phi $ at $U=U\alpha \bigcap U_\beta$  and $ \phi  =
 \phi _{\beta  \alpha}$ is non other then the composition
 pr:$\circ ( \phi ^{-1}_\beta \circ \phi _\alpha )$  of the
 homeomorphism $\phi ^{-1}_\beta \circ \phi _\alpha : U\times K^n
 \to U\times K^n $
and the projection  pr: $U\times K^n \to K^n $. This is  the cause
why the map $\hat \phi $ is  continuous. Hence  according to  lemma1.7
the map $\phi _{\beta alpha}$ is also continuous. \par
Thus \it all the maps of transition \rm $\phi _{\beta  \alpha}$  \it
are the continuous maps \rm .\par
The set of  all the maps $U\to G$ for all the sets U and every group
G is the group with respect to the operations
$$
\phi \to \phi^{-1},   (\phi ,\psi)\to \phi \psi
\tag 39.1
$$
defined with the formulas
$$
\phi^{-1}(b)=  \phi (b) ^{-1} , \quad \phi \psi(b) =\phi (b) \psi(b) ,
\quad b\in U
\tag 39.1a
$$
\remark{\bf Remark} We should not   confuse $\phi^{-1}$ with the
opposite map and $\phi \psi$ with the composition of maps.
\endremark
If U  is the topologic space and G is the topologic  group   and
the maps $\phi$ and $\psi$ are \it  continuous \rm
then one can prove  that the maps $\phi^{-1}$  are also \it
continuous\rm.\par
In particular  we can define the map $ \phi^{-1}_{\beta  \alpha} :
U_ \alpha \bigcap U_\beta \to $GL(n;K) for the map
$ \phi_{\beta  \alpha} : U_ \alpha \bigcap U_\beta \to $GL(n;K) and
the map
$$
(\phi_{\gamma \beta}\vert _U)( \phi_{\beta  \alpha}\vert _U) :U
\to GL(n;K)
\tag 40.1
$$
for the maps $\phi_{\beta  \alpha}$ and $ fi_{\gamma \beta}$ in
the case when $U_ \alpha \bigcap U_\beta \bigcap U_\gamma \neq \emptyset $
where $U= U_ \alpha \bigcap U_\beta \bigcap U_\gamma $. It follows,
straightly  from the definition of the maps $\phi_{\beta  \alpha}$
and the usual features  of  the matrices that
$$
\phi^{-1}_{\beta \alpha}= \phi_{ \alpha \beta }
\tag 41.1
$$
on $U_ \alpha \bigcap U_\beta $ and
$$
\phi_{\gamma \beta}fi_{\beta  \alpha}=  \phi_{\gamma \alpha }
\tag 41.1a
$$
on $U_ \alpha \bigcap U_\beta \bigcap U_\gamma $
for  every indexes $\alpha ,\beta, \gamma $  for  which
$U_ \alpha \bigcap U_\beta $\linebreak $\neq \emptyset $ and
$U_ \alpha \bigcap U_\beta \bigcap U_\gamma \neq \emptyset $
correspondingly.
\definition{\bf Definition  1.16} Let B  be the topologic space,
G be the topologic group and ${\Cal U}= \lbrace U_ \alpha \rbrace $  be the
open covering  of the space B. The family $\boldsymbol\phi  =
\lbrace \phi_{\beta  \alpha} \rbrace $  of the continuous maps
$$
\phi _{\beta  \alpha} : U_ \alpha \bigcap U_\beta \to G
\tag 42.1
$$
defined for the some indexes $\alpha ,\beta $ for  which $U_ \alpha
\bigcap U_\beta  \neq \emptyset $ is called the matrix cocycle over
the group
G of the covering ${\Cal U}$ if it satisfies the relations (41.1,41.1a)
\enddefinition
Thus we see that \it every  vector  fibre bundle \rm  $\zeta =(E,
\pi ,B)$ \it defines the some matrix cocycle \rm
$\boldsymbol\phi =\lbrace fi_{\beta  \alpha} \rbrace $ \it  for
every covering  of  trivialisation ${\Cal U}$\it of the  space \rm B.\par
We shall  call  this cocycle -\it  the gluing  cocycle \rm of the
vector  fibre bundle $\zeta$  and shall denote it  with
the symbol
$\boldsymbol \phi_ \zeta$.
\proclaim{\bf Theorem 1.8} Let B  be the topologic space, ${\Cal U}$
be its  open covering  and$\boldsymbol\phi =
\lbrace \phi_{\beta  \alpha} \rbrace $
be the some matrix cocycle over the group GL(n;K) of the covering
${\Cal U}$. Then exists up to isomorphism  the unique vector  fibre
bundle $\zeta$  of the rank n with the base B,with the covering  of
trivialisation ${\Cal U}$ and with the gluing  cocycle $\boldsymbol \phi$.
\endproclaim
\demo{\it Proof}. Let us prove firstly that this vector  fibre
bundle is unique.The statement that two vector  fibre bundles
$ \zeta =(E,\pi,B)$ and
$ \zeta ' =(E',\pi ',B)$ with the covering  of  trivialisation
${\Cal U}$ have the same gluing  cocycle $\boldsymbol \phi$ means
that there
exist the such
trivialisations
$$
\phi_\alpha :U_\alpha \times K^n \to {\Cal E}_{U_\alpha },
\quad \phi '_\alpha :U_\alpha \times K^n \to {\Cal E}'_{U_\alpha }
\tag 43.1
$$
for  these fibre bundles  that for the some indexes $\alpha ,\beta$
with  $U_ \alpha \bigcap U_\beta  \neq \emptyset $  the following
equality is
correct :
$$
{\phi  _\beta }^{-1}\circ  \phi_\alpha ={\phi' _\beta}^{-1} \circ
\phi '_\alpha :(U_ \alpha \bigcap U_\beta ) \times K^n \to :(U_\alpha
\bigcap U_\beta ) \times K^n
\tag 44.1
$$
Therefore the equality
$$
\phi'_\beta \circ \phi^{-1}_\beta =\phi '_\alpha \circ
{\phi^{-1}} _\alpha :
{\Cal E}_{U_ \alpha \bigcap U_\beta}\to {\Cal E}_{U_ \alpha \bigcap
U_\beta}
\tag 45.1
$$
is also correct.\par
Hence the formula
$$
f(p)= (\phi'_\alpha \circ  {\phi_\alpha }^{-1}) (p)
\tag 46.1
$$
for $p\in {\Cal E}_{U_ \alpha }$, i.e. for $\pi (p)\in U_ \alpha $,
defines correct  the some map $f: E \to E'$ which  is the isomorphism
of the
 vector fibre bundles  $ \zeta \to \zeta '$ ( The isomorphism $\phi $
 of the two fibre bundles  $ \zeta \to \zeta '$  is defined with the
 following
commutative
diagram
$$
\CD
 E  @>\phi >>                                         E '\\
@V\pi VV                                             @V\pi'VV \\
 B  @> {\bar{\phi}}>>  B'
\endCD
\tag 47.1
$$
We set  $ \define\id{\operatorname{id}} {\bar \phi}= \id$ in our case;
we can write down this commutative diagram with  the expression
$\pi ' \circ \phi ={\bar \phi} \circ \pi $
which will have the large important  in the future. Thus $ {\bar \phi}$
is continuous automatic. If $ {\bar \phi} $  is continuous then the
 fibre map
$\phi$ is called  \it the
morphism \rm  $ \zeta \to \zeta '$  of the two fibre bundles. And if
the  both   $\phi$  and $ {\bar \phi} $  are the homeomorphisms then
the morphism
$ \zeta \to \zeta '$  is called  \it the isomorphism \rm of the two
fibre bundles $ \zeta $ and $ \zeta ' $ )\par
Let us prove now that  the vector fibre bundles  $ \zeta $ exists.
Let us consider firstly the disjunctive joining up
$$
{\tilde E} = \bigsqcup_\alpha (U_\alpha \times K^n)
\tag 48.1
$$
of the spaces $U_\alpha \times K^n$. Let us denote the point
(b,{\bf x}) $\in U_\alpha \times K^n$ of the space ${\tilde E}
(b \in U_\alpha $,
{\bf x} $\in  K^n$ ) as (b ,{\bf x})$ _\alpha $ and let us introduce
in E the relation $\sim $ considering that
(b,{\bf x})$ _\alpha \sim
(c,${\bf y})$ _\beta $  for $ b \in U_\alpha ,c \in U_\beta $;
{\bf x},{\bf y} $\in K^n $ when and only when c=b  and
{\bf y}=$\phi _{\beta \alpha }(b)${\bf x} , i.e. when {\bf y} and
{\bf x} belong  to the same orbit of the group GL(n;K). It is follows
immediately from these
relations  that this  is the equivalence relation.  Let ${\Cal E}$ is
the corresponding factor-space of the space ${\tilde E}$ supplied
with  the corresponding factor topology.\par
We want to  remind of our reader what is this  the factor
topology- the very important point of the lot of the topological
theories.\par
If [9] the some topologic space X and the some equivalence relation
E on X and the map $ q:X\to X/E$ are given ( the map q associates
the every point $x\in X$ with  its equivalence class $\lbrack x \rbrack
 \in X/E$ ). If we search for the \it good\rm topology then we demand
quite
reasonable that q should be continuous. There exist the most \it fine
\rm topology  in the class of all topologies with respect  to which the
map q
is continuous ; this is the family of  all sets $U\in X$  such that
$q^{-1} (U)$ is open in X. This topology is called  \it the factor
topology \rm .\par
Now we can continue  prove the theorem. The formula
$$
\pi  \lbrack  b, {\bold x}\rbrack _\alpha  =b,  \quad  b\in B, \quad
{\bold x}\in  K^n
\tag 49.1
$$
where $\alpha $ is the such index that $b\in  U_\alpha $ and
$\lbrack  b, {\bold x}\rbrack _\alpha  $ is the equivalence class
of the point
$(b, {\bold x})_\alpha$ defines correct the continuous surjection
$$
\pi  :{\Cal E} \to B
\tag 50.1
$$
The formula
$$
\phi  _\alpha (b, {\bold x}) =\lbrack  b, {\bold x}\rbrack _\alpha ,
\quad    b\in U_\alpha ,   \quad  {\bold x}\in  K^n
\tag 51.1
$$
for  every $\alpha $  defines the continuous map of the fibres
$$
\phi  _\alpha :  U_\alpha \times K^n  \to :{\Cal E}_{ U_\alpha }
\tag 52.1
$$
where $:{\Cal E}_{ U_\alpha }= \pi  ^{-1} U_\alpha $ is the subspace
of the space ${\Cal E}$ consisted of all points of the class
$\lbrack  b, {\bold x}\rbrack _\alpha  ,b\in U_\alpha ,  {\bold x}\in
K^n $.Furthermore it is easy to see that the  formula
$$
\lbrack  b, {\bold x}\rbrack _\alpha \to (b, {\bold x}),\quad b\in
U_\alpha , \quad    {\bold x}\in  K^n
\tag 53.1
$$
defines correct the continuous map ${\Cal E}_{ U_\alpha } \to U_\alpha
\times K^n$ opposite to the map $\phi  _\alpha $. Therefore
$\phi  _\alpha $ is the homeomorphism of the fibres.This means that
the triad $\zeta =(E, \pi ,B)$  satisfies definition 1.11(in the
position b.)\par
In order to satisfy the positions a.and c.of definition 1.11  let
us note that the fibre ${\bold F}_b$ of the map $\pi$ in the some point
$b \in B$ consists as we proved this above of all points of the
class $\lbrack  b, {\bold x}\rbrack _\alpha $  where $\alpha $ is the
arbitrary index for which $b \in U_\alpha $.If besides that
${\bold x}\in U_\beta$ (this is real because of the standard  local
diffeomorphism $U_\alpha \to {\bold R}^m , m\leq n $) then
$\lbrack  b, {\bold x}\rbrack _\alpha  =\lbrack  b, {\bold y}
\rbrack _\beta $ where
${\bold y}=\phi_{\beta \alpha}(b){\bold x}$. Since  the map
$\phi_{\beta \alpha}(b): {\bold R}^n \to {\bold R}^n $ is linear
the formulas
$$
\lbrack  b, {\bold x}\rbrack _\alpha +\lbrack  b, {\bold y}
\rbrack _\alpha = \lbrack  b, {\bold x}+{\bold y}\rbrack _\alpha,
{\bold x}+{\bold y}\in K^n
\tag 54.1
$$
$$
\lambda \lbrack  b, {\bold x}\rbrack _\alpha =\lbrack  b,
\lambda {\bold x}\rbrack _\alpha,   \qquad {\bold x}\in K^n
\tag 55.1
$$
define correct the structure of the linear space in ${\bold F}_b$.
This proves the positions a.and c of the  definition 1.11.\par
Thus $\zeta$ is the vector fibre bundle and the maps $\phi  _\alpha $
are its trivialisations.Besides that
$$
(\phi ^{-1} _\beta \circ \phi  _\alpha )(b, {\bold x}) =
\phi ^{-1} _\beta \lbrack  b,
{\bold x}\rbrack _\alpha =\phi ^{-1} _\beta \lbrack  b,
\phi_{\beta \alpha}(b){\bold x}\rbrack _\beta =(b,\phi_{\beta \alpha}
(b){\bold x})
\tag 56.1
$$
for every point $(b, {\bold x}) \in U_\alpha \bigcap U_\beta $ and
therefore the gluing  cocycle $ {\boldsymbol \phi}_\zeta $ of  this
vector fibre bundle
is this cocycle $ {\boldsymbol \phi} = \lbrack \phi_{\beta \alpha}
\rbrack $.\qed
\enddemo
The described construction explains  in particular why the
cocycle $ {\boldsymbol \phi}_\zeta $ is called the gluing  cocycle.
The maps
$ \phi_{\beta \alpha}$ formed  this cocycle  by the analogous
reasons  are called \it  the gluing  functions \rm .\par
The theorem 1.8 reduces the vector fibre bundles to  its  matrix
cocycles still  not completely  since the cocycle
$ {\boldsymbol \phi}_\zeta $
depends  on the choice  of the trivialisations $ \phi_\alpha $ and
can turn out  the other  at the other choice  of the trivialisations.\par
However  one can control easy this ambiguity.\par
Let $\lbrace \phi_\alpha :U_\alpha \times K^n \to {\Cal E}_{U_\alpha}
 \rbrace $ and
 $\lbrace \phi' _\alpha:U_\alpha \times K^n \to {\Cal E}_{U_\alpha}
 \rbrace $ be the two system of the trivialisations of the vector fibre
bundle
$\zeta $  over the same covering of trivialisation  ${\Cal U}=\lbrace
 U_\alpha\rbrace $ . Then the formula
$$
\gamma_\alpha (b) =\phi _{ \alpha,b}^{-1} \circ \phi ' _{ \alpha,b} :
 K^n \to K^n ,\qquad b\in U_\alpha
\tag 57.1
$$
for every $\alpha$ defines the some map
$$
\gamma_\alpha : U_\alpha \to GL(n;K)
\tag 58.1
$$
which is connected with the homeomorphism
$$
\phi _ \alpha ^{-1}\circ \phi ' _ \alpha :U \times K^n \to U \times K^n
\tag 59.1
$$
with the relation
$$
(\phi ^{-1}_ \alpha\circ \phi ' _ \alpha )(b){\bold x}) =
(b,\gamma_\alpha (b) {\bold x}) , \quad b\in U_\alpha,
\quad{\bold x}\in K^n
\tag 60.1
$$
therefore  because of the lemma 1.7  it  is the continuous map.\par
According to the construction
$$
\phi '_{\beta \alpha }(b) = {\phi '_{\beta ,b} }^{-1}\circ
\phi ' _{ \alpha ,b} = $$
$$ ={\phi '_{\beta,b}}^{-1}\circ \phi _{\beta,b}\circ {\phi{\beta,b}
}^{-1}\circ $$
$$ \circ \phi_{\alpha,b}\circ \phi_{\alpha,b}^{-1}\circ \phi ' _{
 \alpha,b} = $$
$$=\gamma _\beta ^{-1}(b)\circ \phi_{\beta \alpha }(b) \circ
\gamma _\alpha (b)
$$
i.e.
$$
\phi '_{\beta \alpha } =  \gamma _\beta ^{-1} \phi_{\beta \alpha}
\gamma _\alpha
\tag 61.1
$$
in the group of all continuous maps $U_\alpha \bigcap U_\beta \to
GL(n;K) $ (where of course we mean under  $ \gamma _\alpha $
and $ \gamma _\beta$ theirs  restrictions on $ U_\alpha \bigcap
U_\beta $). Formula  defines the group automorphism of the group
GL(n;K).
\definition{\bf Definition  1.17} We shall say that two cocycles
${\boldsymbol \phi }=\lbrace\phi_{\beta \alpha} \rbrace $
and ${\boldsymbol \phi}' =\lbrace\phi '_{\beta \alpha} \rbrace $
of  the covering ${\Cal U}$  over the group G are cohomological  if
there exist the such continuous maps
$$
\gamma _\alpha : U_\alpha \to G
\tag 62.1
$$
that  the relation (61.1) is fulfilled  for every indexes $\alpha,\beta$
 with $U_\alpha \bigcap U_\beta \neq \emptyset $
\enddefinition
The just proved statement mean in this terminology \it  that the
gluing  cocycles of  the same vector fibre bundle \rm $\zeta $
\it corresponded
 the different trivialisations \rm $\phi _\alpha $ (\it but the same
 covering  of  trivialisation \rm  ${\Cal U}$ ) \it  are cohomological
\rm .\par
The relation of cohomology of cocycles is evident \it  the relation of
equivalence\rm .The corresponding classes are called
\it   the  cohomology
classes \rm of  the covering ${\Cal U}$  over the group G. We shall
denote the cohomology class of  the  cocycle ${\bold \phi }$ with  the
symbol
$ \lbrack {\bold \phi }\rbrack $  and the set of  all  cohomology
classes of  the  covering ${\Cal U}$  over the group G with  the symbol
$H^1 ({\Cal U}:G)$.
\remark{\bf Remark}
 We shall   expound in our further lections the notion of  the
 cohomology classes for the differential forms. This is the other
 notion then
the
cohomology classes for the covering ${\Cal U}$  over the group G,
but the idea of  the equivalence classes underlains  in the  both
theories. We
can also note that the set $H^1 ({\Cal U}:G)$ has not the group
structure in contrast  to cohomology group of  the differential forms
as we shall  make sure in this.
\endremark
The following theorem is now evident.
\proclaim{\bf Theorem 1.9.} The formula
$$
\zeta \to \lbrack {\boldsymbol \phi }\rbrack
\tag 63.1
$$
sets  the bijective accordance between the set of  the
equivalence classes of  the isomorphical  vector fibre bundles
$ \zeta $ of  the
rank n over the space B had the given covering  of  trivialisation
${\Cal U}$  and the set  $H^1 ({\Cal U};G)$.
\endproclaim
We can add to said in the theorem 1.9  the following  remark. \it
The every cocycle \rm ${\boldsymbol \phi }'=\lbrace \phi '_{\beta \alpha}
\rbrace $
\it  cohomological  to  the gluing  cocycle \rm ${\boldsymbol \phi }$
\it  of  the vector fibre bundle\rm  $\zeta$ \it  is also the gluing
 cocycle of  this
 vector fibre bundle\rm (corresponding to the some new trivialisations
 $\phi '_\alpha : U_\alpha \times K^n \to {\Cal E}_{ U_\alpha}$).
Really  if  the cocycle ${\boldsymbol \phi }= {\boldsymbol \phi }_
\zeta $ corresponds  to the  trivialisations $\phi _\alpha : U_\alpha
\times K^n \to {\Cal E}_{ U_\alpha}$
,and the equalities of  the type (61.1) where $\gamma_ \alpha $
are the maps   takes place   then setting
$$
\phi '_\alpha (b,{\bold x})= \phi _\alpha (b,\gamma _
\alpha (b){\bold x}), \quad   b\in U_\alpha , \quad {\bold x}\in  K^n
\tag 64.1
$$
we obtain the trivialisations $\phi '_\alpha : U_\alpha
\times K^n \to {\Cal E}_{ U_\alpha}$ to which corresponds
the cocycle ${\boldsymbol
\phi }'$\par
Exists always \it  the labelled element \rm $\lbrack
{\Cal E}\rbrack $  in the every set $H^1 ({\Cal U}:G)$ which
is the  cohomology class
of the cocycle ${\Cal E}$ consisted of the constant maps
$U_\alpha \bigcap U_\beta \to G$ every of which maps the all set
$U_\alpha \bigcap U_\beta $ in the unit e of the group G.
It is easy to prove that \it the trivial vector fibre
bundle\rm $\theta ^n_B$
\it corresponds to the class \rm $\lbrack {\Cal E}\rbrack $
\it   in the case \rm G= GL(n;K).
\remark{\bf Remark } If the covering ${\Cal U}'$ is \it
inscribed \rm in the covering ${\Cal U}$ (i.e.
 $U' _ \alpha \subset U _  \beta $ for every indexes $\alpha,\beta$ )
 then the operation  of the maps restriction defines the injective map
$H^1 ({\Cal U};G) \to H^1 ({\Cal U}';G)$  which we can  consider
as an \it embedding   \rm . This allows us to introduce in consideration
the joining  up of  the sets  $H^1 ({\Cal U};G)$  over the all open
coverings ${\Cal U}$  of  the space B. This joining  up is denoted  as
$H^1 (B;G)$ and  is in the bijective correspondence with the set of
the equivalence classes of  the local trivial G- fibre bundles ( at
G= GL(n;K) there are the vector fibre bundles  of  the rank n  over
the space B).
\endremark
Thus the latter remark  builds the bridge between the principal (or
the local trivial G- fibre bundles) and the vector fibre bundles.This
bridge
allows us to consider   \it the\rm K-\it theory \rm - the very
beautiful modern theory which will the base of our consideration of
the theory
of the topologic index of  the  elliptical operator.\par
But we must firstly   define the two utterly important notions
had the crucial significance in all topology. There are the notions
of the
 \it functor \rm and the  \it cellular spaces \rm . It  is  better
 to familiarize oneself  with these  notions by the monograph [10]  of
Robert M.Switzer.\par
So, let us consider the  notion  of the functor .
\definition{\bf  Definition  1.18.}The category consists of the :
\roster
\item some class of  objects (for example, spaces,groups, etc.;
\item sets hom (X,Y) of the morphisms defined on X and with theirs
values in Y for every ordered pair of objects X and Y; if
f $\in $ hom
(X,Y)  then we write down $f: X\to Y$ or $X @>f >> Y$ ;
\item map  {hom}
(Y,Z)$\times $ hom(X,Y) $\to $ hom (X,Z)
called the composition  set  for the ordered triad (X,Y,Z) of objects;
if f$\in$ hom(X,Y) ,g $\in $ hom(Y,Z) then the
image of the pair (g,f) in hom (X,Z) denotes as g$\circ $f
\endroster
\enddefinition
The two axioms must be fulfilled by this :
\proclaim{\bf C1} If f $\in$ hom(X,Y) ,g $\in $ hom(Y,Z),h $\in$ hom(Z,W)
then $ h \circ (g\circ f)= (h\circ g)\circ
f$.
\endproclaim
\proclaim{\bf C2} It exists the such morphism  1$_Y \in $hom(Y,Y) for
every object Y that we have $1_Y \circ g =g$ and
$ h \circ 1_Y =h$ for the some morphisms g$\in $ hom(X,Y) and h $\in $
hom(Y,Z).
\endproclaim
\remark{\bf Remark } The considered above morphism of the two fibre
bundles is the one of the examples of morphisms.
\endremark
\definition{\bf Definition  1.19 } The  functor from the category
${\Cal E}$ in the category ${\Cal D}$ is the correspondence which :
\roster
\item compares the object $ F(X) \in {\Cal D}$  to the every object
$X \in {\Cal E}$ ;
\item compares the morphism  F(f) $\in$ hom $_{{\Cal D}} $( F(X), F(Y) )
to the every morphism  f $ \in $ hom$ _{{\Cal E}}$(X,Y).
\endroster
\enddefinition
The two axioms must be fulfilled by this :
\proclaim{\bf F1} The equality  $F (1_X)= 1_{F(X)}$ takes place for
every object $X \in {\Cal E}$.
\endproclaim
\proclaim{\bf F2}
$$
F(g\circ f)= F(g)\circ F(f) \in hom_{{\Cal D}}( F(X), F(Z) )
\tag 65.1
$$
for every f $\in$ hom$_{{\Cal E}}$(X,Y) ,g $\in $hom $ _{{\Cal E}}$(Y,Z).
\endproclaim
The \it dual \rm to functors notion of \it cofunctors \rm is also very
important in the many applications of differential topology.
\definition{\bf Definition 1.20} The cofunctor F*  from the category
${\Cal E}$ in the category ${\Cal D}$ is the correspondence which :
\roster
\item compares the object F*(X) $\in {\Cal D}$  to the every object
$X \in {\Cal E}$ ;
\item compares the morphism
$$
F*(f) \in hom _{{\Cal D}}(F*(Y),F*(X))
\tag 66.1
$$
to the every morphism  f$\in $hom $_{{\Cal E}}$(X,Y).
\endroster
\enddefinition
The two axioms must be fulfilled by this :
\proclaim{\bf CF1} The equality  $F* (1_X)= 1_{F*(X)}$ takes place
for every object $X \in {\Cal E}$.
\endproclaim
\proclaim{\bf CF2}
$$
F*(g\circ f)= F*(f)\circ F*(g) \in hom_{{\Cal E}}( F*(Z), F*(X) )
\tag 67.1
$$
for every f $\in $ hom $_{{\Cal E}}$(X,Y) ,g $\in $ hom$_{{\Cal E}}$(Y,Z)
\endproclaim
Let us now consider the cellular spaces.
\definition{\bf Definition 1.21.} The  cell division K of the some
space X is the family K= $\lbrace e^n_\alpha: n=0,1,2...; \alpha \in J_n
\rbrace $,
of the subsets in X indexed with the nonnegative integer numbers n and
the elements $ \alpha$ from  the set of indexes $J_n$.
The set $e^n_\alpha$ is called the cell of the dimension n.
\enddefinition
Let us set $K^n = \lbrace e^r_\alpha: r\leq n,\alpha \in J_n \rbrace $.
If we  consider the category ${\Cal T}$ \it of all topologic spaces
and all continuous maps \rm then we consider in definition  that
$ K^n =\emptyset$ for n<0. If we  consider the category ${\Cal TR}$
\it of all topologic spaces with the labelled points \rm ( the labelled
point -it is the point $x_0 \in X$ chosen in X for the some purposes,
for example  this is the origin of the co-ordinates)\it and all
continuous maps preserved  the labelled points  \rm then we  consider the
labelled point $x_0 $ as a cell of the dimension $-\infty$ and
$K^n =\lbrace\lbrace x_0 \rbrace \rbrace, n<0 $.The set $K^n $
is called
\it the  \rm n-\it dimensional frame \rm of the division K \par
Let us set $\vert K^n \vert = \bigcup \Sb r\leq n \\ \alpha \in J_n
\endSb e^r_\alpha $. Let us note that $\vert K^n \vert $ is the subspace
in the space X while K is the family of cells  only. Let us introduce
the following denotations :
$$
{\dot e}^n_\alpha =e^n_\alpha \bigcap \vert K^{n-1} \vert
\tag 68.1
$$
$$
e^{\circ n}_\alpha = e^n_\alpha -{\dot e}^n_\alpha
\tag 68.1a
$$
for every cell $e^n_\alpha $.\par
The set ${\dot e}^n_\alpha $ is called \it  the boundary \rm
of the cell $e^n_\alpha $, and $e^{\circ n}_\alpha$  is called
\it  its interior\rm .
One demands that K  should satisfies the following conditions:
\roster
\item
$$
X= \bigcup  \Sb n_\alpha \endSb e^n_\alpha = \vert K \vert
\tag 69.1
$$
\item
$$
e^{\circ n}_\alpha  \bigcap  e^{\circ n}_\beta \neq \emptyset
\quad \Rightarrow n=m, \quad \alpha = \beta
\tag 69.1a
$$
\item  it  exists the surjective map
$$
{f_\alpha }^n: (D^n, S^ {n-1})\to ({e_\alpha} ^n ,{{\dot e}_\alpha }^n)
\tag 69.1b
$$
( $ D^n$ is the n-dimensional unit \it  disk \rm  and $S^{n-1}$
is the (n-1)-dimensional unit sphere \it  bounds this disk \rm )
for  every cell
 $e^n_\alpha $
which maps homeomorphic $D^{\circ n}_\alpha $  on $e^{\circ n}_\alpha $.
\endroster
The map $f^n_\alpha$  is called \it  the characteristic  map \rm of
the cell  $e^n_\alpha $. \par
It is follows from condition (69.1b) that every cell $e^n_\alpha $
is the compact subset in X  therefore it is closed  since we suppose
that X is the
\it  Hausdorff  space \rm ( every compact subset is closed in the
Hausdorff  space,  look for example [9] ).The conditions (69.1),(69.1a)
mean
that X  is the \it   nonconnected joining up \rm of the interiors
$e^{\circ n}_\alpha$ ( one can prove this with the help of the
induction by the frames).
\definition{\bf Definition 1.22 .}Let us  set
$$
dim K= sup\lbrace n: J_n\neq \emptyset \rbrace
\tag 70.1
$$
The number K can be equal to $\infty $ :\par
The cell $e^m_\beta $  is called \it  the direct border \rm of the
cell $e^n_\alpha $ if
$e^{\circ m}_\beta \bigcap e^n_\alpha  \neq \emptyset $.
Thus every cell $e^n_\alpha $ is the direct border itself, and if
$e^m_\beta $  is the other direct border  then $m <n $. The cell
$e^m_\beta $  is
called  \it the  border\rm of the  cell $e^n_\alpha $ if  it exists
the such sequence of the cells
$$
e^m_\beta =e^{m_0}_{\beta_0},e^{m_1}_{\beta_1},  ... ,
e^{m_s}_{\beta_s}=  e^n_\alpha
\tag 71.1
$$
that  every cell $e^{m_i}_{\beta_i}$ is the direct border of
the $e^{m_{i+1}}_{\beta_{i+1}},0\leq i<s$. The cell is called
\it the principal
\rm if it is not the border of any other cell. For example  if
dim  K= n $< \infty$ then all the n-dimensional cells  are principal.
\enddefinition
\remark{\bf Remark } In general $e^{\circ n}_\alpha$ is  not the open
set in the space X. Furthermore  even the interiors of the principal
cells are not
always the open  sets.  It is connected  with the very bad topology
of the some infinite cellular divisions.
\endremark
\definition{\bf Definition 1.23} We shall say that  the structure of
the cellular space is given  on X if it has the cellular division with the
following features :
\roster
\item every closed cell  has the finite number of the direct borders
only;
\item  X has the weak topology generated with the division K , i.e.
the subset $S \subset X$ is closed if and only if  when the
intersection $S\bigcap e^n_\alpha $ is closed in \linebreak
$e^n_\alpha $( for  every n and every $\alpha \in J_n$).
\endroster
\enddefinition
Since the characteristic  map $f ^n_\alpha$ induces  the
homeomorphism between $e^n_\alpha $ and the factor-space
$D^n / \sim$ (the
denotation (X,A) means always that we consider  the pair $A\subset X$
with the closed  subspace $A\in X$ where A is  squeezed  in the
point); $ x\sim y$ if and only if  when $f ^n_\alpha x =f ^n_\alpha y$
then it is follows from the  (2)  of definition 1.23 that
$S \subset X$ is closed if and only if  when the countrimages
$(f ^n_\alpha)^{-1} S$ are closed  in the disk $D^n$ for every n,
$\alpha$.
Besides that  it is evident that the interiors  of the principal
cells are the open  sets  in the cellular space.\par
We shall  call usually X \it the cellular space\rm if one can
introduce on it the some structure K of the cellular space. Let
us note that
if  X permits even the one cellular structure then one can
introduce on it the many of such structures.\par
The conception of the cellular spaces  suggested first by
J.H.C.Whitehead [11] has the broad application in the modern mathematics.
It  is appropriate to  mention here  the simplexes: it  is in fact
the cellular division of the metrical space. We shall discuss  this
notion in detail in
the future,  but we want to mention here  the great role of simplexes
in the theoretical physics, namely in general relativity.It is
\it Regge
calculation\rm ,i.e.the approximation of the some space-time manifold
with this cellular division [12].  The consideration of the continuous
defformations  of the simplexes \linebreak ( of the Planck  size)
led S.Hawking [13] to the hypothesis of the \it space-time foam \rm
with the
configuration of the \it  gravitational instantons \rm  corresponded
to the minimum  of the gravitation's action.This theory has the
great future affecting all the directions of modern theoretical physics.
The author  of  this lections has  his point of  view on the
problem of  the space-time foam  and the gravitational instantons  which
he will state may be in  the visible future. One can add to this
theme that the S.Hawking's discovery  of  the space-time foam  and the
gravitational instantons  was the event in  the   modern  physics
equal  by importance to the Einstein's discovery of the general
relativity. I think that we in  the 21 century shall estimate
the  all significance of S.Hawking's  discovery.
So we described the necessary mathematical apparatus which we shall
apply now for the study of K-theory.\par
First of all we note that the cellular spaces forms the category
${\Cal RW}$ of the cellular spaces  with the labelled points and
the continuous
maps preserved  the labelled points ( the labelled points here is
the additional, but the important for our statement assumption).\par
The general category ${\Cal RT}$ of  \it  all \rm spaces  with the
labelled points and the continuous  maps preserved  the labelled points
will also necessary us  later on.
\definition{\bf Definition 1.24} The homotopy of the  space X in the
space Y is the continuous map $ F: X\times I \to Y$ where I is the
unit
interval [0,1]. The homotopy F defines the map $ F_t:X  \to Y$ given
with the formula $F_t(x) = F(x,t), x \in X$ for every $t \in I$.
Let $f,g :X \to Y$ be  the  some  maps. One says that f homotopic  g
(and one writes down $f \simeq g$) if it exists the such homotopy
$ F: X\times I \to Y$ that $F_0=f $ and $F_1 =g$. In other words the
maps f and g are homotopic  when and only when f is deformed
continuously  in g  by means of the some family $F_t$.Let  A be  the
some subspace in X. The homotopy F is called the homotopy relatively
A (or  the  homotopy rel A) if F(a,t)=F(a,0) for all $a \in A,t\in I$.
\enddefinition
Let us prove now the one small  but  very important theorem about  the
relation of homotopy.
\proclaim{\bf Theorem 1.10 } The relation of homotopy is the relation
of equivalence.
\endproclaim
\demo{\it Proof} The feature $f\simeq f$ is obvious: we must take
F(x,t)= f(x) for all $x\in X, t\in  I$.If $f \simeq g$ and  F is the
homotopy
from f to g then the map $G:X\times I \to Y$ given with the formula
$G(x,t)= F(x,1-t),x\in X, t\in I$ is the homotopy from g to f   therefore
$fg\simeq f$.  Let  now F be  the homotopy from f to g  and G is the
homotopy from g to h. Then the homotopy H defined with the formula
$$
H(x,t)= \cases F(x,t) , & 0\leq t \leq \frac {1}{2},\\
G(x,2t-1) , & \frac {1}{2}\leq t \leq 1
\endcases
\tag 72.1
$$
sets the homotopy from f to h.\qed
\enddemo
Thus the  relation $\simeq $ divides the set of all continuous maps
$ F: X \to Y$  on the  equivalence classes.These  equivalence classes
are called \it the homotopic classes \rm ;the corresponding set of
all homotopical classes is denoted as [X;Y].The  homotopical class
of  the continuous map $ f: X \to Y$ is denoted as [f].\par
The notion of homotopy is the one from the most of important notions
in the modern mathematics. We shall  use it very broadly in our
lections, for example  this  notion is the crucial detail of the proof
of  Pontrjagin theorem about  degree of map.\par
The just introduced  notion of homotopy and theorem 1.10 allow us to
consider the new category ${\Cal RW}'$.  This category is
obtained from the category ${\Cal RW}$,  but  now we consider as a
continuous maps (morphisms!) the homotopic classes
[X;Y].Then we define  the composition $\lbrack g\rbrack \circ
\lbrack f\rbrack $ as $\lbrack g\circ f\rbrack $  for given
$\lbrack f\rbrack \in $ hom (X,Y), g $\in $ hom (Y,Z). It is easy to
check  that the morphism $\lbrack g\rbrack \circ \lbrack f\rbrack $
is defined correct  and the axioms  C1 and C2 are fulfilled
(as a consequences of  theorem 1.10 ).\par
We introduce similarly  the category ${\Cal RT}'$,- the  homotopical
category for  all labelled toplogical spaces  and the continuous maps
which preserve the labelled points. This category has the above features
of the homotopical classes.
\definition{\bf Definition 1.25} The morphism $\phi :\zeta \to \zeta '$
of  the two principal fibre bundles is the such pair of  the maps
$\phi :E \to  E',\phi :B \to B'$ that $\pi '\circ\phi = \phi \circ \pi$
where $\pi,\pi '$ are the corresponding projections of  the principal
fibre bundles
$ \zeta$ and $ \zeta '$ ,and $\phi (e,g)= \phi (e) g$ for all
$g \in G,e\in E$.  The notion of  equivalence of  the two principal
fibre bundles
$ \zeta$ and $ \zeta '$ is introduced on the following way. One say
that the two principal  fibre bundles are  equivalent if  there the
exists the
such morphisms $\phi :\zeta \to\zeta ' ,\psi:\zeta ' \to\zeta$ that
B=B' and $\phi \circ \psi = \psi \circ \phi =1$.
The definition of  the equivalence of  the two vector  fibre bundles
has the same construction.
\enddefinition
The following series of  the theorems  and the definitions continues
this construction.
\definition{\bf Definition 1.26 }One say that the map $\pi : E\to B$
has the feature of  the covering homotopy relatively the space X  if it
exists the such homotopy $F:X\times I \to E$ for every map $f:X\to E$
and the every homotopy $G:X\times I \to B$ that $f=F _0$
and $\pi \circ F=G$. The homotopy F is called the lift  of  the homotopy G.
\enddefinition
\definition{\bf Definition 1.27}
The family of  the trivialisations  of  the principal  fibre bundle
$\zeta $ forms the atlas $\{( U_\alpha ,  \phi _\alpha)\}
 $ ($ \phi_\alpha:
U_\alpha\times G\to \pi{-1}U_\alpha$ is the trivialisations  of  the
principal  fibre bundle $\zeta $.
\enddefinition
\proclaim{\bf Lemma 1.11} Let  $\zeta $ is the principal  fibre bundle
over  the space B.Then one can  compare the only set of cocycles
${\bar \zeta}=( U_\alpha,\phi_{\alpha\beta}) $to the every atlas
$\{( U_\alpha ,  \phi _\alpha)\} $ for which
$$
\phi_{\beta}(b,g)= \phi_{\alpha}(b, \phi_{\alpha\beta}(b)g),
\quad b \in B, \quad g\in G
\tag 73.1
$$
where $\phi_{\alpha\beta}: U_\alpha\bigcap U_\beta \to G$ are now
the  cocycle over  the group G , and this distinguishes this construction
from  the  was for  the  vector fibre bundles with the  linear group
GL(n;K).
\endproclaim
The  proof of this lemma is analogous to the proof of the theorem
1.8 and we shall not produce this proof here.Our reader can
familiarize himself
with this proof  by the monograph [10].
\proclaim{\bf Lemma 1.12 } If  ${\bar \zeta},{\bar \zeta}'$ are
the two sets of  cocycles associated according to lemma 1.11 with
the atlases
$\{( U_\alpha ,  \phi _\alpha)\} ;\{(  U'_\gamma,\phi '_\gamma)\}$
of the principal fibre bundles \linebreak $\zeta =(B,\pi,E,G)$ ,
$\zeta ' =(B',\pi ',E',G)$, and if
$\phi :\zeta \to \zeta '$ is the morphism of  these fibre bundles
then  it exists the only morphism of  the sets  of  cocycles
$r:{\bar \zeta}\to {\bar \zeta}'$ that ${\bar r}={\bar \phi }$
(where ${\bar r}=B\to B' $)and
$$
\phi \circ \phi_\alpha (b, g)=\phi '_\gamma(\bar \phi )(b),
r_{\gamma \alpha}(b)g),\quad b \in U_\alpha\bigcap U_\beta, \quad g\in G
\tag 74.1
$$
\endproclaim
\demo{\it Proof} The maps
$$
\theta _{\gamma \alpha}={\phi '_{\gamma}}^{-1}\circ \phi \circ
( \phi_ \alpha \vert (U_\alpha\bigcap {\bar \phi }^{-1}U'_\gamma)\times G):
(U_\alpha\bigcap {\bar \phi }U'_\gamma)\times G \to U'_\gamma)\times G
\tag 75.1
$$
for the some indexes $\alpha, \gamma$ satisfy the condition
$\pi_{U'_\gamma}\circ \theta _{\gamma \alpha}={\bar \phi }\circ
\pi_{U_\alpha \bigcap{\bar \phi }^{-1}U'_\gamma}$ and, therefore,
have the form $\theta _{\gamma \alpha}(b,g)=({\bar \phi }(b),
h_{\gamma \alpha}(b,g))$ for the some  $h_{\gamma \alpha}:
(U_\alpha\bigcap {\bar \phi }^{-1}U'_\gamma)\times G \to G$.\par
Whence
$$
\phi \circ  \phi_ \alpha (b,g)= \phi '_ \gamma ({\bar \phi }(b),
h_{\gamma \alpha}(b,g)), \quad b\in  U_\alpha\bigcap
{\bar \phi }^{-1}U'_\gamma ,
\quad  g\in  G
\tag 76.1
$$
But then $ \phi '_ \gamma ({\bar \phi }(b), h_{\gamma
\alpha}(b,g)) =\phi \circ  \phi_ \alpha (b,g)= ( \phi \circ
\phi_ \alpha(b,1) ) \circ g =
\phi '_ \gamma ({\bar \phi }(b), h_{\gamma \alpha}(b,1)\circ
g =\phi '_ \gamma ({\bar \phi }(b), h_{\gamma \alpha}( b,1)g)$ hence
$h_{\gamma \alpha}(b,g)= h_{\gamma \alpha}( b,1)\circ g $.
Thus  the condition  will fulfilled  if  we shall  set
$r_{\gamma \alpha}(b) =h_{\gamma \alpha}(b,1)$ for $b
\in U_\alpha\bigcap {\bar \phi }^{-1}U'_\gamma $.  Let ${\bar \zeta }=
\{U_\alpha ,\phi_{\alpha\beta} \}$, ${\bar \zeta } ' =
\{U'_\gamma,  ,\phi '_{\gamma \sigma}\}$.Then
$$
\phi '_\gamma ({\bar \phi }(b),  r_{\gamma \alpha}(b)
\phi { \alpha \beta}(b) g )=\phi \circ \phi_\alpha(b,\phi
{ \alpha \beta}(b) g )
= $$
$$ =\phi \circ \phi_\beta (b,g) = \phi '_\sigma ({\bar \phi }(b),
r_{\sigma\beta}(b) g) = \phi '_\gamma ({\bar \phi }(b), \phi '_\gamma
( {\bar \phi }(b),
\phi '_{\gamma \sigma}({\bar \phi }(b)) r_{\sigma\beta}(b) g)
$$
for all  $ b \in U_\alpha\bigcap U_\beta \bigcap
{\bar \phi }^{-1}U'_\gamma \bigcap {\bar \phi }^{-1}U'_\sigma $  whence
$$
r_{\gamma \alpha}(b) \phi { \alpha \beta}(b) g =
\phi '_{\gamma \sigma }({\bar \phi }(b)) r_{\sigma\beta}(b) g
\tag 77.1
$$
Therefore $r = \{ r_{\gamma \alpha}\}$ is the morphism of  the sets
of cocycles.\qed
\enddemo
\remark{\bf Remark} The  demand
$$
\phi _\alpha (b,g) = \phi _\alpha (b,1) \circ g , \qquad  b
\in U_\alpha, g\in G
\tag 78.1
$$
is the additional demand to definition of  the principal  fibre
bundle. This  demand is correct  when it exists the continuous
homotopy which
connects  the element $g\in G $ with the  unit 1 of the group G,
i.e. when g \it belongs  to the connection's component of the unit \rm 1.
\endremark
\proclaim{\bf Theorem 1.13}Let $ \zeta =(B,\pi,E,G), \zeta ' =
(B',\pi ',E',G)$ are the two principal  fibre bundles with  the atlases
$\{(U_\alpha ,\phi _\alpha )\}$ , $\{ (U'_\gamma,,\phi ' _\gamma)\}$
and the associated sets of  cocycles ${\bar \zeta}, {\bar \zeta} '$.
It exists the  morphism $\phi  \to \phi '$ of  the principal  fibre
bundles induced the  some morphism $r:{\bar \zeta}  \to {\bar \zeta} '$
of  the sets of  cocycles.
\endproclaim
\demo{\it Proof}We set ${\bar \phi}= {\bar r}= B\to B'$  and  define
the map $\phi :E \to E'$ as following: if $e \in  E; e=\phi _ \alpha
(b,g)$  and
$\pi (e) \in  U_\alpha \bigcap {\bar \phi}^{-1}(U' _ \gamma )$ then
$\phi (e)= \phi '  _ \gamma {\bar \phi}(b),r_{\gamma \alpha}(b)g ) $
according
to  lemma 1.12.If  besides that  $\pi (e) \in  U_\beta \bigcap {\bar
\phi}^{-1}(U' _ \sigma)$ then $e=\phi _\beta (b,\phi _{\beta \alpha }
(b) g)$ and
$$
\phi '  _\sigma ({\bar \phi}(b), r_{\sigma \beta}(b) \phi _{\beta
\alpha} (b) g)= \phi '  _\sigma ({\bar \phi}(b),  \phi '  _{\sigma
\gamma }({\bar \phi}(b))
r_{\gamma \alpha}(b)g = $$
$$=\phi '  _ \gamma {\bar \phi} r_{\gamma \alpha}(b), g) =\phi  (e) $$
thus $\phi  (e)$ is  defined correct.Since  the restriction $\phi \vert
\pi ^ {-1}(U_\alpha  \bigcap \phi ^{-1}(U' _ \gamma ))$ is the
 composition
$$
\pi ^ {-1}(U_\alpha  \bigcap {\bar \phi } ^{-1}(U' _ \gamma )) @>
\phi ^{-1}_\alpha >> (U_\alpha  \bigcap {\bar \phi } ^{-1}(U' _ \gamma ))
\times G @>  ({\bar \phi} ,r_{\gamma \alpha})\times  1 >>
$$ $$
({\bar \phi } (U_\alpha )\bigcap U' _ \gamma ) \times G \times G
@> 1 \times \mu >> ({\bar \phi } (U_\alpha )\bigcap U' _ \gamma )
\times G
@>  \phi  '_ \gamma >> {\pi ' ({\bar \phi } }^{-1}(U_\alpha )\bigcap U' _
\gamma )
$$
(where $\mu :  G \times G  \to G $ is the  group multiplication)
then it is evident  that $\phi $ is continuous. We have
$\pi '\circ \phi  (e) =
\pi ' \circ \phi '  _ \gamma {\bar \phi} r_{\gamma \alpha}(b), g)
={\bar \phi }(b) = {\bar \phi }\circ \pi (e)$. Since $eh =
\phi _\alpha (b,g) h =
\phi _\alpha (b,gh)$ then $\phi (eh)= \phi  '_ \gamma ( {\bar
\phi}(b), r_{\gamma\alpha }(b) gh) =\phi  '_ \gamma ( {\bar \phi}(b),
r_{\gamma\alpha }(b) g)
h=\phi (e) h$ for $h \in G, e \in E$.Thus  $\phi $ is the morphism
of  the two  principal  fibre bundles. In conclusion since  according
to
the definition of  $\phi $ takes place the equality
$\phi \circ \phi _ \alpha (b,g)= \phi  '_ \gamma {\bar \phi}(b), r_
{\gamma \alpha}(b), g), b \in
U_\alpha  \bigcap \phi ^{-1}(U' _ \gamma ), g \in  G$ then   $\phi $
generates the morphism $r =\{ r_{\gamma \alpha} \} $ of the sets of
cocycles.\qed
\enddemo
\proclaim{\bf Theorem 1.14} Let ${\bar \zeta} =(U_\alpha  ,
\phi _{\alpha \beta})$ is the some set of  cocycles for the space B
and the topologic
group G .Then  there  exist the such principal  fibre bundle
$ \zeta =(B,\pi ,E,G)$ and the such atlas $\{(U_\alpha  ,\phi _\alpha )\}$
 for$\zeta $ that
${\bar \zeta} $ is the  set of  cocycles associated with this atlas.
\endproclaim
The proof of  this  theorem is also analogous to the proof of  the
theorem 1.8 and we refer our reader again to the  monograph [10].\par
The general definition 1.17 of  the two cocycles  over  the group G and
the  definition 1.25 of  the equivalence classes of  the
principal  fibre bundles generates the following theorem :
\proclaim{\bf Theorem 1.15 }It exists the bijective correspondence
between the equivalence  \linebreak classes of  the principal  fibre
bundles
over the base B and the cohomology classes from the definition 1.17 .
One can describe this correspondence as following:
if $\zeta $ is the some principal  fibre bundle then the fixed
cohomology class$ \lbrack {\boldsymbol \phi }_\zeta \rbrack $  of
the principal fibre bundle
$\zeta $ is compared to the some atlas  of the fibre bundle $\zeta $.
\endproclaim
\demo{\it Proof}If $\phi:\zeta \to \zeta '$ is the equivalence of the
principal  fibre bundles  then  lemma 1.12 supplies us the morphism
$ r(\phi): {\bar \zeta}  \to {\bar \zeta} '$. Since${\bar \phi }=1_B$
then this morphism is the equivalence of the sets of  cocycles.
Therefore the above correspondence is  defined correct.\par
Let now ${\bar \zeta} $ be the some set of  cocycles. According to
theorem 1.14  it exists the such principal  fibre bundle $\zeta $
with  its atlas $\{(U_\alpha  ,\phi _\alpha )\}$ that ${\bar \zeta}$
is the set of  cocycles associated with this atlas. Thus the above
correspondence is
surjective.\par
Let $\zeta , \zeta '$  be the such two principal  fibre bundles that
${\bar \zeta}$ and  ${\bar \zeta}$ are equivalent. Let us denote this
equivalence  as $r: {\bar \zeta} \to {\bar \zeta} '$. According to
theorem 1.13   it exists the  morphism $\phi :\zeta \to \zeta '$ induced r.
In particular ${\bar \phi} =1_B$. Besides that  takes place the
morphism $r^{-1}:{\bar \zeta} ' \to {\bar \zeta}$ seted  with the formula
$r^{-1}=\{ r^{-1}_{\gamma \alpha }\}$.The morphism of the fibre
bundles associated with $r^{-1}$: $\phi ^{-1}: \zeta ' \to \zeta $
is the
contrary  to $\phi$. Really the equalities $\phi ^{-1} \circ \phi
\circ \phi _\alpha  (b,g)= \phi ^{-1}  \phi  '_ \gamma ({\bar \phi}(b),
 r_{\gamma \alpha}(b) g)=
\phi _\alpha (b, r_{\gamma \alpha} ^{-1} (b)r_{\gamma \alpha}(b) g) =\phi
 _\alpha (b, g)$ whence $\phi ^{-1}\circ \phi =1$. One can prove analogous
that $\phi \circ \phi ^{-1}=1$. Therefore,$ \zeta \simeq \zeta '$  and
thus our  correspondence is injective.\qed
\enddemo
\proclaim{\bf Corollary 1.16} Let $\phi : \zeta \to\zeta '$  be the
such  morphism of  the  two principal  fibre bundles over B that
${\bar \phi }=1_B$. Then $\phi$ is the equivalence.
\endproclaim
\demo{\it Proof} It  is obvious that the  morphism  $r (\phi ):
{\bar \zeta}  \to {\bar \zeta}'$ of  the sets of  cocycles associated with
$\phi$ is the equivalence .\qed
\enddemo
Let us consider now the following construction.
Let $\zeta =(B,\pi, E,G)$ is the  principal  fibre bundle over
B and $f: B \to B'$ is the
some map. Let us construct the  principal  fibre bundle $f * \zeta$
 \it induced \rm from the fibre bundle $\zeta$ as following.Let $E' =
\{(b',e) \in B' \times E: f (b')= \pi (e)\}$. Let us define the map
$\pi ': E'\to B'$ supposing $\pi ' (b',e) =b'$. If  $ f ' :E' \to E$
is the  map
given with the formula $ f ' (b',e) = e$ then  the following diagram
is commutative:
$$
\CD
E'                   @> f '>>                                     E\\
@V\pi ' VV                                                 @V\pi VV\\
B'                    @> f >>                        B
\endCD
\tag 79.1
$$
This  commutative diagram is  like to the  commutative diagram (47.1)\par
The action of  the  group G on E'  is  defined  with the rule
$( b',e)h =(b',eh), h \in G $. Let $\{( U_\alpha ,  \phi _\alpha)\}$
is the  atlas for $\zeta$.
Then $\{ f ^{-1}U_\alpha , f ' _\alpha )\}$  is the  atlas for $ f
*\zeta = ( B' , \pi ' ,  E' ,G) $ where
$$
\phi'_\alpha :  f ^{-1} U_\alpha  \times G  \to \pi  ^{-1'}
(f ^{-1} U_\alpha )
\tag 80.1
$$
have the  form $f  ' _\alpha (b ' ,g ) = (b ' , f  _\alpha
(f(b '),g)), b ' \in f ^{-1}U_\alpha , g  \in G $.It is evident
that if $\zeta \simeq  \zeta ' $
then $f *\zeta \simeq f *\  \zeta ' $.\par
We can define the  fibre bundle $f *\zeta$  in the terms of
cocycles as following.
If $ (U_\alpha , \phi _{\beta \alpha })$ is the  set of  cocycles
for $\zeta$ then $ f  ^{-1} U_\alpha ,  \phi _{\beta \alpha}\circ f $
is the  set of
cocycles for  ${\bar \phi }*\zeta$ .\par
This definition "works" both for the principal and the vector fibre
bundles. We must substitute simply the arbitrary topologic  group
for the case of  the principal fibre bundles onto the GL(n;K) group
in the case of  the vector fibre bundles.
\proclaim{\bf Theorem 1.17} Let $ \phi:\zeta \to \eta$ be the morphism
of  the vector  fibre bundles;then it exists the  morphism
$\psi:\zeta \to
{\bar \phi } *\eta $ with ${\bar \psi } = 1_B $. Therefore according
to corollary 1.16 modified on  the  case of  GL(n;K) group takes place
 the equivalence
of  the  fibre bundles $ \zeta  \simeq {\bar \phi } *\eta $ .
\endproclaim
\demo{\it Proof} Since ${\bar \psi } = 1_B $ we can set  ${\bar \phi }
*\eta =(B', \pi ' ,E',G )$ where $ E' =\{(b,e)\in B \times  E_ \eta :
{\bar \phi }
(b)=\pi _ \eta (e) \}$(our reader  must compare these formulas with
 above diagram for their understanding ). Let us define now  the map
$\psi  :E_\zeta  \to E' $ setting $\psi  (e)= (\pi _\zeta  \phi (e))$.
Since $p_ \eta \circ \phi (e)= {\bar \phi } \circ \pi _\zeta (e)$ for
all $e \in  E_ \eta $
then  it  is obvious  that $ \psi  (e) \in E' $.It  easy to check that
$ \pi ' \circ \psi  =\pi _\zeta  $ and $\psi  (eh)=(\pi _\zeta  (eh),
 \phi  (eh) )=
=(\pi _\zeta  (e), \phi  (e)h )= (\pi _\zeta  (e), \phi  (e) )h = \psi
(e) h, h \in G$ ( the equality $ \pi _\zeta (eh) =  \pi _\zeta (e)$
follows from the  definition
of the  projection $\pi $ ). Thus   $\psi  $ is the morphism of  the
fibre bundles  with ${\bar \psi } = 1_B $, i.e.  it is the equivalence of
the fibre bundles.\qed
\enddemo
The following definition which  is necessary now from the considered
construction describes the new type  of  the fibre bundles,-  \it the
associated fibre bundle \rm .Thus  our classification of  the fibre bundles
is replenished now with  the new representative.
\definition{\bf Definition  1.28} One can compare the associated fibre
space  $\zeta \lbrack Y \rbrack $ with the  fibre Y to  the principal
 fibre bundle
$\zeta $ .It constructed  as following . Let  us  define  the right action
G on $ E \times  Y $ supposing $(e,y) g =(eg, g ^{-1} y), g \in G, e
\in E,y\in Y $. Let $ E_Y =  E \times  Y /G$. Let  us  denote
$\{ e ,y  \} $ the image of  the  point  (e,y) in $E_Y $. Then
$ \{eg , y\}= \{e , gy\}$
for $g \in G$ . Let  us  set  the  map $\pi _Y : E_Y \to  B $ with the
formula $\pi _Y \{ e ,y  \}= \pi (e)$. Since $ \pi (eg) = \pi (e)$
for $g \in G $
then  $\pi  _Y $ is defined correct. It is easy to see that it is
continuous. If $\{ (U_ \alpha , \phi _ \alpha ) \} $is the atlas for
$\zeta $, let  us
define  the atlas $\{ (U_ \alpha , \psi _ \alpha ) \} $ for
$\zeta \lbrack Y \rbrack = (B, \pi  _Y,E_Y ,Y) $ supposing
$$
\psi _ \alpha  (b,y) =  \{\phi _ \alpha (b,1), y \}, \quad  b
\in U_ \alpha , \quad  y  \in  Y
\tag 81.1
$$
The map  $\psi _ \alpha  $ is continuous and satisfies  the
condition  $ \pi  _Y \circ \psi _ \alpha = \pi  _{ U_ \alpha } $.
The composition
$$
\pi  ^{-1} ( U_ \alpha ) \times Y @>   \phi _ \alpha ^{-1}\times 1
>>  U_ \alpha \times  G  \times  Y @>  1\times  \rho >> U_ \alpha
\times  Y
\tag 82.1
$$
where $ \rho  : G  \times  Y \to  Y $  is the action of the group G
on Y  generates the map  $\pi _Y ^{-1}\ (U_ \alpha \to U_ \alpha
\times Y $
contrary to  $ \psi _ \alpha $. \par
If $ Y= K ^n $ then  the formula $ r \{ e,v \} + s \{ e',v' \} =
\{e, rv+s g ^{-1}v'\}$ where  e'g =e, $g  \in GL (n;K); r,s \in K $
sets the structure
of  the vector space on  the fibres  of $\zeta \lbrack K ^n \rbrack $.
The all $ \psi _ \alpha $ are linear on  the fibres  relatively this
structure.
\enddefinition
\proclaim{\bf Theorem 1.18.} Let $\zeta , \zeta ' $  be the two
principal  fibre bundles . It exists the bijective correspondence
between the
morphisms $\phi  : \zeta   \to \zeta ' $ and  the sections  s of
the  fibre bundle's  space $\lbrack E' \rbrack $ defined with the rule
 $\phi  \longmapsto s_\phi $
where $ s_\phi  (b) =\{e, \phi (e) \}$ for every $e \in \pi ^{-1}
(b)$ ( the left action  of  the group G on E' is given with the
formula $g e' = e'g^{-1},
e'  \in  E' ,g \in  G $).
\endproclaim
\demo{\it Proof}. Let  us show firstly that the  map $s_\phi $ is
defined correct. Really if ${\tilde e }\in \pi ^{-1} (b)$ is the
some element of the
fibre  over b different from e then it exists always the such
$ g \in G $ that   ${\tilde e } =eg $ and
 $\{ {\tilde e }, \phi ( {\tilde e }\}=
\{ eg , \phi ( eg )\}= \{ eg , \phi  ( e) g \}= \{ e , \phi  ( e)
\} $. Further, $s_\phi $ is  continuous, so long as it  is none other
than the  map $E/G \to
E \times E' /G $ induced with  $(1, \phi )$ and group G is always
topological, i.e. continuous in  our  theory.It is evident
that $\pi _{E'}\circ s_\phi  =1_B $ .\par
Let  us  suppose that $ s: B \to E _{E'} = E \times E'/G $ is the
section for  the projection $\pi _{E'}$.The composition $ s \circ \pi $
ought  to have the form $e  \longmapsto \{ e , \phi _s (e) \}$
with the some map $\phi _s :E \to E' $.If $ g  \in G $ then
$\{ eg , \phi _s (e) g \}=s \circ  \pi  (e)
= s \circ  \pi  (eg) = \{ eg , \phi _s (eg) \}$ whence
$ \phi _s (eg) = \phi _s (e) g $ for $ e \in  E, g  \in G $.
Let  ${\bar \phi} _s  : B \to B' $ be
 the  generated
map $ E/G \to  E'/G $. If  we shall prove that  $\phi _s  $ is
continuous  then it will mean that $\phi _s  $ is  the sought for
morphism of the
principal fibre bundles.\par
Let  $ e \in  E $  and $ U \subset E' $  be the  open neighborhood
of the point $\phi _s  (e) $. Since the action of the  group G on E is
continuous
then there exist the open neighbourhood  U' of the  point $\phi _s  (e) $
in E' and the open neighbourhood  V of the unit  1 of the  group G  such
that  $ U' V \subset U $ . Let  us take the such neighbourhood V' of
the unit  1 of the  group G  that $ (V') ^{-1}V'  \subset  V $.
Let  now  $ (W, \psi ) $ be the  trivialisation  for $\zeta $ contained
$b = \pi  (e)$ ; let  us suppose  that $ e = \psi (b,h), h \in G $.
Then
the set $ W ' = \psi (W \times h V' ) $  is open in E and
$\{ W ' \times U' \} $ is  open in  $E \times E' /G $ .
Thus $ O =W ' \bigcap \pi  ^{-1}
\circ s^{-1} \{ W ' \times U' \} $ is open in  E , $e \in O$,
and besides that
$$
s \circ \pi  (O)\subset  \{ W ' \times U' \}
\tag 83.1
$$
Let us  show that $\phi _s  (O) )\subset  U $. Let $ x \in O$ ;
then $s \circ \pi  (x) \in  \{ W ' \times U' \}$ : let us  take for
certainty $ s \circ \pi  (x) =
\{x',y ' \}, x  \in  W ' , y ' \in  U ' $. Since $ x, x ' \in  W ' $
then $ x' =x g^{-1}$ for  some $ g \in  V$. Therefore
$\{x,\phi _s (x) \}= s \circ
\{x',y ' \}=\{x g^{-1},y ' \} =\{x, y ' g \} $. But from the  fact
that $ y ' \in  U ' , g \in  G $ follows that $ y' g \in  U $ , i.e.
$ \phi _s \in  U $ \qed
\enddemo
Before we shall proceed to the next  lemma we must  give the
definition which will play the important role in our lections.
This is the  long
ago announced \linebreak\it simplex \rm.
\definition{\bf Definition  1.29}[2a,p.335] One say that the (n+1)
points of  the affine space are affine independent  if they  contain
in the no
 (n-1)
plane ,i.e. if the vectors $ {\bold k}_1 - {\bold k}_0 ,...,
{\bold k}_m - {\bold k}_0 $ are linear  independent
( where ${\bold k}_0 , {\bold k}_1
 ,..., {\bold k}_m $ are the radius-vectors ).\par
The set  of the all points of the form
$$
{\bold k} = t_0 {\bold k}_0  +   ... +   t_ m {\bold k}_m
\tag 84.1
$$
where ${\bold k}_0 , {\bold k}_1  ,..., {\bold k}_m $ are  the
some affine independent  points of  the space ${\bold R}^n $ or
in general,
of  the arbitrary linear  space over the field ${\bold R} $ and
$ t_0 ,..., t _m $ are  the such real  numbers that
$$
0 \leq  t_0 \leq  1 ,  ... , 0 \leq  t_0 \leq  1
\tag 85.1
$$
and
$$
t_0  +  ... +t_ m  = 1
\tag 85.1a
$$
(we  identify as ever the points of ${\bold R}^n  $ with their
radius-vectors)  is called the m-dimensional simplex with  the vertexes
${\bold k}_0, ... , {\bold k}_m $ and denoted  with  the symbol
${\bold k}_0 ...  {\bold k}_m $ .At m=0 this  is the point ${\bold k}_0 $,
at m=1 this  is the segment ${\bold k}_0 {\bold k}_1 $, at m=2 this
is the triangle ${\bold k}_0 {\bold k}_1  {\bold k}_2 $,  at m=3 this
is the tetrahedron
${\bold k}_0 {\bold k}_1  {\bold k}_2  {\bold k}_3 $( the 4-dimensional
simplex at m=4  is the  generation of this  construction; it as  we
this already
mentioned is the  base of the Regge  calculation and the  S.Hawking's
space-time  foam  theory ). \par
The numbers $ t_0, ... ,t_ m $ of the  point ${\bold k} $of the
simplex  ${\bold k}_0 ..., {\bold k}_m $ are called  its baricentric
co-ordinates. \par
The point ${\bold k} $ of the  simplex  ${\bold k}_0 ..., {\bold k}_m $
is called the interior point  if $ 0 < t _i < 1 $ for all
i =0, ... ,m .
\enddefinition
\definition{\bf Definition  1.30.} Let A be  the some subspace in X
 and $ i : A \to X $ is embedding. The subspace A is called the retract
of  the space  X if it exists the such map $ r : X \to A $ that
$ r \circ i = 1_ A $.If $A \subset X$ is the retract ( with the
retraction r)
and  besides that $i\circ r \simeq 1_X $ then subspace A is called
the deformation retract in X.In the case when $i\circ r\simeq 1_X $ rel A,
i.e.when the homotopy is constant on A  the subspace A is called the
strong deformation retract  in X.
\enddefinition
\proclaim{\bf Lemma 1.19} Let $\zeta  = (D^n , \pi  ,E,G ),
\zeta ' = (B ' , \pi '  ,E ', G )$  be the  principal  fibre bundles,
$ \phi : ( \pi \times 1) ^ {-1}
( D^n  \times \{0\} \bigcup S ^ {n-1} \times I ) \to E ' $ be the
 morphism of the fibre bundles and  $ F: D^n \times I \to B ' $
 be the some
continuation of the morphism $ {\bar  \phi }$.Then  it exists the
morphism $\Phi : \zeta \times I  \to \zeta ' $  with $ {\bar \Phi }= F $
which  continues
$ \phi $
\endproclaim
\demo{\it Proof}. Let us  choose  the atlases $\{(U_ \alpha ,
\phi_ \alpha ) \}, \{(U '_ \gamma ,\phi_  \gamma ) \} $ of the
fibre bundles
$\zeta  , \zeta ' $.  The sets $ (U_ \alpha  \times I ) \bigcap
F^ {-1} (U '_ \gamma ) $  form the open covering of the space
$ D^n \times I $.\par
Since $ D^n \times I $ is the compact metric space  then it exists
the  such positive number $\lambda >0 $ that the every  subset
$ S \subset  D^n \times I $ of the diameter $ < \lambda $ contains
in the some $ (U_ \alpha  \times I ) \bigcap F^ {-1} (U '_ \gamma )$.
Let us take the  such small triangulation for $ ( D^n , S ^ {n-1}) $
that   every  simplex $\sigma $ from $D^n $ has the diameter
$ < \lambda /2 $, and then let
us divide the segment $ I =\{0= t_0 < t_1 < ... < t_k =1\} $ such
that  every set  $\sigma \times \lbrack t _i , t_{ i+1}\rbrack $
has the diameter
$< \lambda , 0 \leq i < k $ .\par
Let us suppose that  the map  $\Phi $ is already defined  on
$ E \times  \lbrack t _0 , t _i \rbrack \bigcup \pi ^{-1}( S ^ {n-1}
\times  I $. Let us
continue  $\Phi $ by the simplexes on $E \times  \lbrack t _i ,
t _ {i+1}\rbrack  $ with  the help of  induction by dim $\sigma $.
If dim $\sigma =0 $ and  $\sigma \nsubseteq  S ^ {n-1}$ then let us
take  the  such  $\alpha , \gamma $ that
$\sigma \times \lbrack t _i , t_{ i+1}\rbrack  \subset ( U_ \alpha
\times I ) \bigcup F^ {-1} (U '_ \gamma ) $.Since $ \pi  ' \circ \Phi (
\phi_ \alpha
( \sigma ,g ), t _i ) = F (\sigma , t _i ) \in U '_ \gamma $ for
every  $ g \in G $ then it exists  the such map $ f_ \sigma  : \sigma
\times I  \to G $ that
$$
\Phi ( \phi_ \alpha ( \sigma ,g ), t _i ) = \phi '_ \gamma  ( F
( \sigma , t _i ), f_ \sigma  (g))
\tag 86.1
$$
Since $\Phi ( \phi_ \alpha ( \sigma ,g ), t _i ) = \Phi ( \phi_
\alpha ( \sigma ,1 ), t _i ) g $ then $ f_ \sigma  (g) = f_ \sigma
(1) g $ for  all  $ g \in G $.
Let $ g _ \sigma  =f_ \sigma  (1) $ ; then $\Phi ( \phi_ \alpha
( \sigma ,g ), t _i ) = \phi '_ \gamma   ( F ( \sigma , t _i ),
g_ \sigma g $.Let us define
$\Phi $ on $ (\pi \times I ) ^ {-1} ( \sigma  \times \lbrack t _i ,
t _ {i+1}\rbrack $ setting
$$
\Phi ( \phi_ \alpha ( \sigma ,g ), t ) =  \phi '_ \gamma  ( F
( \sigma , t  ) ,  g_ \sigma g ,\qquad t \in \lbrack t _i ,
t _ {i+1}\rbrack
\tag 87.1
$$
Then  $\Phi $ is continues continuous the map $\Phi \vert (
\pi \times I ) ^ {-1} ( \sigma  \times \{t _i \} ) $ and
since I is the continuous
interval  satisfies the conditions
$ \pi ' \circ  \Phi  =F \circ  ( \pi  \times 1) ,  \Phi ( (e,t) h )
 =\Phi  (e,t) h  $ for  all $(e,t) \in  (\pi \times I ) ^ {-1}
 ( \sigma  \times \{t _i \} ),
h \in  G $ .\par
Let us suppose now that the map $\Phi $ is constructed on
$ (\pi \times I ) ^ {-1} ( \sigma ' , \lbrack 0 , t _ {i+1}\rbrack  ) $
for all the simplexes
$ \sigma '  $with dim $ \sigma '  <m $.Let dim $ \sigma =m ,\sigma
\nsubseteq  S ^ {n-1}$. Let us choose again the such $ \alpha , \gamma $
that
$\sigma \times  \lbrack t _i , t _ {i+1}\rbrack \subset (U_ \alpha
\times I ) \bigcap F ^{-1}(U' _\gamma ) $. The map $\Phi $ is defined
already
on  $ (\pi \times I ) ^ {-1}  ( \sigma  \times \{t _i \}\bigcap
{\dot \sigma}  \times \lbrack t _i , t _ {i+1}\rbrack ) $ where
${\dot \sigma} $ is the
boundary of the simplex $ \sigma $ ; let us continue it on
$(\pi \times I ) ^ {-1}  ( \sigma  \times \lbrack t _i , t _ {i+1}
\rbrack ) $ .
As above
we find that the map $\Phi $ has the form
$$
\Phi  ( \phi_ \alpha ( x ,g ), t ) =  \phi '_ \gamma  ( F ( x, t),
f _ \sigma  ( x, t) g )
\tag 88.1
$$
on  $ (\pi \times I ) ^ {-1}  (  \sigma  \times \{t _i \} \bigcap
{\dot \sigma}  \times \lbrack t _i , t _ {i+1}\rbrack ) $ for the suitable
$ f _ \sigma :\sigma  \times \{t _i \} \bigcap  {\dot \sigma}
\times \lbrack t _i , t _ {i+1}\rbrack ) \to G $.Since $ \sigma  \times
\{t _i \} \bigcap
{\dot \sigma} \times \lbrack t _i , t _ {i+1}\rbrack ) $ is the retract for
$ \sigma  \times \lbrack t _i , t _ {i+1}\rbrack $ then it exists  the
continuation ${\bar f _ \sigma } : \sigma  \times \lbrack t _i ,
t _ {i+1}\rbrack
\to G $ of the map $ f _ \sigma $. Let us define $\Phi  $ on $ (\pi
\times I ) ^ {-1}  ( \sigma  \times \lbrack t _i , t _ {i+1}\rbrack ) $
supposing
$$
\Phi ( \phi_ \alpha ( \sigma ,g ), t ) =  \phi '_ \gamma  ( F ( x, t),
f _ \sigma  ( x, t) g )
\tag 89.1
$$
for  $( x, t) \in \sigma  \times \lbrack t _i , t _ {i+1}\rbrack , g
\in  $. It  is obvious that $\Phi  $ is continuous, it is the continuation
of the map
$$
\Phi  \vert  (\pi \times I ) ^ {-1}  (  \sigma  \times \{t _i \}
\bigcap  {\dot \sigma}  \times \lbrack t _i , t _ {i+1}\rbrack )
\tag 90.1
$$
and satisfies the conditions
$$
\pi ' \circ  \Phi  = F \circ  ( \pi \times I ) , \qquad \Phi
( ( e,t ) h ) =\Phi ( e,t ) h
\tag 91.1
$$
for all $( e,t ) \in  (\pi \times I ) ^ {-1}  ( \sigma  \times
\lbrack t _i , t _ {i+1}\rbrack ), h \in  G $. This completes the step of
induction.\qed
\enddemo
\proclaim{\bf Lemma 1.20}Let $\phi  : \zeta   \to \zeta ' $ be the
morphism of  the fibre bundles and $ F: B \times I \to B '$ is the
homotopy
$ F_0 = {\bar \phi } $ . If B  is the cellular space then it exists
the  such morphism of the fibre bundles $\Phi :  \zeta \times I \to
\zeta ' $
that ${\bar \Phi }= F $  and $\Phi \vert E \times \{0\} = \phi $.
\endproclaim
\demo{\it Proof}  Let us  carry out the proof  by induction by the
frames: if $\Phi $ is given  on $(\pi ^{-1}(B ^ {n-1}\times I ) $ then
because  of
theorem  1.18 the  section $ s : B ^ {n-1}\times I \to  E \times I
\times E ' /G $ is defined. We want  to continue  this section on $ B ^ n
\times I $
(the initial step n=0  of  induction  is obvious: or $ B ^ {-1} =
\emptyset $  or $  B ^ {-1} =\{b_0\}$ and F  is the  homotopy rel $ b_0$).
Let $ f ^n _\alpha :  ( D^n , S ^ {n-1}) \to  (B^n ,B ^ {n-1} ) $
be the  characteristic  map of the cell $ e^n _\alpha $. The  formula
$$
(y, t ) \longmapsto (y,  s (f ^n _\alpha (y),t )), \qquad (y, t )
\in  D^n \times \{0\}\bigcup S ^ {n-1} \times I
\tag 92.1
$$
sets the section of the  fibre bundle $ f* ^n _\alpha \zeta \times I )
\lbrack E' \rbrack $  for every $\alpha$ .It  follows  from theorem  1.18
and lemma 1.19 that it exists  the such map $s _ \alpha : D^n \times I
 \to E \Sb n* \P f _ {\alpha \sigma} \endSb  \times I \times E ' /G $
that $\pi _2 \circ  s _ \alpha = F \circ  ( f ^n _\alpha \times 1 ) $
for the map $\pi _2 : E  \times E ' /G  \to E ' /G = B' $. Continuing
s on
$ e ^n _\alpha \times I  $ by the formula
$$
s (f ^n _\alpha (y) , t ) = (f ^ {n '} _\alpha \times _ G 1) \circ
s_\alpha (y,t)
\tag 93.1
$$
$ y \in  D^n , t \in  I $ we find that s is  continuous and it  is
 the section . The continued such section s defines  in one's turn
 the morphism
$\Phi  : (\pi \times I ) ^ {-1}( B^n \times I )  \to E  ' $ .\qed
\enddemo
\proclaim{\bf Theorem 1.21}
Let $\zeta =(B,\pi, E,G)$ be  the principal  fibre bundle and
$f_0,f_1 :B' \to B$ be the two homotopical maps of  the cellular
space B' into the cellular space B.
Then $ f*_0 \zeta  \simeq f*_1 \zeta $ .
\endproclaim
\demo{\it Proof} Let $F: B' \times I \to B$ be the homotopy from
$f_0$ to $f_1$. Since it exists the morphism of  the fibre bundles
$\Phi  : f*_0 \zeta
\times I  \to \zeta $  then  by theorem 1.17 and lemma 1.20  takes
place  the equivalence $ f*_0 \zeta  \times I \simeq F* \zeta  $.
If we define the map
$ i _1 : B '  \to B '  \times I $ with the formula $ i _1 (b ' ) =
(b ' ,1 ), b ' \in B ' $  then $ f_1 = F \circ  i _1$  and
$$
f *_1  \zeta = (F \circ  i _1)* \zeta \simeq i * _1 (F* \zeta )
\simeq i * _1 (f *_0 \zeta \times I ) = $$ $$=
(f *_0 \zeta \times I ) \vert ( B ' \times  \{1\})  \simeq  f *_0 \zeta
$$
\qed
\enddemo
One  can define  the  cofunctor $k_G :{\Cal RW}'  \to  {\Cal RT}'$
for the some toplogical group G as following. Let us denote as
$k_G (X)$ the  set of all equivalence classes of  the (labelled)
principal  fibre bundles over X (i.e. we must take  into account
the labelled points
of  these
fibre bundles in this case!).  Let us also define $k_G \lbrack f
\rbrack : k_G (Y)\to k_G (X)$ for  every homotopic class
$\lbrack f \rbrack$  of  the maps $f:(X,X_0) \to (Y,Y _0) $ supposing
that $k_G \lbrack f \rbrack (\{\zeta\})=\{f*\zeta\}$ where $\{\zeta\}$
is the equivalence class of  the (labelled) principal  fibre bundle
$ \zeta$ .  Theorem 1.21 ensure that this map is defined correct.
The labelled
element  of $ k_G $ is the equivalence class of the trivial fibre
bundle $ ( X ,\pi \times 1, X \times G, G )$.
\definition{\bf Definition  1.31} Let us  denote as $ Z \bigvee X $
the subspace $ Z  \times  \{x_0 \} \bigcup \{z_0 \} \times  X \subset
 Z \times X ;
\{x_0 \} \in X , \{z_0 \} \in  Z $. One  can interpret it as a space
obtained with the identification of the labelled points $ x_0 $ and
$ z_0 $. It is obvious that  $ Z \bigvee X $  is the space with the
labelled point $ (z _0 ,x_0 )$. One calls the such object the bouquet
or the disjoint union of the spaces X and Y.
\enddefinition
Let us  show  now that our cofunctor $k_G $ satisfies the two imortant
axioms which we now shall define.
\proclaim{\bf The  sum axiom} The  morphism
$$
\{i * _\alpha \} : F* ( \bigvee \Sb \alpha \endSb X_\alpha )
{\boldsymbol \to} \prod \Sb \alpha \endSb F* (X_\alpha )
\tag 94.1
$$
is the bijection for  the arbitrary family $\{X_\alpha \}$ of
the cellular spaces from $:{\Cal RW}' $ generated  with the
embedding $ i_\beta :
X_\beta  \to \bigvee \Sb \alpha \endSb X_\alpha $.
\endproclaim
\proclaim{\bf The  Myer- Wjetoris axiom } Let  the cellular triad
$ ( X ; A_1, A_2 ) $, i.e. the cellular space X  with the such its
subspaces  $A_1$
and $A_2 $ that $ A_1 \bigcup A_2 = X $ be given . Then it exists
$ y \in F* (X )$ with $ y \vert A_1 =x_1 , y \vert A_2 =x_2 $  for
every above
$A_1$ and $A_2 $  and $ x_1 \in F* (A_1 ) , x_2 \in F* (A_2 ) $
with $ x_1 \vert A_1 \bigcap  A_2 = x_2 \vert A_1 \bigcap  A_2 $.
\endproclaim
Let us  make sure that the cofunctor $k_G $ satisfies these  axioms.
For  the purpose  to prove this we must give again  the series of
definitions which are connected with the theory  of  the cellular
spaces
and have the strategic  importance for the many directions of
differential topology.\par
Firstly  the above definition of  disjoint union  generates the
following object.
\definition{\bf Definition 1.32.} Let $ (X , x_0 ), (Y , y_0 )
\in {\Cal RT} $. Then  let us define the reduced product
$ ( X  \bigwedge Y , *) $
of the spaces X  and Y , where *  is the  labelled point of the
reduced product $ ( X  \bigwedge Y , *) $ ,as a factor-space
$$
X  \bigwedge Y  = X \times Y / X  \bigvee Y
\tag 95.1
$$
The  labelled point of the reduced product $ ( X  \bigwedge Y , *) $
is the  point $ * =  \pi ( X  \bigvee Y ) $ where
$ \pi : X \times Y \to  X  \bigwedge Y  $  is the  natural projection.
We shall denote the  point $ \pi (x ,y ) $ as $\lbrack x , y \rbrack $.
If the  maps
$$
f : ( X ,x_0 ) \to ( X ' ,x '_0 )
\tag 95.1a
$$
and
$$
g : ( Y ,y_0 ) \to ( Y ' ,y '_0 )
\tag 95.1b
$$
are given then $ f  \times g : X \times Y \to X ' \times Y ' $ moves
$ X  \bigvee Y $ into $ X ' \bigvee Y ' $ and  therefore  induces  the
 map
$ f \bigwedge g : X  \bigvee Y  \to X ' \bigvee Y ' $.
\enddefinition
The one of examples of the reduced products is the  \it cone \rm .
\definition{\bf Definition 1.33} The cone $ ( CX , * ) \in  {\Cal RT}$
over X is the reduced product
$$
( CX , * ) = ( I  \bigwedge X , * )
\tag 96.1
$$
where  the  point  0 of  the segment I is the labelled point of this
segment. According  to  definition 1.31 of  disjoint union  we  can
write down
$$
CX =I \times X / ( \{0 \}\times X \bigcup I \times \{ x_0 \})
\tag 97.1
$$
Let us denote as $ \lbrack t ,x \rbrack $ the image of the point
$ ( t ,x ) \in I \times X $ in CX .The map $ i : X \to CX $ defined
with  the
formula $ i (x) = \lbrack 1 ,x \rbrack $ sets the homomorphism of
the space X  on the image im i ; therefore  we  can identify X with
im i
and consider X as a subspace of the cone CX.
\enddefinition
\definition{\bf Definition 1.34} Let us  construct the cone of map
$ Y \bigcup _f  CX $ for  the  given map $ f : (X ,x_0 ) \to ( Y, y _0 )$
in
${\Cal RT}$.  $ Y \bigcup _f  CX $ is obtained from $ Y\bigvee X $ by
the identification $\lbrack 1 ,x \rbrack \in  CX $ with
$ f (x) \in  Y $
 for all $ x \in  X $.More precisely  we stick the root of  the cone
 CX to  the space Y with  the help of  the map f . It  is evident that
the projection $ q : Y\bigvee X  \to Y \bigcup _f  CX $ defines the
homomorphism between  the spaces Y and q (Y);therefore  we  can consider
Y as a subspace in $ Y \bigcup _f  CX $.
\enddefinition
\definition{\bf Definition 1.35} Let X  be the some space and $ g :
 S ^ {n-1}\to X $ is the some map. Then  according  to  definition
 1.34
one can form the cone   $ X \bigcup _g  C S ^ {n-1}$ of  the map g.
The obtained as a result  the new space is called the  space X with
the stuck n-dimensional cell .The map g is called the stuck map of  the
 cell. Restricting the  natural projection $ q : X \bigvee C S ^ {n-1}
\to X \bigcup _g  C S ^ {n-1}$ on $ C S ^ {n-1}$ we obtain the map
$ f: C S ^ {n-1} \to X \bigcup _g  C S ^ {n-1}$ ,which  is the
homeomorphism on the interior of the cone $ C S ^ {n-1}$. One  can
call this map  the characteristic  map of  the cell .This  definition
of  the characteristic  map
conforms to the above its definition  by our first acquaintance with
the cellular spaces. Since $ C S ^ {n-1} \cong D ^{n} $  then
one  can consider f   as a  map
$$
f : (D ^{n} ,S ^ {n-1} )  \to (X \bigcup _g  C S ^ {n-1}, X )
\tag 98.1
$$
Let us note that $ f \vert  S ^ {n-1} = g $.
\enddefinition
\definition{\bf Definition 1.36} Let X  be the some space  and
$ A \subset X $ is  its subspace. The structure of  the relative cellular
space
on  the pair ( X , A) is the such sequence $ A = ( X , A) ^ {-1}
\subset ( X , A) ^ 0 \subset ... \subset  ( X , A) ^ n
\subset  ( X , A) ^ {n+1}
\subset ... \subset X $, that $( X , A) ^ n  $  is obtained from
$( X , A) ^ {n-1} $ by the sticking of  the n- dimensional cells,
$ n \geq 0,
X = \bigcup \Sb n \geq -1\endSb ( X , A) ^ n  $ and  X is provided
with the weak topology : $ S \subset X $ is closed when and only
when  $ S \bigcap ( X , A) ^ n  $ is closed in $ ( X , A) ^ n  $ .\par
Let us call the pair ( X , A) the relative cellular space  if it can
be provided with the some structure of  the relative cellular space.\par
Let us  set dim ( X , A) =n ,if $ ( X , A) ^ n  = X $ and
$ ( X , A) ^ {n-1}\neq X $. Let us note that if  $ A = \{x_0 \}$
then X is the cellular space .
\enddefinition
\proclaim{\bf Lemma 1.22.}Let  (X; A, B ) be the cellular triad
with the above features. Then it exists the  such open set
$ A' \supset A $
and the homotopy
$ H : X \times I \to X $ which satisfies the following conditions :
\roster
\item $H_0 =1 _X $ ;
\item H is immovable on A ;
\item $ H_1 ( A ' ) \subset  A $ ;
\item $ H ( B \times I ) \subset  B $
\endroster
\endproclaim
\demo{\it Proof} Let the cell $ A ^{-1} =A $ and $ H ^{-1}:
X \times I \to X $ be the stationary homotopy:$ (H ^{-1}(x,t ) =
x $ for all
$ t \in I $.
Let us  suppose  by induction  that we constructed already the
such open neighbourhood $ A ^ k $ of the subspace A in
$ ( X , A) ^ {k+1}$,
that  $ A ^ {k-1}= A ^ k \bigcap ( X , A) ^ k $ and the such homotopy:
$ H^ k : X  \times I \to X $  that
\roster
\item $H_0 ^ k = H_1 ^ {k-1}$ ;
\item $H ^ k $ is the stationary homotopy on $( X , A) ^ k $ ;
\item  $H_1 ^ k ( A ^ k ) \subset A $;
\item  $H ^ k (B  \times I ) \subset  B $.
\endroster
Let us  denote as $ \{e ^ {n+1}_\gamma \}$ the set  of all (n+1)-
dimensional cells from ( X , A) ,and as $ f ^ {n+1}_\gamma $ the
characteristic
map of the cell $ e ^ {n+1}_\gamma $.Let $ D^ {n+1}_0  =\{x \in D^
{n+1}: \vert  x \vert  \leq  1/2 \}$.Then $ D^ {n+1} - D^ {n+1}_0 $  is the
open neighbourhood of the sphere $ S ^n $. The homotopy : $ K :
D^ {n+1} \times I \to D^ {n+1} $ setting with the formulas
$$
K (x, t ) = \cases (1+t )x, & \vert  x \vert  \leq  \frac {1}{1+t }\\
\frac {x}{\vert  x \vert }, & \vert  x \vert  \geq  \frac {1}{1+t }
\endcases
\tag 99.1
$$
for $ x \in D^ {n+1}, t \in I $ shows us  that $ D^ {n+1} - D^ {n+1}_0 $
is tightened in the point on $ S ^n $. Let now
$$
U ^ {n+1}_\gamma = \{ f ^ {n+1}_\gamma (y) : f ^ {n+1}_\gamma ( \frac
{y}{\vert  y \vert }) \in A^ {n+1},  y \in  D^ {n+1} - D^ {n+1}_0 \}
\tag 100.1
$$
It is  easy  to see that $U ^ {n+1}_\gamma $ is the open subset  in
$ e ^ {n+1}_\gamma $.Therefore if one sets $ A^ n = A^ {n+1}\cup
\bigcup \Sb \gamma \endSb U ^ {n+1}_\gamma $ , then one can check  that
$ A^ n $ is the open neighbourhood of the subspace A in
$ ( X , A) ^ {n+1}$, and $ A^ n \bigcap ( X , A) ^ n = A^ {n-1}$.
Let us define homotopy $ H^ n : ( X , A) ^ {n+1}\times I \to X $  as
$$
H^ n (x,t ) = \cases H^ {n-1} _1(x ) , & x \in ( X , A) ^ n \\
H^ {n-1} _1 ( f ^ {n+1}_\gamma ( K (y, t ) ),  & x \in f ^
{n+1}_\gamma (y) , y \in  D^ {n+1} , t \in I
\endcases
\tag 101.1
$$
Then $ H^ n $ is continuous and satisfies the equality
$ H^ n  _0 = H^ {n-1} _1 $ on $ ( X , A) ^ {n+1}$, and this allows
us to continue it on
X with the conservation  of  this equality (as a weak topology!).
The fulfilment of  the conditions (1),(2),(3) is evident. Since $
H^ {n-1}$
satisfies the condition   then $ H^ n $ also satisfies it. Therefore we
constructed by induction by k  the all neighbourhoods $ A^k $ and
the all homotopies $ H^ k $  for $ k \geq  -1 $.\par
Let us set  now $ A ' = \bigcup \Sb k \geq  -1 \endSb A^k $.
The set A' is  open , since $ A' \bigcap e ^ m _\gamma = A^{m -1}\bigcap
e ^ m _\gamma $  for all m, $ \gamma $.In conclusion, let us set
the homotopy $ H: X \times I \to X $ as
$$
H(x,t ) = \cases H^{r -1} (x , (r+1)( r (t-1) +1)),   & \frac {r-1}{r}
\leq t   \leq  \frac {r}{r+1}, x \in X \\
H^ r (x , 1), & t =1,  x \in ( X , A) ^ r
\endcases
\tag 102.1
$$
It is easy to see that  homotopy  H is continuous and $ H _0 = H ^
0 _0 = 1 _X $. Besides that H is immovable in  subspace A ,
 $ H _1 ( A' ) \subset A $ and $ H ( B \times I ) \subset B $.\qed
\enddemo
\proclaim{\bf Lemma 1.23}Let ( X, A, B ) be the triad with
$ X = A ^ \circ \bigcup B ^ \circ $ ($  A ^ \circ $ and
$B ^ \circ $ are the
open sets ) ; $\zeta $ be  the n-dimensional vector  fibre
bundle ( the principal fibre bundle correspondingly) over A, $\zeta ' $
be  the n-dimensional vector  fibre bundle ( the principal
fibre bundle correspondingly) over  B and $ \phi : \zeta \vert A
\bigcap B \to
\zeta ' \vert A  \bigcap B  $  be the equivalence  of its restrictions
on $ A  \bigcap B $ (in general  if  $ \eta = (B ,\pi , E, K^n ) $ and
$ A\subset B $  then $\eta \vert A = (A, \pi  \vert  \pi  ^{-1}(A) ,
\pi  ^{-1}(A) , K^n ) $ ). Then it exists the such n-dimensional vector
fibre bundle ( the principal fibre bundle correspondingly)  $ \eta $
over  X that $\eta \vert A \simeq \zeta  $.
\endproclaim
\demo{\it Proof} We work  in the  both cases in fact with the set of
cocycles. Let $ \{ U _ \alpha $,\linebreak $\phi _ {\alpha \beta } \},
\{ U  '_ \gamma ,\phi ' _ {\gamma \delta } \}$ are the sets of cocycles
for  $ \zeta , \zeta ' $ .Let us denote as $ r _ {\gamma \alpha }:
U _ \alpha \bigcap U  '_ \gamma \to  G $ ( or GL (n ,K ) )  the function
determined  the equivalence $  \zeta \vert A  \bigcap B  \simeq \zeta '
\vert A  \bigcap B  $ ; it means  that
$$
r _ {\delta \beta }(b) \phi _ {\beta\alpha }(b) = \phi ' _ {\delta
\gamma }(b)  r _ {\gamma \alpha }(b)
\tag 103.1
$$
for  all $ b \in U _ \alpha  \bigcap U  _ \beta \bigcap U  '_ \gamma
\bigcap U  '_  \delta $ .  Then it is easy to check  that
$\{A ^ \circ \bigcap U _ \alpha  , B ^ \circ \bigcap  U  '_ \gamma \}$
is the open covering of X  and
$$
\phi _ {\beta\alpha } : A ^ \circ \bigcap U _ \alpha \bigcap  U  _
\beta \quad \to  G
\tag 104.1
$$
$$
\phi ' _ {\delta \gamma }: B ^ \circ  \bigcap  U  '_ \gamma \bigcap
U  '_  \delta \quad  \to  G
\tag 104.1a
$$
$$
r _ {\gamma \alpha }:  A ^ \circ \bigcap  B ^ \circ  \bigcap
U _ \alpha \bigcap U  '_ \gamma  \quad  \to  G
\tag 104.1b
$$
forms the set of cocycles on X. For example  if  $ b \in U _ \alpha
\bigcap U  _ \beta \bigcap U  '_ \gamma $  then
$$
r _{\gamma \beta }(b) \phi _ {\beta\alpha }(b) = r _{\gamma \alpha }(b)
\tag 105.1
$$
In order to prove this  it is necessary to set  $ \delta = \gamma $
in the above equation. Thus we obtain the principal fibre bundle
(or the vector fibre bundle ) $\eta $ over X.  It is obvious that
$ \eta  \vert  A \simeq \zeta , \eta  \vert  B \simeq \zeta ' $.\qed
\enddemo
\proclaim{\bf Theorem 1.24} Cofunctor $ k_G $ satisfies the sum and
Myer-Wjetoris axioms.
\endproclaim
\demo{\it Proof} Let ( X, A, B ) is the some cellular triad.
According to lemma 1.22  there exist the open neighbourhood  A' of the
set A and
the homotopy $ H :  X \times I \to X $ of the neighbourhood  A' on A
transfered B itself . If  $ j : (A , A \bigcap B ) \to  (A' , A'
\bigcap B ) $
is  the pair's embedding  then $ j* : k_G (A ' ) \to  k_G (A  ) $
and $ ( j \vert A \bigcap B ) * : k_G ( A' \bigcap B ) \to  k_G ( A
\bigcap B ) $
are the bijections. Let us suppose that  the elements $ x_1 \in k_G
( A ) , x_2 \in k_G ( B ) $  for which $ i* _1 (x_1 ) = i* _2 (x_2 ) $
  jn set
$ k_G ( A \bigcap B )$ (where $ i _ 1, i _ 2 $ are the maps  from theorem
 1.21 ) are given. Let us shoose $ x '_1 \in k_G ( A' ) $  such that
 $ j* (x '_1 ) = x_1 $. Then we have $ i ' * _1 (x '_1 ) = i ' * _2
 (x '_2 ) $ in set $ k_G ( A' \bigcap B ) $  ( here $ i '  _1 : A'
 \bigcap B \to  A' , i '  _2 : A' \bigcap B \to  B $ are the
natural embeddings ). Therefore  the elements $ x '_1, x '_2 $ are
represented  with the such fibre bundles $\zeta _1 $ on A' , $\zeta _2 $
on
B  that $\zeta _1 \vert (A' \bigcap B ) = \zeta _2 \vert
(A' \bigcap B ) $.  According to lemma 1.23  there exists such a fibre
bundle $ \eta $ over X
that $  \eta  \vert  A '  \simeq \zeta _1 ,  \eta  \vert  B  \simeq
\zeta _2 $ ( $ A'  \bigcup B = X $ ). If  $y = \{ \eta \} \in  k_G ( X) $
is the
equivalence class  of  $  \eta $, then $ j* ' _1 y = x_1 , j* ' _2 y = x_2 $
where $  j ' _1 : A '  \to  X , j ' _2 : B  \to  X $ are the embeddings.
Thus $ k_G $ satisfies the Myer- Wjetoris axiom. \par
Let $ \{X _ \alpha , x _ \alpha \} $  be the some family of the cellular
spaces with the labelled points. Let us denote as $ \zeta _ \alpha =
(X _ \alpha , \pi  _ \alpha , E_ \alpha ,G ) $  the principal fibre bundle
represented $ y _ \alpha $ for every $ y \alpha \in  k_G ( X _ \alpha ) $.
Let us define the  fibre bundle  $ \zeta = \bigvee \Sb \alpha \endSb X _
 \alpha  ,\pi  ,E , G ) $ over $ \bigvee \Sb \alpha \endSb X _ \alpha  $
identifying those points $ e \in  E_ \alpha , e ' \in  E_ \beta $ in $
\bigvee \Sb \alpha  \endSb E_ \alpha $ for which it exists  the such
element
$ g \in  G $ that $ e = e _ \alpha g , e ' = e _ \beta g $  ( $ e _
\alpha \in  E_ \alpha ,e _ \beta \in  E_ \beta $ are the labelled points ).
Let E
be  the result  of this  identification and $ \pi  : E \to  \bigvee \Sb
\alpha \endSb X _ \alpha  $ is  the natural projection.Let us set  on E
the action of  the group G supposing $ \{e \}h =\{eh\}, h \in  G $ .
It is easy to show that this action is defined correct , and
$ \pi ( \{e\}) = \pi ( \{e\}) $ if  and only if $ \{e' \}=\{eh\}$ for
some $ h \in  G $.If $ \{U ^ \alpha _\beta  ,\phi ^ \alpha _\beta \} $
is the such atlas
of the fibre bundle $ \zeta _ \alpha $  that  $ \phi ^ \alpha _\beta
( x _ \alpha , 1 ) = e  _ \alpha $ always  when $ x _ \alpha \in
U ^ \alpha _\beta  $
then one can construct the atlas of the fibre bundle $ \zeta $ from $
\{U ^ \alpha _\beta  ,\phi ^ \alpha _\beta \} $ .  It is evident that
the
map $ i * _ \alpha : k_G ( \bigvee \Sb \alpha \endSb X _ \alpha  ) $
transfers $ \{\zeta \}$ in $ y_ \alpha  $. Therefore $ \{* _ \alpha \} :
k_G ( \bigvee \Sb \alpha \endSb X _ \alpha  ) \to \prod \Sb \alpha
\endSb k_G ( X _ \alpha ) $  is surjective. Similarly  if $ \zeta , \eta $
are  the such fibre bundles over $ \bigvee \Sb \alpha \endSb X _ \alpha $ ,
 that $ \zeta \vert X _ \alpha \simeq \eta \vert X _ \alpha $  for all
$ \alpha $  then gluing together these equivalences on $ \bigvee \Sb
\alpha \endSb X _ \alpha $  ( they preserve the labelled points ) we
obtain the equivalence $ \zeta \simeq \eta $. Thus , the map $ \{ i * _
 \alpha \}$ is injective , whence $  k_G $ satisfies the sum axiom .\qed
\enddemo
\definition{\bf Definition 1.37} Let F* (Y) be  the some cofunctor on
the space Y. The element $ u \in  F (Y) $ is called the n-universal
element if
$$
T _u : \lbrack S ^q , s_0 ; Y, y_0 \rbrack \to F* (S ^q )
\tag 106.1
$$
is the isomorphism for q <n  and the epimorphism for q = n.
The element  u is called the universal element if it is n-universal
for every
$ n \geq 0 $ .
\enddefinition
\remark{\bf Remark } The set  of  homotopigal classes $ \lbrack S ^q ,
 s_0 ; Y, y_0 \rbrack $ plays  the decisive role in  modern differential
 and
algebraic topology with  their applications in theoretical physics. It
is turn  out that this set  has \it the group structure \rm the unit
of  which is the class of the stationary homotopies. It exists the
standard denotation in all the physic-mathematical literature for
this group which is called  \it the \rm q-\it dimensional homotopigal
group or the \rm q-\it dimensional spheroid  \rm . We shall
denote it henceforth as $ \pi _q (Y,y_0 ) $. We shall discuss the
group structure of  $ \pi _q (Y,y_0 ) $ later on , and now  our reader
can postulate for himself this fact.
\endremark
\definition{\bf Definition 1.38} The directed set is the  such set
with the relation $ \leq $ of partial putting in order : $ ( \Lambda ,
\leq  )$
that there exist  the such  elements $ \gamma , \alpha , \beta  \in
\Lambda $ that  $ \alpha \leq \gamma , \beta \leq \gamma $.
\enddefinition
\definition{\bf Definition 1.39} The inverse ( or the projective )
system of the abelian groups is the  such family $\{G _\alpha \}$
of the Abelian groups  indexed with the elements of the directed
set $\Lambda $ and the homomorphisms $ j ^ \beta  _\alpha : G _\beta
\to G _\alpha ,\alpha \leq \beta  $, that  $ j ^ \beta  _\alpha \circ
j ^ \gamma _\beta  = j ^ \gamma _\alpha $ by $ \alpha  \leq  \beta  \leq
\gamma $ and $ j ^ \alpha  _\alpha =1$ for every $ \alpha  \in \Lambda $.
If the some inverse system $ ( G _\alpha ,j ^ \beta  _\alpha ,
\Lambda )$ is given  then the  inverse ( or the projective ) limit is
the group $ \lim ^ 0 G _\alpha = inv lim  G _\alpha $
defined as following:\par
$ inv lim G _\alpha $ is the subgroup  of the  group
$\prod \Sb \alpha  \in \Lambda \endSb  G _\alpha $ consisted  from all such
$ f \in \prod \Sb \alpha  \in \Lambda \endSb  G _\alpha $   that
$j ^ \beta  _\alpha ( f ( \beta ) ) =  f ( \alpha )$ for all
$ \alpha , \beta $ with
$\alpha \leq  \beta $. The restrictions of  the projections
$ \pi  _\alpha : \prod \Sb \alpha  \in \Lambda \endSb  G _\alpha
\to G _\alpha $
onto the subgroup  $inv lim  G _\alpha $  generates  the homomorphisms
$$
\pi  _\alpha : inv lim  G _\alpha \to G _\alpha
\tag 107.1
$$
satisfied the conditions $ j ^ \beta  _\alpha \circ \pi
_\beta = \pi  _\alpha $ for every $ \alpha , \beta \in \Lambda $ with
$ \alpha \leq \beta $.
\enddefinition
\proclaim{\bf Lemma 1.25} If $ ( X , x_0 ) \in {\Cal RW }$ and
 $ \{x_0 \}= X _{-1}\subset X _0 \subset ...  \subset X _n
  \subset ...  \subset X $
is the such sequence of the cellular subspaces X  that
$ X = \bigcup \Sb n \endSb $ then the map
$$
\{i * _n \}:  F *( X)  \to inv lim F * ( X _n )
\tag 108.1
$$
is surjective.
\endproclaim
\demo{\it Proof}Let us consider  the so called infinite telescope
$ X ' = \bigcup \Sb n \geq -1\endSb \lbrack n-1 , n \rbrack ^ +
\bigwedge X _n $, where $\lbrack n-1 , n \rbrack ^ + $ is  the real
segment with the natural ends and the labelled point + .
and let  us set
$$
A_1 = \bigcup \Sb k \geq -1\\ (k =2n+1, n\in {\bold N}) \endSb
\lbrack k-1 , k \rbrack ^ +\bigwedge X_k \subset X '
\tag 109.1
$$
$$
A_2 = \bigcup  \Sb k \geq 0 \\ (k =2n, n\in {\bold N}) \endSb
\lbrack k-1 , k \rbrack ^ + \bigwedge X_k \subset X '
\tag 109.1a
$$
Then
$$
X '= A_1 \bigcup A_2 , \qquad A_1 \bigcap A_2 = \bigvee \Sb k\endSb X_k
\tag 110.1
$$
$$
A_1 \simeq \bigvee \Sb k=2n+1, n\in {\bold N}\endSb X_k  \qquad
A_2 \simeq \bigvee \Sb k=2n ,n\in {\bold N}\endSb X_k
\tag 110.1a
$$
(the latter relations are correct because of the fact that every
segment is tighten in the point).As it follows from the sum axiom
 there exist the such $y_1 \in F* (A_1)$ and $y_2 \in F* (A_2)$ for
 the some element $ \{x_k\}\in inv lim F(X _k)$ that $ y_1 \vert
X _k = x_k $, k is odd ,and $ y_2 \vert X _k = x_k $, k is even.
Let us consider the  element $ y_1 \vert A_1 \bigcap A_2 $.If k is odd
then $ y_1 \vert X _k = x_k $ ;if k is even then $ y_1 \vert X _k =
j* _k ( y_1 \vert X_{k+1})= j* _k (x_{k+1})= x_k $  where $j _k :
X _k \to X_{k+1} $ is the natural embedding.The analogous feature
is correct also for the element $ y_2 \vert A_1 \bigcap A_2 $. Therefore
$ y_1 \vert A_1 \bigcap A_2 =y_2 \vert A_1 \bigcap A_2 $.Therefore
with the help of the Myer- Wjetoris axiom  we obtain the element
$ y ' \in F(X ') $ with $y ' \vert A_1 =y_1 ,y ' \vert A_2 =y_2 $.
Then $y '\vert X _k = x_k ,k \geq -1 $.But $X '\simeq X $; therefore
exist the such element $y\in  F* (X) $ that $y \vert X _k =x_k ,k
\geq -1 $.\qed
\enddemo
\proclaim{\bf Lemma 1.26} Let $ ( Y, y_0 ) \in  {\Cal RW }$ and $ u
_n  \in  F* (Y) $ be the some n-universal element. Then it exists
 the cellular
space $Y '$ obtained from Y by  the sticking of the (n+1)- dimensional
cells and the such (n+1)- universal element $ u _ {n+1} \in   F* (Y ' ) $
that $ u _ {n+1} \vert Y = u _n  $.
\endproclaim
\demo{\it Proof}. Let  us take the one copy $ S ^{n+1}_ \lambda  $ of
the sphere $ S ^{n+1}$ for every $ \lambda  \in  F* ( S ^{n+1} ) $
and let us
form the cellular space $ Y \bigvee ( \bigvee \Sb \lambda \endSb
S ^{n+1}_ \lambda ) $. If  $ n \geq 0 $  then let  us choose  the
map-representative $ f : ( S ^ n , s_0 ) \to (Y, y_0 ) $ for every
class $ \alpha \in  \pi _n (Y,y_0 ) $  with $ T _ {u_n} (\alpha ) =0 $
and let  us stick
according to  definition 1.36 the (n +1)-dimensional cell
$ e ^{n+1}_ \alpha $ to $ Y \bigvee ( \bigvee \Sb \lambda \endSb
S ^{n+1}_ \lambda ) $
by this map. As a result we obtain the cellular space \linebreak $Y'$.
According to  sum axiomit exists  the  such element $v \in
F* (Y \bigvee ( \bigvee \Sb \lambda \endSb S ^{n+1}_ \lambda ) $ that
 $ v \vert Y = u_n , v \vert S ^{n+1}_ \lambda = \lambda $ . If
$$
g : \bigvee \Sb \alpha \endSb S ^ n  _\alpha \to Y \bigvee  \bigvee
\Sb \lambda \endSb S ^{n+1}_ \lambda
\tag 111.1
$$
is the " large sticking map " for the (n +1)-dimensional cells
$ e ^{n+1}_ \alpha $   then takes  place the exact sequence
$$
F* ( \bigvee \Sb \alpha \endSb S ^ n  _\alpha ) @<< g * < F*
( Y \bigvee  \bigvee \Sb \lambda \endSb S ^{n+1}_ \lambda )
@<< j * <  F* ( Y ' )
\tag 112.1
$$
It is easy to see that $ g * (v) \vert  S ^ n  _\alpha = T _ {u_n}
(\alpha ) =0 $ for all $ \alpha $. Since the map
$$
\{i * _\alpha \} : F * ( \bigvee \Sb \alpha \endSb S ^ n  _\alpha )
\to \prod  \Sb \alpha \endSb F * ( S ^ n  _\alpha )
\tag 113.1
$$
is bijective  according to  sum axiom   then  g* (v) =0 .
Thus it exists  the  such element $ u _{n+1}\in F* ( Y ' ) $
that $ u _{n+1}
\vert  ( Y \bigvee ( \bigvee \Sb \lambda \endSb S ^{n+1}_ \lambda ) =v $
( the  exact sequence   with g* (v) =0 give us the epimorphism).
In particular  $ u _{n+1}\vert  Y = u_n $  and $ u _{n+1}\vert
S ^{n+1}_ \lambda = \lambda $.\par
Now we ought to show that the element $ u _{n+1}$ is the (n+1)-
universal element. With this purpose let us consider the  commutative
diagram
$$
\sarrowlength =0.42\harrowlength
\commdiag{ \pi _q (Y, y_0 )& \mapright^{\fam6 i * }
&\pi _q (Y ', y_0 )\cr
&\arrow(1,-1)\lft{\fam6 T _{u_n}} \quad \arrow(-1,-1)\rt
{\fam6 T _{u_ {n+1}}}\cr
                                   & F*( S ^ q ) &\cr }
\tag 114.1
$$
Since  the space $Y'$  is obtained from   the space Y  by the
sticking of  the (n+1) -dimensional cells  then i*   is  the isomorphism
for $q<n $ (embedding )   and  it is  the  epimorphism for q= n because
of commutativity. Therefore  $ T {u_n}$ is  also the  isomorphism
for q< n     and  the  epimorphism
for q= n because of  the   commutative diagram  (114.1). Let us suppose
that   $ T {u_{n+1 }} ( \beta )=0 $  for some $\beta \in
\pi _n (Y ', y_0 ) $.
Since  i*   is  the epimorphism for q= n  then it exists  the  such
element $ \alpha \in  \pi _n (Y, y_0 ) $  that $ i* ( \alpha ) =
\beta $. But then
$ T _{u_n} ( \alpha ) = T {u_{n+1}} i* ( \alpha ) = T {u_{n+1 }}
( \beta ) =0 $.
Therefore   it exists  the cell $ e ^ {n +1}_  \alpha $  in $Y ' $
stuck up  the  element $  i*  ( \alpha ) $ , i.e. $  i*  ( \alpha ) =
\beta =0 $. It means   that $ T {u_{n+1 }} $ is  the   monomorphism
for q= n and   therefore  the isomorphism   for $ q \leq n $.
In conclusion  $  T {u_{n+1 }} ( \lbrack i  _ \lambda  \rbrack ) =
i* _ \lambda  (u  _{n+1 } ) =  \lambda  $, where $ i _ \lambda
S ^{n+1}_ \lambda  \to Y ' $  is  the embedding . Thus $ T {u_{n+1 }} $
is  the  epimorphism for q= n +1.\qed
\enddemo
\proclaim{\bf Corollary 1.27}It exists  the  cellular  space $Y '$ for
the every $ (Y, y_0 ) \in  {\Cal RW}$ and $ v \in  F*( Y ) $ contained
Y as a
cellular subspace  and the universal element $ u \in  F*( Y ) $ with $
 u \vert  Y = v $.
\endproclaim
\demo{\it Proof}Supposing $ Y _{-1} = Y , u_{-1} = v $ and applying
lemma 1.25 we find  by induction the sequence
$ Y = Y _{-1} \subset  Y _0 \subset  Y _1 \subset  ... \subset  Y _n
\subset  ... $  of the  cellular  spaces  in which $ Y _n $ is obtained
from
$ Y  _{n-1} $  by  the sticking of the  n -dimensional cells  and also
the sequence of the  n -universal elements $ u _n \in  F*( Y _n ) $
satisfied the condition  $ u _n \vert  Y  _{n-1}= u_{n-1}$. Let us set
now $ Y ' = \bigcup \Sb n \geq -1 \endSb Y_n $ and let us provide
$Y '$ with the weak topology. Then  according to lemma 1.26  it exists
the such $ u\in  F*( Y ') $ ,that $u \vert Y _n = u _n $. Let us
consider the  commutative
diagram
 $$
\sarrowlength =0.42\harrowlength
\commdiag{ \pi _q (Y, y_0 )& \mapright^{\fam6 i * } &\pi _q (Y ', y_0 )\cr
&\arrow(1,-1)\lft{\fam6 T _{u_n}} \quad \arrow(-1,-1)\rt {\fam6 T _u}\cr
                                   & F*( S ^ q ) &\cr }
\tag 115.1
$$
Since  the space $Y'$  is obtained from  the space $Y _n $ by  the
sticking of the cells of dimensions greater than n  then $ i*_n $ is the
isomorphism for q<n. Therefore $ T _u : \pi _q (Y ', y_0 ) \to F*( S ^ q )$
is the isomorphism for all q , i.e. u is the universal element.\qed
\enddemo
\proclaim{\bf Corollary 1.28}
It exists the cellular space $ (Y,y_0 )\in {\Cal RW}$ and the universal
element $ u \in F*( Y ) $.
\endproclaim
\demo{\it Proof} Let us set $ Y = \{ y_0\} $ in the previous corollary,
and let v is the labelled element in F*( Y ).Then $Y '$ from
above corollary is the sought for cellular space  and $ u \in F*( Y ') $
is the  sought for element.\qed
\enddemo
\definition{\bf Definition 1.40}Let (X,A)and (Y,B) be the relative
cellular spaces.The map $ f:(X,A)\to (Y,B) $ is called the cellular
map ,if $ f((X,A)^n )\subset (Y,B)^n $ for every \linebreak $ n \geq -1 $.
\enddefinition
\proclaim{\bf Lemma 1.29}Let $ f:(Y,y_0 ) \to (Y ',y '_0 )$  be the
cellular map and $ u \in F*( Y ), u '\in F*( Y ')$ be the such
universal elements that $ f* u' =u$. Then f generates the isomorphism
$$
f_* :\pi _q (Y, y_0 )\to \pi _q (Y ', y_0 ), \qquad q \geq 0
\tag 116.1
$$
\endproclaim
\demo{\it Proof} The following commutative diagram  take place for every
$q\geq 0 $ :
 $$
\sarrowlength =0,5\harrowlength
\commdiag{ \pi _q (Y, y_0 )& \mapright^{\fam6 f_* } &\pi _q (Y ', y_0 )\cr
&\arrow(1,-1)\lft{\fam6 T _u \cong} \quad \arrow(-1,-1)\rt {\fam6
T_{u '}\cong}\cr
& F*( S ^ q ) &\cr }
\tag 117.1
$$
The required statement follows from this commutative diagram.\qed
\enddemo
\definition{\bf Definition 1.41}The map $ f:X \to Y $ is called the
n-equivalence,$ n \geq 0 $ if the map $f _*:\pi _r (X,x_0)\to
\pi _r (Y,y_0) $ is bijective at $r<n$ and is surjective at r=n.
The map f is called the weak homotopical equivalence if it is the
n-equivalence for all $n\geq 0 $
\enddefinition
\definition{\bf Definition 1.42}Let us define the space $M_f$ for
the some map $ f:X \to Y $ from category ${\Cal T} $ by identification
the points $[1,x]$ and f(x) in $( I \times X) \bigcup Y $ for all
$x \in X$.The space $M_f$ is called the inreduced cylinder of the map f.
\enddefinition
The following definitions which are necessary for the proof of the
next lemma represent themselves  also the very important notions for the
topologist.
\definition{\bf Definition 1.43}Let us denote as $ X^Y $ the set of
all continuous maps $f: X\to Y$ of the two topologic spaces X and Y.
Let us furnish this set with the compact-open topology  taking as a
prebase [9]of this topology  the all sets of the type
$N_{K,U}=\{ f:f(K)\subset U\}$  where $K\subset X$ is compact and
$U\subset Y$ is open.
\enddefinition
\definition{\bf Definition 1.44}Let us define the space of loops
$ (\Omega Y,\omega _0) \in {\Cal RT}$ for $(Y,y_0)\in {\Cal RT}$ as
$$
\Omega Y = (Y,y_0)^{(S^1,s_0)}
\tag 118.1
$$
with the constant loop $\omega _0 $ ($\omega _0 (s)=y_0$ for all
$s \in S^1 $)as a labelled point.\par
Let us form now the iterant spaces of the loops $ \Omega ^n Y$
defining them by \linebreak induction
:$ \Omega ^n Y =\Omega (\Omega ^{n-1}Y),
 n\geq 1,
\Omega ^0 Y = Y$.
\enddefinition
\definition{\bf Definition 1.45}Let us denote as ${\Cal T}^2$ the
category of the topologic pairs (X,A)  where A is the subspace in
X and the continuous maps $f:(X,A)\to (Y,B)$.\linebreak Similarly
let us denote as $ {\Cal RT} ^2$ the category of the labelled pairs
$(X,A,x_0),
x_0\in A \subset X$ and the continuous maps $f:(X,A,x_0)\to (Y,B,y_0)$
preserved the labelled points.
\enddefinition
\definition{\bf Definition 1.46} The way in the topologic space X is
every continuous map $w:I\to X$. One call very often w the way
from w(0) into w(1) or the way connected  the points w(0) and w(1).
Thus $X^I$ is the space of all ways in X with the open-compact topology.
Let us introduce on X the relation $\sim$ supposing $x\sim y$ if and
only if it exists the way $w:I\to X$ from x into y.It is evident that
$\sim$ is the equivalence relation. On the other hand it is easy to
see that these definition of way, therefore the definition of the
relation $\sim$ also is likeness to the definition 1.24 of homotopy.
Thus the relation $\sim$ is the homotopic  relation. Let us denote
 the set of the equivalence classes of the relation $\sim$ as $\pi _0 (X)$.
 The elements of set $\pi _0 (X)$ are called the components of the
linear connection or the 0-components of the space X. If $\pi _0 (X)$
contains the one element only  then the space X is called the
linear connected or the 0- connected.
\enddefinition
\definition{\bf Definition 1.47}Let us denote as
$$
P (X,x_0,A)= (X,x_0,A)^{(I,0,1)}
\tag 119.1
$$
for the given pair $(X,A,x_0) \in {\Cal RT} ^2$ ,- the space of all
ways begun in the point $x_0$ and ended in the subspace A. The formula
$\pi(w)=w(1)$ sets the continuous map
$$
\pi :P (X,x_0,A)\to A
\tag 120.1
$$
\enddefinition
\definition{\bf Definition 1.48} Let us define the n-dimensional
relative homotopic set \linebreak $\pi _n(X,A,x_0)$, $n\geq 1$ for
the pair
$(X,A,x_0) \in {\Cal RT} ^2$ supposing
$$
\pi _n(X,A,x_0)=\pi _{n-1}(P (X,x_0,A),\omega _0)=\pi _0(\Omega ^{n-1}P
(X,x_0,A),\omega _0)
\tag 121.1
$$
\enddefinition
\remark{\bf Remark}We can consider $pi _0(X,x_0)$ as a labelled set
whose labelled element is the 0-component of the point $x_0$. The reader
can see also the obvious co-ordination between definition of
$\pi _n(X,A,x_0)$ and formula (106.1).
\endremark
\proclaim{\bf Lemma 1.30}The map $f:(D^n,S^{n-1})\to (X,A,x_0)$ defines
the 0-element in $\pi _n(X,A,x_0)$ when and only when f is
homotopic relatively   $S^{n-1}$ to the such map $f'$ that
$f'(D^n)\subset A$.
\endproclaim
\demo{\it Proof}Let $f\simeq f'$ rel $S^{n-1}$ and $f'(D^n)\subset A$.
Let us denote as H the homotopy connected f and  $f'$. Let us define
now the homotopy $G:(D^n \times I,S^{n-1}\times I ,s_0\times I )
\to (X,A,x_0)$ supposing
$$
G(x,t)= \cases H(x,2t),& 0\leq t\leq 0,5 \\
        f'((2-2t)x+( 2t-1)s_0),& 0,5 \leq t\leq 1
\endcases
\tag 122.1
$$
Then $G _0 =f,G _1 =x_0$.\par
On the contrary  let us suppose that $G:(D^n \times I,S^{n-1}
\times I ,s_0\times I )\to (X,A,x_0)$ sets the 0-homotopy of map f.\par
Let us define the new homotopy $H:D^n \times I \to X $ with the formula
$$
H(x,t)= \cases G(\frac {2x}{2-t},t),& 0\leq \vert x \vert \leq 1-0,5t \\
               G(\frac {x}{\vert x \vert,2-2\vert x \vert }),& 1-0,5t
 \leq \vert x \vert\leq 1
\endcases
\tag 123.1
$$
where $x \in D^n ,t\in I$. Then $H_0=G _0 =f$;$H(x,t)= G(x,0)=H(x,0)$
for all $ x \in S^{n-1},t\in I$ and $H_1(D^n)\subset A$.\qed
\enddemo
\definition{\bf Definition 1.49}The pair $(X,A)\in {\Cal T}^2$is called
the 0-connected pair if every component of the linear connection
of space X crosses A. The pair $(X,A)$ is called the n-connected pair
if it is the 0-connected pair and $\pi _r (X,A,a)=0$ for
$1\leq r \leq n$ and all $a\in A$. The space X is called the n-connected
space if $\pi _k (X,x)=0$ for $0\leq k \leq n$  and all $x\in X$.
\enddefinition
Let us introduce the notion dual to the notion of fibre bundle.
It is easy to see that we can depict the general fibre bundle
with the following  commutative diagram (this is in fact the
commutative diagram illustrated definition 1.26):
$$
\commdiag{ X &\mapright^f & E\cr
\mapdown\lft {\fam6 i _0}&\arrow (3,2)\lft  {\fam6 F}&\mapdown\rt { \pi}\cr
X\times I&\mapright^G & B }
\tag 124.1
$$
We come to the dual notion of \it cofibre bundle \rm $i:A \to X$
if we invert the all arrows in (124.1) and substitute $X\times I$
to the
dual it space $ Y^I$ (according the map of the two cellular spaces:
$f: X\to Y $)and in conclusion substitute $i _0$ with the map $\pi_0$
where $ \pi _0(w)=w(0)$ for the some way w. Remembering that one can
consider G as a map  A$\times $I$\to$Y ,and F as a map
X$\times $I$\to $Y :
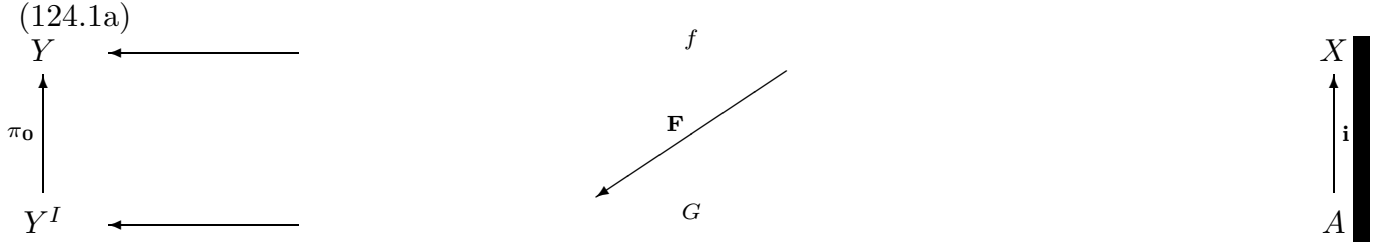
$$
\commdiag{Y&\mapleft ^f & X\cr
\mapup \lft {\fam6 \pi _0}&\arrow (-3,-2)\lft{\fam6 F}&\mapup
\rt{\fam6 i}\cr
Y^I & \mapleft ^ G & A}
\tag 124.1a
$$
we obtain the following definition:
\definition{\bf Definition 1.50}One say that the embedding $i:A\to X$
has the feature of the prolongation of homotopy relatively the
space Y,if it exists the homotopy $F: X \times I \to Y$ for every map
$f:X\to Y$ ,the homotopy $G:A \to Y$ (look (124.1a))and the
restriction $f\vert A$  which
continues G. The embedding i is called the cofibre bundle if it has the
feature of the prolongation of homotopy relatively the  space Y.
\enddefinition
\remark{\bf Remark}One can show  that if the some map $i:A\to X$ has
the feature of the prolongation of homotopy relatively the some
space Y then it is the homeomorphism of the space A on the some closed
subspace of the space X, i.e.it is the embedding A in X.
\endremark
\proclaim{\bf Lemma 1.31}Let $f: Z\to Y $  be the n-equivalence. Then
it exists the such map $ h' : D^r \to Z $ for the maps $g: S^{r-1}
\to Z$ and $ h : D^r \to Y $ with $ h \vert S^{r-1} = f\circ g $  that
$h ' \vert S^{r-1} =g $ and $f\circ h '\simeq h $ rel $ S^{r-1}$.
\endproclaim
\demo{\it Proof} One can depict the conditions of this lemma with the
following  commutative diagrams :
$$
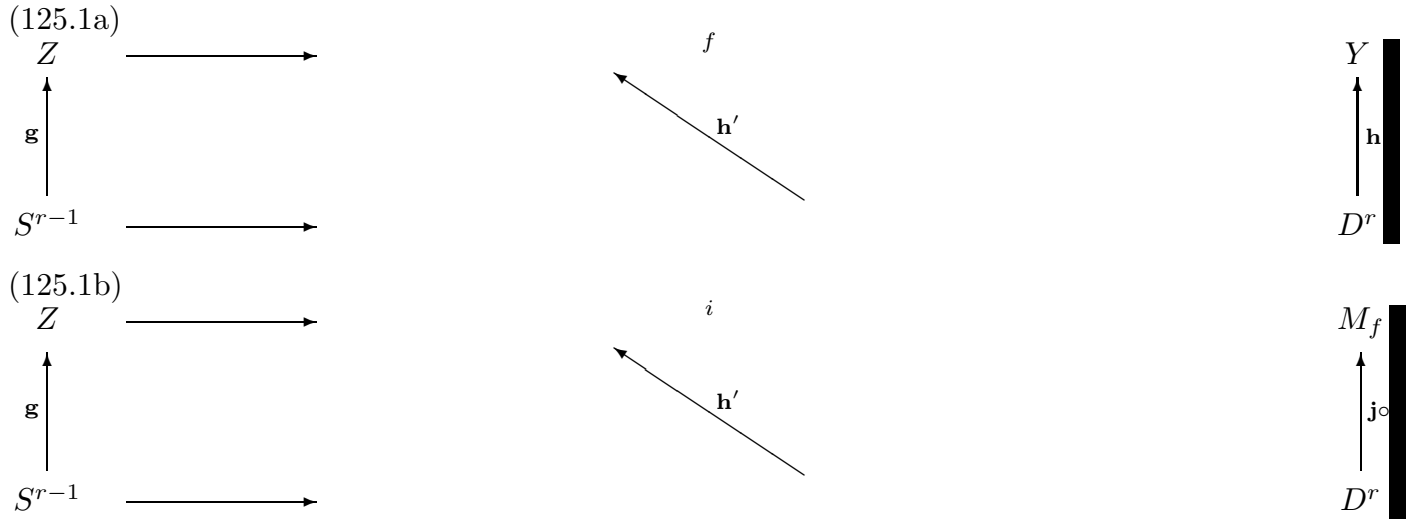
\commdiag{ Z &\mapright^f & Y\cr
\mapup\lft {\fam6 g }&\arrow (-3,2)\rt {\fam6 h'}&\mapup\rt  {\fam6 h}\cr
S^{r-1} &\mapright & D^{r} }
\tag 125.1a
$$
$$
\commdiag{ Z &\mapright^i & M_f\cr
\mapup\lft {\fam6 g }&\arrow (-3,2)\rt {\fam6 h'}&\mapup\rt
{\fam6 j\circ h}\cr
S^{r-1} &\mapright & D^{r} }
\tag 125.1b
$$
Let $ M_f$  be the inreduced cylinder of the map f with the accompanied
maps $ i:Z \to M_f$,$j: Y \to M_f$,$r:M_f \to Y$ and
$ H:M_f\to M_f$ .The idea of the proof consist in the replacement of the
map $f:Z \to Y$ on the embedding $i:Z \to M_f$ and the
application of lemma 1.30 to the n-connected pair $(M_f,Z)$ (125.1b).
The only problem here is the absence of the strict equality
$j\circ h \vert S^{r-1} =i\circ g$; indeed $j\circ h
\vert S^{r-1}=j\circ f \circ g=j\circ r\circ i\circ g \simeq i\circ g$.
The homotopy in this line has the form $H \circ(ig \times \alpha ),
\alpha (t)=1-t$.Since $S^{r-1}\subset D^{r}$ is the cofibre bundle
(compare (125.1a)and (124.1a)),one can continue $H\circ(ig \times \alpha )$
 up to the homotopy $H': D^{r} \times I\to M_f$ with
$H'_0= j\circ h$.Let $\hat h=H'_1$:then $h \vert S^{r-1}=H'_1\vert
S^{r-1}=H_0\circ i\circ g =i\circ g$. Therefore one can apply
lemma 1.30 to the map $\hat h:(D^{r},S^{r-1})\to (M_f,Z)$ (the
necessary part of this lemma).As a result we obtain the map
$h':D^{r} \to Z$ homotopic  rel $S^{r-1}$to the map $\hat h$ and
such that $h'\vert S^{r-1}=g$.Thus $f \circ h'=r\circ i h'
\simeq r \circ h'\simeq r \circ j\circ h\simeq h$.\qed
\enddemo
The proof of the next lemma is connected with classical \it Zorn
lemma \rm .It will useful for us to recall its matter [14,p.36].
\definition{\bf Definition 1.51} Let M be the set with the partial
putting in order. The every its subset A in which one can
associate  the some two elements (by the relation $ \leq $)is called
the chain.The chain is called the maximal chain if it is not contained
as a subset in the no other chain belonged M. Let us call the element
a of the set M with the partial putting in order- the upper border
of the subset $M'\subset M$ if the every element $a'\in M'$ is
subordinated a, i.e.$ a'\leq a$.
\enddefinition
\proclaim{\bf Lemma 1.32}(the Zorn lemma).If the every chain in
the set M with the partial putting in order has its upper border
then every
element from M is subordinated to the some maximal element.
\endproclaim
\proclaim{\bf Lemma 1.33}Let $f:Z\to Y$  be the some n-equivalence
and (X,A)is the relative cellular space with dim (X,A)$\leq n$.
Then it exists the such  map $ h':X \to Z$ for the some maps
$g:A \to Z,h:X \to Y$ with $h\vert A=f\circ g$ that $h'\vert A=g$
and $f\circ h'\simeq h$ rel A (the case $n=\infty$ is not excepted).
\endproclaim
\demo{\it Proof}Let ${\Cal T}$ be  the set of all such triads
$(X',k,K)$ that $A\subset X'\subset X$,$k:X'\to Z$is the map with
$k\vert A =g$ and $K:X'\times I$ is the homotopy connected
$ f\circ k$ with $ h\vert X'$ which is immovable on A. Let us set on
${\Cal T}$the relation $ \leq $ of the partial putting in order
supposing that $(X',k,K)\leq (X'',k',K')$ if and only if
$A\subset X'\subset X''\subset X$,$k'\vert X'=k$,$K'\vert X'\times I=K$.
It is evident that the triad $(A,g,K_0)$ (where $K_0$ is the
constant homotopy A$\times$ I$\to $Y)is contained in ${\Cal T}$,whence
${\Cal T}\neq \emptyset $.Let us show that the set ${\Cal T}$
satisfies the conditions of the Zorn lemma.\par
Let ${\Cal R}$ is the some subset with the partial putting in order in
${\Cal T}$.Let us set $W=\bigcup X'$, where the joining up is by all
subspaces $X '$  for which $(X',k,K)\in {\Cal R}$. Let us set the map
$h'\to Z $ and the homotopy H$'$:W$\times $I$\to Y$
supposing $h'(x)=k(x)$
and $H'(x,t)=K(x,t)$ if $x\in X'$ for the some $(X',k,K)\in {\Cal R}$.
Since the space X has the weak topology then $h'$ and$H'$ are
defined correct and continuous.It is evident that $(W,h',H')\in {\Cal T}$
and $(X',k,K)\leq(W,h',H')$for all $(X',k,K)\in {\Cal R}$.\par
Thus according to Zorn lemma the set ${\Cal T}$ has the maximal element
$(X',k,K)$.\linebreak Therefore we should show only that
 $X'=X$.If$X'\neq X$,
then let us consider the set of cells from $X-X'$ and let us take one
from them which has the minimal dimension. Let us denote it as e. Let
$\phi:(D^k,S^{k-1})\to (X,X')$ is the characteristic map of the cell e
and  $\psi:=\phi\vert S^{k-1}:S^{k-1}\to X'$is the stuck map.
Applying lemma 1.31 to the pair $k\circ \psi:S^{k-1}\to Z,h \circ
\phi:D^k\to Y$ we find the map $\theta:D^k\to Z$ and the homotopy
$\Theta :D^k\times I\to Y$ connected $f\circ \theta$ and $h \circ
\phi$. Let us continue k to the map $k':X'\bigcup e\to Z$ and K to the
homotopy $K':(X'\bigcup e)\times I\to Y$ supposing
$$
k'=\cases k(x),& x\in X'\\
          \theta(y),& x=\phi (y),y\in D^k
\endcases
\tag 126.1a
$$
$$
K'(x,t)=\cases K(x,t),& x\in X'\\
        \Theta(x,t)& x=\phi (y),y\in D^k
\endcases
\tag 126.1b
$$
Then $(X'\bigcup e,k',K)\in {\Cal T}$ and $(X',k,K)< (X'\bigcup e,k',K)$,
this contradict the maximum  of the triad $(X',k,K)$.Therefore
$X'=X$, whence one can take $h'=k$.\qed
\enddemo
\proclaim{\bf Lemma 1.34} Let Y be the space with the universal element
$ u \in F*( Y )$ , $(X,A, x_0)$ be the cellular pair
( $ A \subset X $), $ g:( A, x_0)\to (Y,y_0)$ be the cellular map and
$ v\in F*( Y )$ be the element satisfied the condition
$ v \vert A = g* (u) $. Then it exists the such cellular map
$ h:(X,x_0)\to (Y,y_0)$  that $ h\vert A = g $ and v= h*(u).
\endproclaim
\remark{\bf Remark}We consider here the spaces X,A,Y as a connected spaces.
\endremark
\demo{\it Proof} Let us denote as T the cellular space obtained from
$ (I^+ \bigwedge A) \bigvee X\bigvee Y$ by identification
$ [0,a] \in I^+ \bigwedge A $ and $[1,a] \in I ^+ \bigwedge A $ with
$ g(a)\in Y$.Let $A_1,A_2\subset T$ are the cellular spaces set as
$A_1=([0;1/2]^+\bigwedge A)\bigcup X ,A_2=([1/2;1]^+\bigwedge A)\bigcup Y$.
\linebreak Then $A_1\bigcup A_2=T$,$A_1\bigcap
A_2=\{1/2\}\times A \cong A$.
Besides that it exists the strong deformation retracts $f:A_1\to X$ and
$A_2\to Y$.Therefore there exist the such elements
${\bar v}\in F*(A_1),{\bar u}\in F*(A_2)$,that ${\bar v}\vert X=v,
{\bar u}\vert Y=u$.It is obvious that ${\bar v}\vert A_1\bigcap A_2=
f*(v \vert A)=f*g*(u)={\bar u}\vert A_1\bigcap A_2$. Therefore according
to the Myer- Wjetoris axiom it exists the element $w\in F*(T)$
with $w\vert X=v,w\vert Y=u$.\par
According to lemma 1.26 we can embed T in the some cellular space
$Y'$ and find the universal element $u'\in F*(Y')$with $u'\in F*(Y')$
$u'\vert T=w$.Let $j:Y\to Y'$ is the embedding; then $j*(u')=u'\vert
Y= w\vert Y= w$.Therefore according to the lemma 1.29 the induced
homomorphism of the homotopic groups
$$
j*:\pi _*(Y,y_0 )\to \pi _*(Y',y_0 )
\tag 127.1
$$
is the isomorphism.\par
Let nou ${\bar g}:X \to Y'$ is the embedding X in $Y'$. Then
${\bar g}\vert A \simeq j\circ g$,and the homotopy connected these two maps
 is the
composition $I ^+\bigwedge A \subset T\to Y'$. Since $A\subset X$
is the cofibre bundle relatively Y then it exists the
map
${\ddot g} :X\to Y'$ with ${\ddot g}\vert =j\circ g$ and ,therefore
${\ddot g}\simeq {\bar g}$:
$$
\commdiag{Y &\mapright^{\fam6 j}& Y'\cr
\mapup \lft {\fam6 g}& \arrow (-3,2)\rt {\fam6 h}& \mapup \rt
{\fam6 {\ddot g}}\cr
&A\mapright^{\fam6 \subset}& X}
\tag 128.1
$$
It follows from the proofs of the lemmas 1.31,1.33 that it exists
the such map $h:X \to Y$ that $h\vert A=g$, $j\circ h \simeq{\ddot g}
\simeq {\bar g}$.Therefore
$$
h*(u)=h*j*(u')={\bar g}*(u')=v
\tag 129.1
$$
\qed
\enddemo
Let us call the space Y figured in F*( Y ) as a \it classified space
\rm or a \it space of representation \rm of the cofunctor F*( Y )
\definition{\bf Definition 1.52}The two objects X and Y are called the
equivalent objects if there exist the such morphisms f $\in$ hom
(X,Y) and g $\in $ hom (Y,Z) that $ g \circ f =1_X $ and $ f \circ g =1_Y $.
 We call the morphisms f and g the equivalences in this case.
\enddefinition
\definition{\bf Definition 1.53}Let ${\Cal E}$,  ${\Cal D}$ are the two
categories and $F* ,G* : {\Cal E} \to {\Cal D} $ are the cofunctors
from  ${\Cal E}$ into ${\Cal D}$. Let us consider then the morphism T
(X)$ \in $ hom $ _{\Cal D} $ (G*(X), F*(X)). If the latter is the
equivalence in  category D then T is called the natural equivalence,
and cofunctors F* and  G* are called the natural equivalent.
\enddefinition
\proclaim{\bf Theorem 1.35} (The Brown theorem ).If
$ F* :{\Cal RW} '\to {\Cal RT} $ is the cofunctor satisfied the sum
and the Myer- Wjetoris
axioms then it exists the such  classified space $ (Y,y_0 )\in
{\Cal RW}$ and the such universal element $ u \in F*( Y ) $ that
$ T _u [-;Y,y_0 ]\to F* $ is the  natural equivalence.
\endproclaim
\demo{\it Proof} Because of corollary 1.28 it exists the cellular
space  $ (Y,y_0 )\in {\Cal RW}$ and the  universal element
$ u \in F*( Y ) $.Therefore we must prove only that
$$
T _u [-;Y,y_0 ]\to F*
\tag 130.1
$$
is the bijection for all $ (X,x_0 )\in {\Cal RW}$.\par
a. Let $ v \in F* (X)$.Then let us set $A=\{x_0\}$ in lemma 1.34 and
let us take as $g:(A,x_0)\to (Y,y_0)$ the only map transferred A into
the labelled point $y_0$.Then it exists the map $h:(X,x_0 )\to (Y,y_0)$
with $v=h*(u)=T _u([h])$.Therefore $T _u$ is surjective.\par
b.Let us suppose that $T _u[g_0]=T _u[g_1]$ for the two maps
$g_0,g_1:(X,x_0 )\to (Y,y_0)$. We can consider $g_0$ and $g_1$ as a
cellular
maps without of loss of generality. Let $X'=X\bigwedge I^+$ and
$A'=X\bigwedge \{0,1\}^+$. Let us set the map $g:(A',*)\to (Y,y_0)$
supposing $g[x,0]=g_0(x),g[x,1]=g_1(x),x\in X$. Besides that let
$p:X'\to X$ is the map defined with the equality $p[x,t]=x$ for all
$[x,t]\in X'$ and $v=p*g*_0 u=F*(X')$. Then
$$
v\vert X\bigwedge \{0\}^+=g*_0 u=g* u\vert X\bigwedge \{0\}^+
\tag 131.1a
$$
and
$$
v\vert X\bigwedge \{1\}^+=g*_0 u= T _u[g_0]=T _u[g_1]=g*_1u=
 g*u\vert X\bigwedge \{1\}^+
\tag 131.1b
$$
Therefore $g* u= v \vert A$, and, because of lemma 1.34 there exists
the map $h:X'\to Y$ with $ h\vert A =g$. It is easy to see that h is the
homotopy connected $g_0$ and $g_1$.Therefore $T _u$ is injective.\qed
\enddemo
\definition{\bf Definition 1.53}Let again ${\Cal E}$,  ${\Cal D}$ be
the two categories and $F* ,G* : {\Cal E} \to {\Cal D} $ are the
cofunctors from  ${\Cal E}$ into ${\Cal D}$.The natural transformation
from F to G is the correspondence which asssociate the morphism
T (X)$ \in $ hom $ _{\Cal D} $ (G*(X), F*(X))to every $X\in {\Cal E}$
such  that the equality
$$
T(X)\circ G*(f)=F*(f)\circ T(Y)
\tag 132.1
$$
take place for every morphism $f\in $ hom $_{\Cal D}$(X,Y), i.e.
the following diagram :
$$
\commdiag{F*(X)&\mapleft^{\fam6 F*(f)}& F*(Y)\cr
\mapup \lft {\fam6 T(X)} &&\mapup \rt {\fam6 T(Y)}\cr
G*(X)&\mapleft^{\fam6 G*(f)}& G*(Y)}
\tag 133.1
$$
is commutative.\par
The inversion of all arrows in diagram (133.1)gives us the natural
transformation of the two functors.
\enddefinition
Let us return now to cofunctor $k_G$. The proved above fact that it
satisfies the sum and the Myer- Wjetoris axioms means that
Brown theorem is valid for this cofunctor. And this means that there exists
such a cellular space $(BG,*)$ \linebreak ( defined with precision
of the homotopy type) and the principal G-fibre bundle
$\zeta _G=(BG,\pi,E_G,G)$ for  every topologic group G that the given
with the
formula $T[f]=\{f*_\zeta\}$ the natural transformation
$$
T_G:[-;BG,*]\to k_G(-)
\tag 134.1
$$
is the natural equivalence. Thus the principal G-fibre bundles over
the cellular space X are classified with the homotopical classes of the
maps $f:(X,x_0)\to (BG,*)$.\linebreak Therefore  the n-dimensional
vector fibre bundles over the field K are classified with the homotopic
classes of the maps $(X,x_0)\to (BGL (n,K),*)$.\par
Let us now denote the set $\{f*_\zeta\}$ of the equivalence classes
for the n-dimensional vector fibre bundles over the field K as
Vect$_n$(BGL (n,K)).This designation corresponds partial to analogous
designation in monograph [15]of M.F.Atiyah which we shall follow in
many respects later on.\par
Now we ought to recall the some information from the group theory [16,p.9].
\definition{\bf Definition 1.54}The abstract set G  is called the monoid
if the binary operation $(a,b)\to ab$ called the multiplication
 is defined on G and (ab)c=a(bc) for all $a,b,c\in G$, i.e.the
 multiplication is associative.The element ab is called the product
 of the elements a,b.The element (ab)c=a(bc)is denoted abc.\par
The monoid G is called the semigroup  if it exist the unit element
in G, i.e.the element $1\in G$ that 1a=a1=a  for all $a\in G$ (sometimes
one calls 1 the neutral element).
\enddefinition
\remark{\bf Remark}Thus the semigroup differs from the group \it
with the absence of the contrary element \rm .
\endremark
\definition{\bf Definition 1.55}Let $\zeta=(B,\pi,E,F)$ and
$\zeta '=(B',\pi ',E',F')$ be the two arbitrary fibre bundles.
The product of
$\zeta$ and $\zeta '$is the fibre space $\zeta \times \zeta ' =
(B\times B',\pi\times\pi ',E\times E',F\times F')$.The product of the
principal G-fibre bundle with the principal $G'$-fibre bundle is
provided on a natural way with the structure of the
principal $G\times G'$-fibre bundle. For example if ${\bar
\zeta}=\{U_\alpha,\phi _{\beta\alpha}\}$and ${\bar \zeta}'=\{V_\gamma,
\phi ' _{\sigma \gamma}\}$ are the  sets  of the transition
functions for $\zeta$ and $\zeta'$ correspondingly then
$$
{\bar\zeta }\times{\bar\zeta }'=\{U_\alpha\times V_\gamma ,
\phi _{\beta\alpha}\times\phi ' _{\sigma \gamma}\}
\tag 135.1
$$
is the set  of the transition functions for $\zeta \times \zeta ' $.
Similarly if $\zeta ,\zeta '$are the vector fibre bundles with the
fibres $K^n,K^m$,then $\zeta \times \zeta '$is prowided on a natural
way with the structure of the vector fibre bundle with the fibre
$K^n\times K^m =K^{n+m}$.For example if $\{U_\alpha,\phi_\alpha\}$
and $\{V_\gamma ,\psi_\gamma\}$are the atlases for the vector fibre
bundles
$\zeta$ and $\zeta '$ then $\{U_\alpha\times V_\gamma ,\phi_\alpha
\times \psi_\gamma \}$is the atlas for the vector fibre bundle
$\zeta \times \zeta '$. Besides that it is evident that take place the
equivalence
$$
\zeta [K^n]\times \zeta '[K^m]\simeq (\zeta \times \zeta ' )[K^{n+m}]
\tag 136.1
$$
for the GL (n,K)- fibre bundle $\zeta$ and GL (m,K)- fibre bundle
$\zeta '$.
\enddefinition
\definition{\bf Definition 1.55}If $\zeta ,\zeta '$are the vector
fibre bundles over the some base B then one can form the new
fibre bundle $\zeta \bigoplus \zeta '$ supposing  $\zeta \bigoplus
\zeta '=\Delta *(\zeta \times \zeta ' )$ where $\Delta :B\to B\times B'$
is the diagonal map. It is called the Whitnev sum of the vector fibre
bundles $\zeta$and $\zeta '$.The fibre $\zeta \bigoplus \zeta '$
over the point $b\in B$is  the direct sum of the fibres of the fibre
bundles $\zeta ,\zeta '$ over this point.
\enddefinition
When the Whitnev sum is already defined we can consider now the Whitnev
sum of the representative of the equivalence classes for
the n-dimensional vector fibre bundles over the field K, i.e. in fact
the Whitnev sum of the elements of Vect$_n$(BGL (n,K)). And now we
shall prove that \it the Whitnev sum of the above equivalence classes
is the multiplication (more precisely the addition)operation and
that \rm Vect$_n$(BGL (n,K)) \it has the structure of the abelian
semigroup with respect to the Whitnev sum\rm .\par
But this is almost evident fact.The definition 1.55, namely the
formula( 136.1),\linebreak provides us the feature of associativity
for Whitnev sum.\par
Which equivalence class is the unit element for the investigated
semigroup ? It  is easy  to see that \it the equivalence class
 of the trivial vector fibre bundles from the example \rm 1 is the
 unit element for our semigroup. Really  we deal in this example with
one chart U=B and the one trivialisation $ \phi :B\times K^n \to B
\times V$. Therefore this is indeed the unit element. The formula ( 136.1)
then provides again the semigroup operation (as a Whitnev sum ) with
above trivial class.Thus we proved that the Whitnev sum
of the elements of Vect$_n$(BGL (n,K)) forms the structure of the
semigroup .\par
In conclusion we ought  to prove the abelian nature of this semigroup.
But this provide the following two lemmas.
\proclaim{\bf Lemma 1.36}Let $\zeta _1 ,\zeta _2$are the fibre spaces
 over $B_1$ and $B_2$ correspondingly, and $\tau :B_1\times B_2 \to
B_1\times B_2 $ is the map given with the formula $\tau (x,y)=(y,x)$.
Then $\tau *(\zeta _2\times \zeta _1 )\simeq \zeta _1 \times\zeta _2 $.
If $\zeta _1 ,\zeta _2$are the principal (the vector fibre bundles)
correspondingly  then takes place the equivalence of the
principal (the vector fibre bundles) correspondingly.
\endproclaim
\proclaim{\bf Corollary 1.37}Let $\zeta _1 ,\zeta _2$ be the vector
fibre bundles over the same base B. Then $\zeta _1 \bigoplus\zeta _2
\simeq \zeta _2 \bigoplus\zeta _1$.
\endproclaim
\demo{\it Proof}Since $\tau \circ \Delta =\Delta$ then
$$
\zeta _1 \bigoplus\zeta _2= \Delta *(\zeta _1 \times\zeta _2 )\simeq
\Delta *\tau *(\zeta _2\times \zeta _1)=
\Delta *(\zeta _2\times \zeta _1)=\zeta _2 \bigoplus\zeta _1
\tag 137.1
$$
\qed
\enddemo
The inverse element for the given element is absent in the construction
of \linebreak Vect$_n$(BGL (n,K)). \par
Thus we proved the construction of the abelian semigroup on the set
Vect$_n$(BGL (n,K)) of the equivalence classes  of the vector fibre
 bundles. How we
can construct the \it group  structure \rm by given semigroup Vect$_n$(
BGL (n,K))? This is the very interesting task by itself and has many
applications   which we  shall discuss later on.We shall follow now
the monograph [15]of M.F.Atiyah and the monograph [2]of M.M.Postnikov.
\par
Let [2,p.413] the some abelian semigroup M be given.We shall  construct
now  \it the  group of differences \rm GM for the  abelian
semigroup M. This
group consist of the formal differences of the form a-b where $a,b \in M$.
We consider the two such differences a-b and $a_1-b_1$as a equal
differences when  it exists the
such element $c\in M$ that $a+b_1+c=a_1+b+c$. The addition in the  group
of differences is defined
as following:
$$
(a-b)+(c-d)=(a+c)-(b+d)
\tag 138.1
$$
The just defined group of differences is called very often \it the
Grothendieck \linebreak group \rm. This French mathematician utilised
widely the
construction of the  group of differences  and attracted the general
attention to this group. Althouth the above construction was well-known
long before Grothendieck.\par
The formula
$$
\chi :a \mapsto (a+c)-c,\qquad c\in M
\tag 139.1
$$
defines correct the some map
$$
\chi :M \to GM
\tag 139.1a
$$
In definition $\chi (a)= \chi (b)$ when and only when it exists the such
element $c\in M$ that a+c =b+c.\par
Let  the some other group G be given. Then  it turns out that it exists
the only homomorphism $Gf:GM \to G$ for the some homomorphism
$f:M \to G$,
that the diagram
$$
\sarrowlength =.42 \harrowlength
\commdiag{M &\mapright^{\fam6 \chi}& GM \cr
&\arrow (1,-1)\lft {\fam6 f}\quad \arrow (-1,-1)\rt{\fam6 Gf}\cr
& G }
\tag 140.1
$$
is commutative. This was the general scheme for the construction of the
 group of differences. We shall denote in our next statement of the
matter the abelian group of differences corresponding the abelian
semigroup Vect$_n$(BGL (n,K)) with the letter K :K(A)for the some
abelian semigroup A.\par
M.F.Atiyah proposed  the alternative scheme of the definition for the
group of differences. Let $\Delta :A\to A\times A$ be the diagonal
 homomorphism of the semigroups,and let K(A) be the set of the conjugate
 classes of the semigroup $\Delta (A)$ in the  semigroup
$A\times A$.
The set K(A)is the factor-semigroup, but  the permutation of the
co-ordinates  in $A\times A$ generates \it the inverse element \rm in K(A),
therefore K(A)\it is the group\rm . Let us define $\alpha _A: A\to K(A)$
as a composition of the embedding $a\to(a,0)$ with the natural
projection $A\times A \to K(A)$ (for the simplicity we suppose that A
contains zero element).The pair $(K(A),\alpha _A)$ is  therefore
 the functor of the semigroup A, and if $\gamma :A\to B$ is the
 homomorphism of the semigroups then we have the commutative diagram
$$
\commdiag{A &\mapright^{\fam6 \alpha _A}& K(A)\cr
\mapdown\lft {\fam6 \gamma}&&\mapdown \rt{\fam6 K(\gamma)}\cr
B &\mapright^{\fam6 \alpha _B}& K(B)\cr}
\tag 141.1
$$
If B is the group then $\alpha _B$ is the isomorphism.\par
If A is besides that   the \it semiring \rm (i.e.the multiplication,
distributive with respectto  the addition is defined on A) then K(A)
obviously is the ring.\par
It is easy to see then the Whitnev sum as a semigroup operation is the
addition in the ring (the abelian nature of this addition is one of
the demands to the \linebreak ring ).The direct product (136.1) then is
the multiplication in this ring. Thus \linebreak Vect$_n$(BGL (n,K))\it
 has the structure of the semiring\rm and the above scheme allow us to
 construct the ring  which we shall call following M.F.Atiyah as
K(Vect (B)).This is
on the other hand is the group of differences (this is provided  with
the M.F.Atiyah's scheme ).\par
If E $\in $Vect $_n$(BGL (n,K)),then let us denote as [E] the image of
the vector fibre bundle E in the ring K(Vect (B)).We shall write down
very often E instead of [E].\par
Using the M.F.Atiyah's scheme we can see that if X is the some space
(for example the base B of the vector fibre bundle) then
the every element of the group K(Vect (X)) has the form [E]-[F] where
E,F are the vector fibre bundles over X. Let G is the such vector
fibre bundle that the vector fibre bundle $F\bigoplus G$ is trivial.
Let us denote the trivial vector fibre bundle of the dimension n as
$\underline {n}$.
Let $F\bigoplus G =n$. Then  $[E]-[F]=[E]+[G]-([F]+G])=
[E\bigoplus G]-[\underline {n}]$. Thus  every element of the group
K(Vect (X))
has the form $[H]-[\underline {n}]$.\par
Let E,F are the such vector  fibre bundles that [E]=[F].
Then the existence of the such vector  fibre bundle G that
$E\bigoplus G \cong
F\bigoplus G $ follows from the M.F.Atiyah's scheme.
Let G' be the such vector fibre bundle that $G\bigoplus
G'\cong \underline {n}$.\linebreak Then
$E\bigoplus G\bigoplus G'\cong F\bigoplus G\bigoplus G'$,
therefore  $E\bigoplus \underline {n} \cong F\bigoplus \underline {n}$.
If the two
vector fibre bundles become equivalent after the addition to every
of them of the suitable trivial vector fibre bundles, then these
vector fibre bundles are called \it the stable-equivalent vector fibre
bundles \rm. Thus[E]=[F] if and only if E and F are stable-equivalent.
\par
Thus we described  in outline the K-theory - the very beauteful and
elegant thing. The K-theory will serve us as a base for the theory of the
topologic index of the elliptical operator.\par
In the conclusion of our first lecture let us acquaint with the one
more thing. This is the elementary information about \it the sheafs\rm.
The sheafs are also one of the elements of the theory of the topologic
index of the elliptical operator. We follow in the statement of the
sheafs theory  the monograph [17] of F.Hirzebruch. These objects as
our reader this will see soon are very alike on the fibre bundles.
\definition{\bf Definition 1.56} The sheaf $\bold {\Omega}$ (of the
abelian groups)over the topologic space X is the triad $\bold {\Omega }
= (S,\pi, X)$ satisfied the following conditions
\roster
\item S and X are the topologic spaces and $\pi S\to X$ is the onto
continuous map;
\item every point $\alpha \in S$ has the  open neighbourhood N in S
such that $\pi \vert N$ is the homeomorphism between N and the open
neighbourhood of $\pi (\alpha)$ in X.
\endroster
The counterimage $\pi^{-1}(x)$ of the point $x\in X$ is called the
stalk over x and denoted as $S_x$. Every point of S belong to the unique
stalk. Condition (2) means that $\pi$ is the local  homeomorphism and
implies that the topology of S generates the discrete topology on every
stalk.\par
(3)Every stalk has the structure of the abelian group. The group
operation associate the sum $\alpha +\beta \in S_x $ and the difference
$\alpha -\beta \in S_x $to the points $\alpha ,\beta \in S_x$.
The difference depends continuously on $\alpha$ and $\beta$.\par
The word ''continuously" in (3)means that  if $S \bigoplus S$ is
the subset $\{(\alpha ,\beta )\in S\times S;\pi (\alpha)=\pi(\beta)\}$ of
$S\times S$ with the induced topology the map $S \bigoplus S \to S$
defined by $(\alpha ,\beta )\to ( \alpha-\beta)$ is continuous. The
conditions (1)-(3) imply that the zero element $0_x$ of the abelian
group $S_x$ depends continuously on x, i.e. the map $X \to S$ defined
by $x \to 0_x$  is continuous.Similarly the sum  $\alpha +\beta $
depends continuously on $\alpha ,\beta$.
\enddefinition
\remark{\bf Remark 1}Our reader can see easy comparing the definition
1.56 and 1.4 (with acccount of the continuity of map $\pi$ ) that S
has the structure of the total space over X with the
extra structure of the abelian group, the stalk $S_x$ is the  fibre
over $x\in X$ , and thus the sheaf $\bold {\Omega}$ is \it the one of
examples
of the fibre bundles\rm .
\endremark
\remark{\bf Remark 2}What is this - the discrete topology,about what
one mentioned in definition 1.56? (we hope that our reader is acquainted
with the bases of topology; then  he will easy understand this matter)\par
Let [9]X is the arbitrary set and $ {\Cal O}$is the family of all its
subsets. We \it call \rm X the  \it open \rm set. Also the empty set
will in definition  the open  set . Its subsets $ {\Cal O}$ subdivide
on the open and on the closed subsets. And we postulate that the joint and
the crossing of the open subsets are the open subsets in X. Every set
$F\subset X$ is called the closed subset in X if  its supplement X/F is
the open subset in X. In this case the pair $(X,{\Cal O})$ is called
\it the topological space\rm .\par
We introduce now the additional "technical"demand for the subsets
$ {\Cal O}$. Let every set $A \subset X $ be open and closed
simultaneously :\it the open-closed sets \rm . Every set contained
the some point $x \in X$ is its neighbourhood. The family of all
one-point
sets of the set X forms \it the base \rm of the topological space
$(X,{\Cal O})$, i.e. every  different from empty open set is represented
as a joint  of the some subfamily of the family of all one-point sets.
It is evident that this base has the minimal capacity (the number of
elements) among the other bases.\par
The set of all \it cardinal \rm (natural) numbers of the form
$\vert {\Cal B} \vert $ where ${\Cal B}$ is the  base of the topological
space
$(X,{\Cal O})$ has always its minimal element since every set of cardinal
numbers is ordered with the relation $<$. This minimal
cardinal number is called \it the weight \rm of the topological space
$(X,{\Cal O})$ and denoted as $w(X,{\Cal O})$. Therefore in our case
the weight  of $(X,{\Cal O})$ is the weight of the family of all one-point
sets in other words \it it is the capacity of X \rm .\par
The family $ {\Cal B} (x)$ of all neighborhoods of the point x is called
\it the base of the topological space $(X,{\Cal O})$ in the point
\rm
x if it exists the element $U\in {\Cal B} (x)$ for every neighborhood V
of x that $x\in V\subset U$.In our case the family consisted of
unique set $\{x\}$ \it is the base of the topological space $(X,{\Cal O})$
in the point x \rm.\par
The \it character of the point \rm x in the topological space
$(X,{\Cal O})$ is the minimal cardinal number of the form
$\vert{\Cal B} \vert $. This cardinal number is denoted as
$\chi (x,(X,{\Cal O}))$.\it The character of the topological space
$(X,{\Cal O})$
\rm is the \it supreme \rm  of all cardinal numbers
$\chi (x,(X,{\Cal O}))$. This cardinal number is denoted as
$\chi ((X,{\Cal O}))$.If
$\chi ((X,{\Cal O})) \leq \aleph _0$ ,i.e. it is \it the countable
\rm then one say that the topological space $(X,{\Cal O})$ satisfies
\it the first countable axiom \rm ; it means that the countable base
in every point $x\in X$. Thus we draw the conclusion that $(X,{\Cal O})$
\it satisfies the first countable axiom in our case\rm.\par
Further since every subset A  is open-closed in our model every subset
$A\subset X$ coincides with its closure and with its interior. The such
topological space  is called \it the discrete space\rm and ${\Cal O}$is
called \it the discrete topology \rm .\par
It is evident that we can construct the above topology also for every
stalk in S.\par
\endremark
\remark{\bf Remark 3}The condition (3)can be modified to give the
definition of the sheaf for any other algebraic structure on each stalk.
It
sufficient to demand  the continuous nature of the algebraic operations.
It  will very often happen that each stalk of S is the K-module
(for the same ring K). In this  case (3) must be modified to include the
condition :\it the module multiplication associates the point
$k\alpha \in S_x$ to $\alpha \in S_x, k\in K$,and the map $S \to S$
defined by $\alpha \to k\alpha $ is continuous for every $k\in K$. Later
on we shall tacitly assume that \it all sheaves are the sheaves of the
abelian groups  or the K-modules \rm .
\endremark
\definition{\bf Definition 1.57} Let $\bold {\Omega}=(S,\pi, X)$ and
$ \tilde{\bold {\Omega}}=({\tilde S },{\tilde \pi }, X)$ be the two
sheaves over the some space X. The homomorphism $h:\bold {\Omega} \to
\tilde{\bold {\Omega}}$ is defined if
\roster
\item h is the continuous map from S to ${\tilde S }$;
\item $\pi ={\tilde \pi }\circ h $,i.e.h maps the stalk $S_x$ to the
stalk ${\tilde S}_x$ for each $x\in X$ ;
\item the restriction
$$
h_x :S_x\to {\tilde S}_x
\tag 142.1
$$
is the homomorphism of the abelian groups.
\endroster
By (1)and (2) h is the local homeomorphism from S to ${\tilde S}$.
\enddefinition
\definition{\bf Definition 1.58}The presheaf over the some space X
consists of the abelian group $S_U$ for every open set $U \subset X$ and
the homomorphism $r^U_W :S_U \to S_V$ for each pair of open sets
$U,V \subset X$ with $V\subset U$. These groups and homomorphisms
satisfy
the following properies:
\roster
\item if U is empty then $ S_U=0$ is the zero group;
\item the homomorphism $r^U_U: S_U \to  S_U$ is the identity.
If $W \subset V\subset U$ then $r^U_W =r^V_W \circ r^U_V$.
\endroster
\enddefinition
\remark{\bf Remark } By (1)it suffices to define $S_U$ and $r^U_W$
only for non-empty open sets U,V.
\endremark
Every presheaf over X determines the some sheaf over X by the following
construction :\par
a. Let $S_x$ is \it the direct limit \rm of the abelian groups
$S_U,x\in U \subset U $  with respect to the homomorphism $r^U_W$.
In  other
words U \it runs through all open neighborhoods of \rm x. Each element
$f\in S_U$ determines the element $f_x\in S_x$ called the \it germ
\rm of X at x.Every point of $S_x$ is the germ. If U,V are the open
neighborhoods of x and $f\in S_U,g\in S_V$ then $f_x=g_x$ if and only if
there exist the open neighborhood W of x such that $W\subset U$,
$W\subset V$ and $r^U_Wf =r^V_W g$.\par
b. The  direct limit $S_x$ of the abelian groups $S_U$ is itself the
 abelian group. Let S be the union of the groups $S_x$ for different
$x\in X$ and let $\pi :S\to X$ map  the points of $S_x$ to x. Then
S is the set in which the group operations (3) from the definition 1.57 are
defined. \par
c. The topology of S is defined by means of basis. The element
$f\in S_U$ defines the germ $f_y \in S_y$ for every point $y \in U$.
The points
$f_y ,y\in U$ form the subset $f_U \subset S$. The sets $f_U$( U runs
through all open sets of X ,and f runs through all elements of$S_U$)
form the required basis for the topology of S.\par
It is easy to see that by a, b, c the triad ${\bold \Omega}= (S,\pi, X)$
\it is the sheaf of the abelian groups over \rm X. This sheaf is
called \it the  sheaf constructed from the presheaf \linebreak
$\{S _U,r^U_W \}$ \rm.\par
Let  $\varTheta=\{S _U,r^U_W\}$ and ${\tilde \varTheta }=
\{{\tilde S}_U, {\tilde r}^U_W \}$ are the presheaves over X.
The homomorphism h
from $\varTheta$ to ${\tilde \varTheta }$ is the system
$\{h _U\}$ of the homomorphisms $h _U :S _U {\tilde S}_U$
which commute with the
homomorphisms $r^U_W,{\tilde r}^U_W $,i.e.${\tilde r}^U_W
\circ h _U =h _V r^U_V $ for $V\subset U$.\par
The homomorphism h is called the \it monomorphism \rm (\it
epimorphism,\linebreak isomorphism \rm correspondingly) if
every homomorphism
$h _U$ is the monomorphism  ( epimorphism, isomorphism
correspondingly).$\varTheta $ is the subpresheaf of
${\tilde \varTheta }$ if for
each open set U the group $ S_U$ is the subgroup of
${\tilde S}_U$ and $r^U_V$ is the restriction of ${\tilde r}^U_V$ on
$S_U$ for
$V\subset U$.If $\varTheta $ is the subpresheaf of ${\tilde \varTheta }$
then the \it quotient presheaf \rm ${\tilde \varTheta }/\varTheta$
is defined. The latter assigns the factor-group ${\tilde S}_U/S_U$.
If h is the homomorphism from the presheaf $\varTheta $ to the presheaf
${\tilde \varTheta }$ then the kernel of h and the image of h
are defined in the natural way. The kernel of h is the subpresheaf of
$\varTheta $ and associates to every open set U the kernel of $h_U$.
 The image of h is the subpresheaf of ${\tilde \varTheta }$
and associates to every open set U the image of $h_U$.
\definition{\bf Definition 1.59}Let ${\bold \Omega}= (S,\pi, X)$
and $\tilde{\bold {\Omega}}=({\tilde S },{\tilde \pi }, X)$ be the sheaves
over the some space X.  The homomorphism $h :{\bold \Omega} \to
\tilde{\bold {\Omega}}$ is  defined if
\roster
\item h is the continuous map from S to ${\tilde S}$;
\item $\pi ={\tilde \pi }\circ h $,i.e.,h maps the stalk $S_x$ to
the stalk ${\tilde S }_x$ for each $x \in X$;
\item the restriction
$$
h_x :S_x \to{\tilde S }_x
\tag 143.1
$$
is the homomorphism of the abelian groups for each $x \in X$
\endroster
\enddefinition
Let ${\bold \Omega}= (S,\pi, X)$ and $\tilde{\bold {\Omega}}=
({\tilde S },{\tilde \pi }, X)$ are the sheaves constructed from the
presheaves
$\varTheta $ and ${\tilde \varTheta }$.The homomorphism $h:\varTheta
 \to {\tilde \varTheta }$ generates the homomorphism from
${\bold \Omega}$ which is also denoted as h. In order to define this
homomorphism it is sufficient to define the homomorphism $h_x$ as this
was in definition 1.59 : if $\alpha \in S_x$ is the germ at x of the
element $f \in S_U$ then $h_x (\alpha )$  is the germ at x of the
element $h_U(f)\in S_U$. This rule gives the well defined homomorphism
$h_x :S_x \to {\tilde S }_x$ called \it the direct limit \rm of the
homomorphisms $h_U$.\par
The \it section \rm of the sheaf ${\bold \Omega}(S,\pi, X)$ over the
open set U is the continuous map $s:U \to S$ for which
$\pi \circ s U \to U $ is the identity (compare with the section of
fibre bundle!).By (3)of definition 1.56 the set of all sections of
${\bold \Omega}$ over U is the \it abelian group \rm which we denote
as $\Gamma ( U,{\bold \Omega})$. The zero element of this group is the
zero section $x \to O_x$.If s is the section of S over U the image set
$s( U)\subset S$ cuts each stalk $S_x, x\in U$ to exactly one
point.\par
Let now associate the group $\Gamma ( U,{\bold \Omega})$ of  sections
of ${\bold \Omega})$ over U to every open set  $U \subset  X$.
where if
U is empty $\Gamma ( U,{\bold \Omega})$ is the zero group.  If
$V \subset U$ let $r^U_V:\Gamma ( U,{\bold \Omega})\to
\Gamma ( V,{\bold \Omega})$ is the homomorphism which associates
to each section of ${\bold \Omega}$ over U its restriction to V
(if V is
empty we put $r^U_V=0$).The presheaf $\{\Gamma( U,{\bold \Omega}),
r^U_V\}$ is called \it the canonical  presheaf \rm of the sheaf
${\bold \Omega}$.By our scheme of construction of the sheaf from the
presheaf the presheaf $\{\Gamma( U,{\bold \Omega}),r^U_V\}$ defines
the sheaf ; this is again the sheaf ${\bold \Omega}$.In fact by (1)
and (2)of definition 1.56 every  point $\alpha \in S$ belongs to at
least one image set s(U) where s is  the section of ${\bold \Omega}$
over the some open set U. If $s,s'$ are sections over $U,U'$ with
$\alpha \in s(U)\bigcap s(U')$then s agrees with $s'$ in the some open
neighborhood of $x=\pi (\alpha )$.Therefore \it germs at x of the
sections of ${\bold \Omega}$ \it over the open neighborhoods of x are
in one-to-one correspondence with the points of the stalk
$S_x$\rm .\par
Let ${\bold \Omega}$ be the sheaf  constructed from the presheaf
$\varTheta =\{S_U,r^U_V\}$.The element $f\in S_U$ has the germ $f_x$ at x
for every $x\in U$.Let $h_U(f)$ is the section $x\to f_x$ of
${\bold \Omega}$ over U. This defines the homomorphism $h_U :S_U\to
\Gamma( U,{\bold \Omega})$ and whence the homomorphism h from
$\varTheta$ to the canonical  presheaf of ${\bold \Omega}$. In general h
\it neither the monomorphism nor the epimorphism \rm [18, $\S1$,
Propositions 1,2].The homomorphism $\{h_U\}$ from
$\varTheta$ to the canonical  presheaf of ${\bold \Omega}$ induces
the identity isomorphism $h:{\bold \Omega} \to{\bold \Omega}$.
\definition{\bf Definition 1.60}${\bold \Omega}'=(S',\pi *,X)$ is
the subsheaf of ${\bold \Omega}=(S,\pi ,X)$ if
\roster
\item S' is the open set in S;
\item $\pi '$ is the restriction of $\pi $ to $S'$ and maps $S'$
onto X;
\item the stalk $ \pi * ^{-1} (x)= S'\bigcap \pi ^{-1}(x)$ is the
subgroup of the stalk $ \pi ^{-1}(x)$ for all $x \in X$.
\endroster
The condition (1) is equivalent to \par
(1*) Let $s(U)\subset S$ be the image set of the section of
${\bold \Omega}$ over U and $ \alpha \in s(U)\bigcap S'$.\linebreak
Then U contains the open neighbourhood V of $\pi\alpha $ such that
$ s(x) \in S'$ for all $x \in V$.\par
Conditions (1*) and (2) imply that $ \pi *$ is the local  homomorphism
and (3) implies that the group operations in
${\bold \Omega}'$ are continuous. Therefore the triad $(S',\pi *,X)$
is the sheaf. The inclusion of $S'$ in S defines the monomorphism from
${\bold \Omega}'$ to ${\bold \Omega}$ called \it the embedding \rm of
${\bold \Omega}'$ in ${\bold \Omega}$.
\enddefinition
The \it zero sheaf \rm o over X  can be defined up to isomorphism as
a triad $(X,\pi ,X)$ where $\pi$ is the identity map and each stalk
is the zero group. The zero sheaf is the subsheaf of every sheaf
${\bold \Omega}$ over X :let $S'$ is the set $0{\bold \Omega}$ of zero
elements  of the stalks of ${\bold \Omega}$,i.e. $0{\bold \Omega}=s(X)$
where s is the zero element of $\Gamma(X,{\bold \Omega})$.\par
Let ${\bold \Omega}= (S,\pi ,X)$ and $\tilde{\bold \Omega}=
({\tilde S},{\tilde\pi} ,X)$ be the sheaves over X and $h:{\bold \Omega}
\to \tilde{\bold \Omega}$ is the homomorphism. If $S'=
h ^{-1}(0(\tilde{\bold \Omega}))$and $\pi *= \pi \vert S'$ then
$(S',\pi *,X)$ gives
us  the subsheaf $h ^{-1}(0)$ called \it the kernel \rm of  h.
The stalk of the sheaf $h ^{-1}(0)$ over x is the kernel of the
homomorphism
$h_x :S_x \to {\tilde S}_x$. If ${\tilde S}'=h(S)$ and ${\tilde\pi}* =
{\tilde\pi}\vert S'$ then $({\tilde S}',{\tilde\pi}' ,X)$ gives
the subsheaf $h({\bold \Omega})$ of $\tilde {\bold \Omega}$ called
\it the image \rm of h .The stalk of the sheaf $h({\bold \Omega})$ over
x is the image of the homomorphism $h_x$.\par
Let $\{A_i\}$ be the sequence of the groups (or the presheaves or sheaves)
and $\{h_i\}$ is the sequence of the homomorphisms $h_i:A_i
\to A_{i+1}$.(The index i takes all integral values between the two bounds
$n_0,n_1$ which may also be $-\infty $ or $\infty $.Thus $A_i$
is defined for $n_0 <i<n_1$ and $h_i$ is defined for $n_0 <i<n_1-1$ .)
The sequences $ A_i,h_i$ as usually are the \it exact sequences \rm
if the kernel of each  homomorphism is equal to the image of the
previous homomorphism. If all $A_i$ are the presheaves $\{S^{(i)}_U\}$
over the topologic space X then the exactness means that it exists
the  exact sequence of the groups
$$
...\to S^{(i)}_U \to S^{(i+1)}_U \to S^{(i+2)}_U \to ...
\tag 144.1
$$
for every open set $U \subset X$.\par
If the $A_i$ are the sheaves  over X then the exactness means that
the stalks of the sheaves $A_i$ form the exact sequence at every point
$x \in X$. Since the direct limit of the exact sequences is again the
exact sequence [19,Charpter 8]we have
\proclaim{\bf Lemma 1.38}Let us consider the exact sequence
$$
...\to  \varTheta_n \to \varTheta_{n+1} \to \varTheta_{n+2} \to ...
\tag 145.1
$$
of the presheaves over X. Then the generated  sequence of the sheaves
$ {\bold \Omega}_i$ constructed from $ \varTheta_i$ is the
exact sequence of the sheaves over X.
\endproclaim
For example let
$$
0 \to {\bold \Omega}' @> h' >> {\bold \Omega}@> h >> {\bold \Omega}''\to 0
\tag 146.1
$$
is the exact sequence  of the sheaves ${\bold \Omega}'= (S',\pi *,X),
{\bold \Omega}= (S,\pi ,X)$ and ${\bold \Omega}''= (S'',\pi **,X)$
over X.\par
The first 0 denotes the zero subsheaf of ${\bold \Omega}'$,the first
arrow  denotes the embedding of 0 in ${\bold \Omega}'$.Therefore
the exactness implies that $h'$ is the monomorphism and can be regarded
as the embedding of the subsheaf ${\bold \Omega}'$
in ${\bold \Omega}$. The final 0 denotes the zero subsheaf of
${\bold \Omega}''$, the final arrow denotes the trivial homomorphism
which maps
each stalk of ${\bold \Omega}''$ to its zero element. Therefore
exactness implies that h is the epimorphism. The exact sequence (146.1)
gives the corresponding exact sequence of the stalks  over x:
$$
0 \to S_x' @> h'_x >> S_x @> h _x>> S''_x \to 0
\tag 147.1
$$
for every $x \in X$.\par
The group $S''_x$ is  isomorphic to the factor-group $S_x/S'_x$. It
is easy to check that $S''$ has the factor-topology (recall  our remark in
theorem 1.8): the subsets of $S''$ are open if and only if their
counterimages under h are the open sets in S. This shows that the sheaf
${\bold \Omega}$ and its subsheaf ${\bold \Omega}'$ given at most
one sheaf ${\bold \Omega}''$ for which the sequence (146.1) is exact.
It is
possible to prove the existence of such the ${\bold \Omega}''$ by
defining first the presheaf for ${\bold \Omega}''$.\par
Let ${\bold \Omega}'$ be the subsheaf of ${\bold \Omega}$ and U is
the open set of X. The group $\Gamma(U,{\bold \Omega}')$ of the sections
of ${\bold \Omega}'$ over U is then  the subgroup of $\Gamma(U,{\bold
\Omega})$,the group of the sections of ${\bold \Omega}$ over U. We
define $ S''_U= \Gamma(U,{\bold \Omega})/\Gamma(U,{\bold \Omega})'$
such that it exists the sequence
$$
0 \to \Gamma(U,{\bold \Omega}')\to \Gamma(U,{\bold \Omega})\to S''_U \to 0
\tag 148.1
$$
If V is the open set contained in U the restriction homomorphism
$\Gamma(U,{\bold \Omega})\to \Gamma(V,{\bold \Omega})$ maps the subgroup
$\Gamma(U,{\bold \Omega}')\subset \Gamma(U,{\bold \Omega})$ to the
subgroup $\Gamma(V,{\bold \Omega}')\subset \Gamma(V,{\bold \Omega})$
and induces the homomorphism $r^U _V:S''_U \to S''_V$. The presheaf
$\{S''_U, r^U _V\}$ is the quotient of the canonical presheaf of
${\bold \Omega}$.Let ${\bold \Omega}''$ is the sheaf constructed from
the presheaf $\{S''_U, r^U _V\}$.The exact sequence (148.1) of
presheaves generates  by  lemma 1.38 the exact sequence of sheaves.
We collect our results in the following theorem :
\proclaim {\bf Theorem 1.39}Let ${\bold \Omega}$ be the sheaf over the
topologic space X and ${\bold \Omega}'$ is the subsheaf of
${\bold \Omega}$ with the embedding $h':{\bold \Omega}'\to{\bold \Omega}$.
It exists the sheaf ${\bold \Omega}''$ over X  unique up to
isomorphism  for which it exists the exact sequence
$$
0 \to {\bold \Omega}' @> h' >> {\bold \Omega}@> h >> {\bold \Omega}''\to 0
\tag 149.1
$$
The homomorphism $h_x$ at each point $x \in X$ gives the isomorphism
between the factor-group $S_x/S'_x$ and the stalk $S''_x$ of
${\bold \Omega}''$ over x.
\endproclaim
\remark {\bf Remark} We obtain the exact sequence
$$
0 \to \Gamma(U,{\bold \Omega}')\to \Gamma(U,{\bold \Omega})\to
\Gamma(U,{\bold \Omega}'') \to 0
\tag 150.1
$$
from (149.1)
\endremark
By (148.1) $S''_U$ is the subgroup of $\Gamma(U,{\bold \Omega}'')$
consisting of all sections ${\bold \Omega}'')$ over U which are images of
the sections of $\Gamma(U,{\bold \Omega})$ over U.\par
Let ${\bold \Omega}=(S,\pi ,X)$ be the sheaf over X and let Y be the
subset of X. If the subset $\pi ^{-1}(Y)$ of S is given the generated
topology of the triad $(\pi ^{-1}(Y),\pi \vert \pi ^{-1}(Y),Y)$ defines
in the natural way the sheaf ${\bold \Omega}\vert Y$ over Y called
the \it restriction \rm of ${\bold \Omega}$ to Y.
\proclaim{\bf Theorem 1.40} Let Y be the closed  subset of the topological
space X and ${\bold \Omega}=(S,\pi ,X)$ be the sheaf over Y.
It exists the sheaf $\hat {\bold \Omega}$ over X unique up to isomorphism
such that $\hat {\bold \Omega}\vert Y={\bold \Omega}$ and
$\hat {\bold \Omega}\vert (X-Y)=0$.The groups $\Gamma(U,\hat{\bold \Omega})$
 and $\Gamma(U \bigcap Y,{\bold \Omega})$ are
isomorphic for
any open set $U \subset X$ ($\hat {\bold \Omega}$ is called the trivial
extension of ${\bold \Omega}$ to X).
\endproclaim
\demo{\it Proof} The uniqueness follows immediately from the properties
of $\hat {\bold \Omega}$:if \linebreak $\hat {\bold \Omega}=
({\hat S},{\hat \pi},X)$ then ${\hat S}=S \bigcup ((X-Y)\times 0),
{\hat \pi}(\alpha)=\pi (\alpha)$ for $\alpha \in S,
{\hat \pi} (a\times 0)=a$ for $a\in X-Y$ and therefore the stalk
${\hat S}_x= {\hat \pi}{-1}(x)$ is equal to $ \pi{-1}(x)$
for $x\in Y$ and equal to the zero group for $x\in X-Y$. The sets
$s(U\bigcap Y)\bigcup ((U \bigcap (X-Y))\times 0)$, for arbitrary open
sets $U \subset X$ and arbitrary sections s of ${\bold \Omega}$ over U,
define the basis for the topology of ${\hat S}$.This completes the
construction of $\hat {\bold \Omega}$.One can also define $\hat
{\bold \Omega}$ by means of presheaf : let us associate the group
${\hat S}_U =\Gamma (U \bigcap Y,{\bold \Omega})$ to each open set
$U \subset X$ and the restriction homomorphism
$r^U_V :\Gamma (U \bigcap Y,{\bold \Omega}\to \Gamma (V
\bigcap Y,{\bold \Omega})$ to each pair of open sets U,V with
$V\subset U$. Since Y
is closed  each point $x\in X-Y$ has the open neighborhood U for which
$U\bigcap Y$ is empty and ${\hat S}_U=0$.Therefore the sheaf
$\hat {\bold \Omega}$ constructed from the presheaf
$\{ {\hat S}_U,r^U_V \}$ has $\hat {\bold \Omega} \vert Y={\bold \Omega}$
and
$\hat {\bold \Omega} \vert (X- Y)=0$.In fact $\{ {\hat S}_U,r^U_V \}$ is
the canonical presheaf of $\hat {\bold \Omega}$. \qed
\enddemo
\remark{\bf Remark}.Let us suppose that the stalk of ${\bold \Omega}$ has
the non-zero element at the some boundary point of Y.
Then ${\hat S}$ is \it the non-Hausdorff space.
\endremark
\it The examples of sheaves.\rm \par
\example {\bf Example 1}Let X be the topologic space and A be the abelian
 group. The \it constant sheaf \rm over X  with stalk A is defined
by the triad $X \times A, \pi, X)$ and is also denoted by A.
 Here $ \pi :X \times A \to X$ is the projection from the product
 $X \times A$
where A has the discrete topology. The sum and difference of points
 (x,a) and $(x',a')$ in $X \times A$ are equal to $(x,a\pm a')$. The
reader can compare easy this case with the case of the trivial fibre
bundles.
\endexample
\example {\bf Example 2} Let X be the topologic space. Let us associate
the additive group $S_U$ of \it all continuous complex valued
functions \rm to U ,-  the non-empty open set in X.  The homomorphism
 $r^U_V: S_U \to S_V$ is defined for $V \subset U$ by taking
the restriction on V of each function defined on U. Let ${\bold C}_c$
 be the sheaf constructed from the presheaf $\{S_U,r^U_V\}$.Then
${\bold C}_c$ is called \it the sheaf of germs of local complex valued
 continuous functions \rm .
\endexample
The sheaf of germs of local \it never zero\rm complex valued continuous
functions is defined similarly: let us associate the abelian group
$S*_U$ of never zero complex valued continuous functions to U ,-  the
 non-empty open set in X. The group operation is the ordinary
multiplication. It exists the homomorphism $S_U \to S*_U$ which
associates the function $e^{2\pi i f} \in S*_U$ to each function $f\in S_U$.
This defines the homomorphism $\{S_U,r^U_V\} \to \{S*_U,r^U_V\}$ of
the presheaves and hence the
homomorphism ${\bold C}_c \to {\bold C}*_c $
the sheaves. The kernel of the homomorphism ${\bold C}_c \to
{\bold C}*_c $ is the subsheaf of${\bold C}_c$ isomorphic to the constant
sheaf
over X with the stalk,-  the additive group ${\bold Z}$ of integer
numbers. Every point $z_0$ of the multiplicative group ${\bold C}*_c$
of non-zero complex numbers has the open neighbourhood in which the
single branch can be chosen for log z. If k is the germ of ${\bold C}*_c$
then $(2\pi i)^{-1}$ log k is the germ of ${\bold C}_c$ which maped
to k under ${\bold C}_c \to {\bold C}*_c $. Therefore it exists the
exact sequence of sheaves over X
$$
0 \to {\bold Z}\to {\bold C}_c \to {\bold C}*_c \to 0
\tag 151.1
$$
\example {\bf Example 3}Let now X be the n-dimensional differentiable
manifold. We adopt now the following definition ([20,$\S$ 1] and
[21]).X is the Hausdorff space with the countable basis (look remark
2 to definition of sheaves).The certain real valued functions at each
point $x \in X$ are distinguished and called \it the differentiable
\rm at x. Every function is defined on some open neighborhood of x and
the following axiom is satisfied :
\proclaim {\it Axiom} It exists the open neighborhood U of x  and the
 homeomorphism g from U onto the open subset of ${\bold R}^ n$ such
that, for all $y \in U$, if f is the real valued  function defined on
the neighborhood V of y and $h=g\vert U \bigcap V$,then f is
differentiable at y if and only if $fh^{-1}$ is
$C^{\infty}$-differentiable at g(y).
\endproclaim
Here $fh^{-1}$ is the real valued  function defined on the open
 neighborhood of g(x) in ${\bold R}^ n$. It is $C^{\infty}$-differentiable
 at
g(x) if all the partial derivatives are continuous in the some
neighborhood of g(x).\par
The homeomorphism g which satisfies this axiom  is called \it the
admissible chart  \rm of the differentiable manifold X.\par
If X is the differentiable manifold ,and U is the open set of X,
let $S_U$ be the additive group of complex valued  functions differentiable
on U(the complex valued  function is differentiable if and only if
its real and imaginary parts are differentiable). The presheaf
$\{S_U,r^U_V\}$ defines the sheaf ${\bold C}_b$ :\it the sheaf of
germs of local complex valued differentiable functions\rm. Similarly
the sheaf ${\bold C}*_b$ of germs of local never zero complex valued
differentiable functions  is defined. It exists the exact sequence
of sheaves over X
$$
0 \to {\bold Z}\to {\bold C}_b \to {\bold C}*_b \to 0
\tag 152.1
$$
\endexample
\example {\bf Example 4}Now let X be the n-dimensional complex manifold.
The definition is analogous to the definition of differentiable
manifold [22].X  is the Hausdorff space with the countable basis (as the
 metrical space!).The certain complex valued  functions at each
point $x \in X$ are distinguished and called \it the holomorphic \rm or
 \it the complex analytic \rm at x. Every function is defined
on some open neighbourhood of x and the following axiom is satisfied :
\proclaim {\it Axiom} It exists the open neighbourhood U of x  and the
homeomorphism g from U onto the open subset of ${\bold C} _n$ such
that for all $y \in U$ if f is the complex  valued  function defined on
the neighbourhood V of y and $h=g\vert U \bigcap V$ then f is
holomorphic at y if and only if $fh^{-1}$ is holomorphic at g(y).
\endproclaim
The homeomorphism g which satisfies this axiom  is called \it the
admissible chart  \rm of the complex manifold X. The admissible charts
of the n-dimensional complex manifold X can be used in a natural way
for definition of 2n-dimensional differentiable manifold with the
same underlying space X.\par
If X is the complex manifold let $S_U$ be the additive group of (complex
valued) functions holomorphic on U. This group defines
the sheaf ${\bold C}_ \omega $:\it the sheaf of germs of local holomorphic
functions  \rm .Similarly the sheaf ${\bold C}*_\omega $
of germs of local never zero  holomorphic functions is defined. It
exists the exact sequence of sheaves over X
$$
0 \to {\bold Z}\to {\bold C}_ \omega \to {\bold C}*_ \omega \to 0
\tag 153.1
$$
\endexample
\remark {\bf Remark}
The sheaves ${\bold C}_c,{\bold C}_b,{\bold C}_ \omega $ can also be
regarded as a sheaves of the ${\bold C}$-modules. All sheaves in the
exact sequences (151.1- 153.1) are however regarded as the sheaves of
the abelian groups. The presheaves used for construction of
${\bold C}_c,{\bold C}*_c,{\bold C}_b ,{\bold C}*_b ,{\bold C}_ \omega ,
{\bold C}*_ \omega $ are all canonical presheaves. For example
$\Gamma (U,{\bold C}_c)$ is the additive group of all complex valued
continuous functions defined on U.
\endremark
\Refs
[1] Sh. Kobayashi and K. Nomizu \it  Foundations of differential
geometry \bf vol. 1.\rm Interscience publications.
New York London ,1963\par
[2]M.M. Postnikov  \it Lektsii \it po \it geometrii.\it Semestr
\rm 4 \it Differentsialnaja geometrija \rm Moskow, Nauka,
1988 \par
[2a]M.M. Postnikov  \it Lektsii \it po  \it geometrii.\it Semestr
\rm 3 \it Gladkie mnogoobrazija  \rm Moskow, Nauka, 1987 \par
[3]D.M. Gitman and I.V. Tjutin \it Kanonicheskoe kvantovanie polej so
svjazjami \rm Moskow, Nauka, 1986 \par
[4]D.V. Volkov, V.I. Tkach \it Pisma JETF, \rm 1980, \bf vol.32. \rm ,
vipusk 11, pp.681-684.; D.V. Volkov, V.I. Tkach ,TMF,1982,
\bf vol.52.\rm N2,pp.171-180 \par
[5]D.V. Volkov,D.P. Sorokin, V.I. Tkach \it Pisma JETF \rm ,1983,\bf vol.38.
\rm, vipusk 8,pp.397-399; \linebreak D.V. Volkov,
D.P.Sorokin, V.I.Tkach
,TMF, 1984, \bf vol.61.\rm N2,pp.241-253 \par
[6]I.P. Volobujev,Yu.A. Kubyshin,J.M. Mourao,G. Rudolph  \it Fizika
elementarnix chastiz i atomnogo jadra \rm ,1989,\bf vol. 20.\rm, vipusk 3
 \par
[7]A.S. Schwarz \it Kvantovaja teorija polja i topologija \rm Moskow,
Nauka,1989 \par
[8]A.I. Maltsev \it Osnovi linejnoj algebri,\rm Moskow,Nauka,1970\par
[9]R. Engelking \it  General topology \rm Panstwowe Wydawnictwo Naukowe,
Warszawa,1977\par
[10]Robert M. Switzer \it  Algebraic topology- homotopy and homology
\rm Springer-Verlag, Berlin-Heidelberg-New York, 1975\par
[11]Whithehead J.H.C.\it Combinatorial homotopy\rm ,Bull. Amer. Math. Soc.,
1949, \bf vol.55
 \rm, p.p. 213-245, 453-496 \par
[12]Regge T. ,Nuovo Cimento, \bf 19\rm p.558, 1961\par
[13] Hawking S.W., Nuclear Phys.,\bf B 114\rm ,p.349,1978 \par
[14] A.N. Kolmogorov,S.V. Fomin \it Elementi teorii funktsij i
funktsionalnogo analiza \linebreak
\rm Moskow, Nauka, 1976\par
[15]\it K-theory \rm , lektures by M.F. Atiyah, notes by D.W. Anderson,
Harward Uniwersity,\linebreak Cambridge, Mass,1965 \par
[16]D.P. Jelobenko,A.I. Stern \it Predstavlenija grupp Li \rm Moskow, Nauka,
1983\par
[17]F. Hirzebruch \it topologic methods in algebraic geometry \rm
(with additional \linebreak section by A. Borel) , Springer-Verlag,
Berlin-Heidelberg-New York,1966 \par
[18]Serre J.-P. \it Faisceaux algebraiques coherents \rm Ann. Math.
\bf 61 \rm,p.197-278, 1955 \par
[19]Eilenberg S. and Steenrod N.\it Foundations of algebraic topology
\rm Princeton Mathematical Series \bf 15 \rm, Princeton University
Press,1952 \par
[20]Rham G. $_{De}$ \it Varietes differentiables \rm Act. Sci. et Ind.
1222, Paris: Hermann, 1954  \par
[21]Lang S. \it Introduction  to differentiable manifolds  \rm New
York: Interscience, 1962 \par
[22]Well A. \it Varietes kaehlerieness \rm Act.Sci.et Ind.1267,
Paris: Hermann,1958 \par
\endRefs
\enddocument